%% file: amuld.tex
\documentclass[a4paper,11pt]{article}
\pdfoutput=1 

\usepackage{jheppub} 

\usepackage[T1]{fontenc} 
\usepackage[utf8]{inputenc}
\usepackage{graphicx}
\usepackage{amssymb}
\usepackage{amsmath}
\usepackage{xcolor}
\usepackage{booktabs}
\usepackage{placeins}
\usepackage{siunitx}

\usepackage{hyperref}
\hypersetup{%
	pdftitle    = {The hadronic vacuum polarization contribution to the muon g-2 at long distances},
}%
\usepackage[noabbrev]{cleveref}

\input{definitions}

\title{\boldmath The hadronic vacuum polarization contribution to the muon $g-2$ at long distances}

\author[a,b]{Dalibor Djukanovic,}
\author[c]{Georg von Hippel,}
\author[d]{Simon Kuberski,}
\author[c,a,b]{Harvey B.\ Meyer,}
\author[a,b]{Nolan Miller,}
\author[c]{Konstantin Ottnad,}
\author[c]{Julian Parrino,}
\author[e]{Andreas Risch,}
\author[c,a,b]{Hartmut~Wittig}

\affiliation[a]{Helmholtz-Institut Mainz, Johannes Gutenberg-Universit\"at Mainz, 55099 Mainz, Germany}
\affiliation[b]{GSI Helmholtz Centre for Heavy Ion Research, 64291 Darmstadt, Germany}
\affiliation[c]{PRISMA$^+$ Cluster of Excellence and Institut f\"ur Kernphysik, Johannes Gutenberg-Universit\"at Mainz, 55099 Mainz, Germany}
\affiliation[d]{Theoretical Physics Department, CERN, 1211 Geneva 23, Switzerland}
\affiliation[e]{Department of Physics, University of Wuppertal, Gaussstr. 20, 42119 Wuppertal, Germany}

\emailAdd{simon.kuberski@cern.ch}

\abstract{
We present our lattice QCD result for the long-distance part of the
hadronic vacuum polarization contribution, $\aLD$, to the muon $g-2$
in the time-momentum representation. This is the numerically dominant,
and at the same time the most challenging part regarding statistical
precision.
Our calculation is based on ensembles with dynamical up, down and
strange quarks, employing the O($a$)-improved Wilson fermion action with
lattice spacings ranging from $0.035-0.099$\,fm.
In order to reduce statistical noise in the long-distance part of the
correlator to the per-mille level, we apply low-mode averaging and
combine it with an explicit spectral reconstruction.
Our result is $\aLD = 423.2(4.2)_{\rm stat}(3.4)_{\rm syst}\times
10^{-10}$ in isospin-symmetric QCD, where the pion \mbox{decay} constant is
used to set the energy scale.
When combined with our previous results for the short- and
intermediate-distance window observables and after including all
sub-dominant contributions as well as isospin-breaking corrections, we
obtain the total leading-order hadronic vacuum polarization
contribution as 
$\ahvp=724.5(4.9)_{\rm stat}(5.2)_{\rm syst}\times 10^{-10}$. 
Our result displays a tension of 3.9~standard deviations
with the data-driven estimate published in the 2020 White Paper, but
leads to a SM prediction for the total muon anomalous magnetic moment
that agrees with the current experimental average.}

\dedicated{\normalfont CERN-TH-2024-196, MITP-24-080}

\begin{document} 
\maketitle
\flushbottom

\input{sec_intro}

\input{sec_setup}
\input{sec_results}
\input{sec_conclusion}

\acknowledgments{We thank Andrew Hanlon, Ben Hörz, Daniel Mohler,
  Colin Morningstar and Srijit Paul for the collaboration on the
  data generation and initial analysis for the spectral reconstruction
  of the isovector contribution. 
  We also thank Volodymyr Biloshytskyi, Dominik Erb, Franziska Hagelstein and Vladimir Pascalutsa
  for an ongoing collaboration on computing electromagnetic corrections to hadronic vacuum polarization.
  Additionally, we thank Laurent Lellouch for discussions on isospin-breaking effects.
  Calculations for this project have been performed on the HPC
  clusters Clover and HIMster-II at Helmholtz Institute Mainz and
  Mogon-II and Mogon-NHR at Johannes Gutenberg-Universität (JGU)
  Mainz, on the HPC systems JUQUEEN and JUWELS and on the GCS
  Supercomputers HAZELHEN and HAWK at Höchstleistungsrechenzentrum
  Stuttgart (HLRS).
  The authors gratefully acknowledge the support of the Gauss Centre
  for Supercomputing (GCS) and the John von Neumann-Institut für
  Computing (NIC) projects HMZ21, HMZ23 and HINTSPEC at JSC, as well
  as projects GCS-HQCD and GCS-MCF300 at HLRS. We also gratefully
  acknowledge the scientific support and HPC resources provided by
  NHR-SW of Johannes Gutenberg-Universität Mainz (project NHR-Gitter).
  This work has been supported by Deutsche Forschungsgemeinschaft
  (German Research Foundation, DFG) through Project HI~2048/1-2
  (Project No.\ 399400745) and through the Cluster of Excellence
  ``Precision Physics, Fundamental Interactions and Structure of
  Matter'' (PRISMA+ EXC 2118/1), funded within the German Excellence
  strategy (Project No.\ 39083149). This project has received funding
  from the European Union's Horizon Europe research and innovation
  programme under the Marie Sk\l{}odowska-Curie grant agreement
  No.\ 101106243.
  We are grateful to our colleagues in the CLS initiative for sharing
  ensembles.
  The \texttt{pyerrors} package \cite{Joswig:2022qfe} relies on
  \texttt{numpy} \cite{harris2020array} and \texttt{autograd}
  \cite{maclaurin2015autograd}.  Plots have been generated with
  \texttt{matplotlib} \cite{Hunter:2007} and \texttt{gnuplot}
  \cite{gnuplot}. 
}

\appendix

\input{app_scheme}
\input{app_blinding}
\input{app_lma}
\input{app_spec}
\input{app_tables}

\bibliographystyle{jhep_collab}
\bibliography{biblio}

\end{document}

%% file: definitions.tex
\renewcommand{\vec}[1]{\boldsymbol{#1}}
\newcommand{\be}{\begin{equation}}
\newcommand{\ee}{\end{equation}}
\newcommand{\ba}{\begin{eqnarray}}
\newcommand{\ea}{\end{eqnarray}}
\newcommand{\la}{\label}

\newcommand{\half}{{\textstyle\frac{1}{2}}}

\newcommand{\txts}{\textstyle}

\newcommand{\MeV}{{\rm{MeV}}}

\newcommand{\cv}{c_{\rm V}}

\newcommand{\ahvp}{a_\mu^{\rm hvp}}

\newcommand{\bea}{\begin{eqnarray}}
\newcommand{\eea}{\end{eqnarray}}
\newcommand{\eq}[1]{eq.\,(\ref{#1})}

\newcommand{\psib}{\overline{\psi}}

\newcommand{\amu}{a_\mu^{\mathrm{hvp}}}
\newcommand{\amul}{a_\mu^{\mathrm{hvp,ud}}}

\newcommand{\loc}{{\scriptstyle\rm L}}
\newcommand{\cons}{{\scriptstyle\rm C}}
\newcommand{\mloc}{ {\scriptscriptstyle\rm L} }
\newcommand{\mcons}{ {\scriptscriptstyle\rm C} }

\newcommand{\Xpi}{y}
\newcommand{\XK}{z}

\newcommand{\Xa}{\mathbf{a}}

\def\af#1#2{a_\mu^{{#1},{#2}}}

\def\Gf#1#2{G^{({#1},{#2})}}
\newcommand{\aLD}{(a_\mu^{\rm hvp})^{\rm LD}}
\newcommand{\aID}{(a_\mu^{\rm hvp})^{\rm ID}}
\newcommand{\aSD}{(a_\mu^{\rm hvp})^{\rm SD}}

\def\ahvpf#1#2{a_\mu^{{#1},{#2}}}

\def\aIDf#1#2{(a_\mu^{{#1},{#2}})^{\rm ID}}
\def\aLDf#1#2{(a_\mu^{{#1},{#2}})^{\rm LD}}

%% file: sec_intro.tex
\section{Introduction \label{s:intro}}

For many years the tension between the experimentally measured muon
anomalous magnetic moment $a_\mu\equiv\frac{1}{2}(g-2)_\mu$ and its
theoretical prediction has been one of the most promising hints for
physics beyond the Standard Model. The largest share of the
uncertainty in the Standard Model prediction arises through the
leading-order hadronic vacuum polarization (HVP) contribution,
$\ahvp$. In the traditional data-driven method, which forms the
basis for the consensus value reported in the 2020 White Paper by the
Muon $g-2$ Theory Initiative \cite{Aoyama:2020ynm}, one obtains
$\ahvp$ from a dispersion integral over the experimentally measured
hadronic cross section, $e^+e^-\to\rm hadrons$
\cite{Davier:2017zfy,Keshavarzi:2018mgv,Colangelo:2018mtw,Hoferichter:2019mqg,Davier:2019can,Keshavarzi:2019abf}. However,
since the publication of the White Paper, this approach has been
challenged on two fronts (see, e.g.,
\cite{Wittig:2023pcl,Kuberski:2023qgx}): Firstly, the cross section
for the dominant channel $e^+e^-\to\pi^+\pi^-$ measured recently by
CMD-3 \cite{CMD-3:2023alj,CMD-3:2023rfe} is significantly enhanced
relative to all other experiments, yielding an estimate for $\ahvp$
that is largely compatible with the latest direct measurement of
$a_\mu$ reported by the E989 experiment
\cite{Muong-2:2021ojo,Muong-2:2023cdq}. Secondly, lattice QCD
calculations have produced precise results for $\ahvp$
\cite{Borsanyi:2020mff} and the so-called intermediate window
observable $\aID$
\cite{Borsanyi:2020mff,Lehner:2020crt,Wang:2022lkq,Aubin:2022hgm,Ce:2022kxy,ExtendedTwistedMass:2022jpw,FermilabLatticeHPQCD:2023jof,RBC:2023pvn}
indicating a strong tension with estimates derived from $e^+e^-$ cross
sections published prior to CMD-3. Tracing the origin(s) of these
tensions and their possible resolution is the subject of intense
research.

In this paper, we report the results of a new precision calculation of
$\ahvp$ in lattice QCD. The main ingredient, which we describe in full
detail in the following sections, is the fully blinded calculation of
the long-distance window observable $\aLD$ in isospin-symmetric QCD (isoQCD),
for which we obtain
\begin{align}
  \aLD=(423.2\pm4.2_{\rm stat}\pm3.4_{\rm syst})\times10^{-10}\,.
\end{align}
When combined with our previous results for the short- and
intermediate-distance window observables
\cite{Kuberski:2024bcj,Ce:2022kxy}, we obtain the total light-quark
connected contribution as
\begin{align}
  (\amu)^{ud,\,\rm conn}
  =(675.7\pm4.1_{\rm stat}\pm3.7_{\rm syst})\times 10^{-10}\,,
\end{align}
which disagrees with the corresponding data-driven evaluation
\cite{Boito:2022dry,Benton:2024kwp} by more than five standard deviations.
After including all sub-leading contributions and accounting for
isospin-breaking corrections we finally arrive at
\begin{align}
  \ahvp=(724.5\pm4.9_{\rm stat}\pm5.2_{\rm syst})\times10^{-10}\,.
\end{align}
This result differs from the White Paper estimate for $\ahvp$ by
3.9~standard deviations, whilst being compatible with the current
experimental average for $a_\mu$. It is also higher than the lattice
estimate by the BMW collaboration \cite{Borsanyi:2020mff}, as well as
their recent update \cite{Boccaletti:2024guq}, which partly relies on
the data-driven method.

This paper is organized as follows: In section~\ref{s:setup} we
describe the details of our calculation, including our noise-reduction
strategy, the extrapolation to the physical point and the
determination of finite-volume corrections. Our main results 
are presented in section~\ref{s:res}, including the long-distance
window observables and the total HVP contribution in isospin-symmetric QCD,
as well as the correction for isospin-breaking effects that must be
added to arrive at our final result for $\ahvp$. After presenting our
conclusions, we discuss additional computational details in several
appendices, including our choice of hadronic scheme
(appendix~\ref{a:scheme}), our blinding strategy
(appendix~\ref{a:blinding}), the use of low-mode averaging as
noise-reduction strategy (appendix~\ref{a:lma}) and the spectral
reconstruction of the long-distance tail of the vector correlator
(appendix~\ref{a:spec}). Detailed results for the long-distance
contributions and ancillary information on individual ensembles are
collected in appendix~\ref{a:tab}.

%% file: sec_setup.tex
\section{Setup \label{s:setup}}
\subsection{Basic definitions}
We employ the standard time-momentum representation (TMR)
\cite{Bernecker:2011gh} to express the HVP contribution $\ahvp$ as the
Euclidean time integral of the spatially summed correlator $G(t)$ of
the electromagnetic current, $J_\mu^\gamma$, convoluted with an
analytically known kernel function, i.e.
\begin{align}\label{eq:TMRdef}
  \ahvp&=\left(\frac{\alpha}{\pi}\right)^2\int_0^{\infty}dt\,
  \tilde{K}(t;\,m_\mu)G(t)\,,\quad  
  \delta_{kl}G(t)=-\int d^3x\,\left\langle
  J_k^\gamma(t,\vec{x})J_l^\gamma(0) \right\rangle \\
  J_\mu^\gamma&=\frac{2}{3}\bar{u}{\gamma_\mu}u
  -\frac{1}{3}\bar{d}{\gamma_\mu}d -\frac{1}{3}\bar{s}{\gamma_\mu}s
  +\frac{2}{3}\bar{c}{\gamma_\mu}c+\ldots \,.
\end{align}
The electromagnetic current can be written conveniently with the help
of matrices $T^m$ acting in flavour space. Adopting the notation
\begin{align}\label{eq:current}
  J_\mu^m\equiv\overline\psi\,T^m\gamma_\mu\,\psi,\qquad
  \overline\psi=(\bar{u},\,\bar{d},\,\bar{s},\,\bar{c},\,\bar{b}) 
\end{align}
we describe the $(u,\,d,\,s)$ flavour sector by setting
\begin{align}
  T^m={\textstyle\frac{1}{2}}\lambda^m\oplus
    \boldsymbol{0},\quad m=1,\ldots,8\,,
\end{align}
where $\lambda^m$ denote the Gell-Mann matrices, and $\boldsymbol{0}$ the null matrix of size $2\times2$.
The charm and bottom quark currents are defined by $T^{\rm c}={\rm
  diag}(0,\,0,\,0,\,1,\,0)$ and $T^{\rm b}={\rm
  diag}(0,\,0,\,0,\,0,\,1)$, respectively. The generic correlator
$G^{(m,\,n)}$ of the currents $J_\mu^m$ and $J_\mu^n$ is then given by
\begin{align}
  \delta_{kl}G^{(m,\,n)}(t)=-\int d^3x \,\left\langle J_k^m(t,\vec{x})J_l^n(0)
  \right\rangle \,.
\end{align}
Setting $m=\gamma$ corresponds to identifying $T^m$ with the physical
quark charge matrix, i.e. $T^\gamma={\rm
  diag}(\frac{2}{3},\,-\frac{1}{3},\,-\frac{1}{3},\,\frac{2}{3},\,-\frac{1}{3})$,
which yields the following decomposition of the electromagnetic current
correlator $G(t)\equiv G^{(\gamma,\,\gamma)}(t)$:
\begin{align}
  \Gf{\gamma}{\gamma}=\Gf{3}{3} +\frac{1}{3}\Gf{8}{8}
  +\frac{4}{9}\Gf{\rm{c}}{\rm{c}}_{\rm conn}
  +\frac{2}{3\sqrt{3}}\Gf{\rm{c}}{8}
  +\frac{4}{9}\Gf{\rm{c}}{\rm{c}}_{\rm disc}
  +\frac{1}{9}\Gf{\rm{b}}{\rm{b}}_{\rm conn} +\ldots\,.
\end{align}
Here the subscripts denote the quark-connected and -disconnected
contributions, and the ellipsis stands for contributions too small to
be relevant in this work. Correspondingly, we can define separate TMR
integrals for each correlator $\Gf{m}{n}$ as
\begin{align}\label{e:def_amu}
  \af{m}{n}&=\left(\frac{\alpha}{\pi}\right)^2\int_0^{\infty}dt\,
  \tilde{K}(t;\,m_\mu)\Gf{m}{n}(t)\,.
\end{align}
In this way, we recover the isovector contribution as $\af{3}{3}$,
while the isoscalar contribution is given by $\frac{1}{3}\af{8}{8}$. 

Our focus in this paper is the long-distance window observable,
$\aLD$, first defined in ref.\,\cite{Blum:2018mom}, which is obtained
by multiplying the integrand in \eq{eq:TMRdef} with an additional
factor $\Theta(t,\,d_1,\,\Delta)$:
\begin{align}
  \aLD&=\left(\frac{\alpha}{\pi}\right)^2\int_0^{\infty}dt\,
  \tilde{K}(t;\,m_\mu)\,G(t)\,\Theta(t,\,d_1,\,\Delta)\,,
\end{align}
where 
\begin{align}
  \Theta(t,\,d_1,\,\Delta)=\half\left(1+\tanh[(t-d_1)/\Delta]\right)
\end{align}
is a smoothed step function at $t\approx d_1$ with width
$\Delta$. With this convention the standard long-distance window is
defined for $d_1=1.0$\,fm and $\Delta=0.15$\,fm. In the same manner we
identify the long-distance window observables for the flavour
decomposition as
\begin{align}
  \aLDf{m}{n}&=\left(\frac{\alpha}{\pi}\right)^2\int_0^{\infty}dt\,
  \tilde{K}(t;\,m_\mu)\,\Gf{m}{n}(t)\,\Theta(t,\,d_1,\,\Delta)\,.
\end{align}
For completeness, we list the short- and intermediate-distance
window observables, i.e.
\begin{align}
  \aSD&=\left(\frac{\alpha}{\pi}\right)^2\int_0^{\infty}dt\,
  \tilde{K}(t;\,m_\mu)\,G(t)\,[1-\Theta(t,\,d_0,\,\Delta)]\,, \\
  \aID&=\left(\frac{\alpha}{\pi}\right)^2\int_0^{\infty}dt\,
  \tilde{K}(t;\,m_\mu)\,G(t)\,[\Theta(t,\,d_0,\,\Delta)
    -\Theta(t,\,d_1,\,\Delta)]\,, 
\end{align}
where the standard choice is $d_0=0.4$\,fm. The generalization of
these observables to the flavour decomposition is obvious.

As detailed in \cref{a:blinding}, we have fully blinded our analysis by using 
five modified versions of the kernel function, which only converge in 
the continuum limit, differing by a multiplicative factor.
After finalizing the analysis, we first performed the unblinding step, and only
then switched to the true kernel function, $\tilde{K}(t;\,m_\mu)$, to produce 
the results and figures presented in this work.
In the following, all results for HVP contributions to $a_\mu$
are quoted in units of $10^{-10}$ unless otherwise specified.

\subsection{Gauge ensembles}
We perform our calculation on a subset of the $2+1$-flavour CLS 
ensembles \cite{Bruno:2014jqa, Bali:2016umi} which feature a 
tree-level Symanzik improved Lüscher-Weisz gauge action and
non-perturbatively $\mathrm{O}(a)$ improved Wilson quarks \cite{Bulava:2013cta}.
The RHMC algorithm is used to simulate the strange quark component,
and a small twist in the Dirac operator stabilizes the simulations 
of light quark masses in large volumes. 
The target action, $2+1$-flavour QCD,
is restored by the inclusion of the appropriate reweighting factors
\cite{Clark:2006fx, Luscher:2012av, Mohler:2020txx, Kuberski:2023zky}.
We focus on the chiral trajectory where the sum of the
bare sea quark masses is held constant. Starting from the SU(3) symmetric
point where $m_\pi = m_K \approx 420$\,MeV, the kaon mass approaches
its physical value from below when the pion mass is lowered towards its
physical value since the combination
$m_K^2 + \textstyle{\frac{1}{2}}m_\pi^2$
is approximately constant along each chiral trajectory.
We also include ensembles with a close-to-physical strange quark mass 
and pion masses around $220$\,MeV to stabilize the interpolation to
physical quark masses. 
Three ensembles at physical values of the light quark mass enter our
interpolation, thereby allowing us to tightly constrain the chiral behaviour. 

Compared to our 2019 computation \cite{Gerardin:2019rua}, we have
significantly extended the set of gauge ensembles which now covers six
lattice spacings from about $0.01$\,fm down to $0.039$\,fm.  Eight
gauge ensembles with pion masses ranging from $225$\,MeV down to
$131$\,MeV have been added or significantly extended to allow for a
safe interpolation to physical quark masses and to constrain possible
mass-dependent cutoff effects. Replacing several ensembles with
relatively small spatial box sizes by new ensembles with larger
volumes greatly strengthens our confidence in controlling finite-size
effects, both by reducing the size of the correction and by offering
an explicit check that we can quantitatively describe the finite-size
effects that are found in the data.
Further details concerning the set of ensembles and their inclusion in 
our work can be found in
refs.~\cite{Gerardin:2019rua,Ce:2022eix,Ce:2022kxy,Kuberski:2024bcj}.
\input{./tables/tab_ens}

\subsection{The electromagnetic current on the lattice}

We use two different discretizations, i.e.\ the local ($\loc$) and the
point-split ($\cons$) variant to realize the vector current of
\eq{eq:current} on the lattice
\begin{align}
	J_{\mu}^{(\mloc),a}(x) &= \psib(x) \gamma_{\mu} T^a \psi(x) \,,\\
	J_{\mu}^{(\mcons),a}(x) &= \frac{1}{2} \left(
	\psib(x+a\hat{\mu})(1+\gamma_{\mu}) U^{\dag}_{\mu}(x)
	T^a \psi(x) \right. \nonumber \\
	& \qquad - \left. \psib(x) (1-\gamma_{\mu} ) U_{\mu}(x)
	T^a \psi(x+a\hat{\mu}) \vphantom{U^{\dag}}\right)  \,,
	\label{eq:consvec1}
\end{align}
where $U_{\mu}(x)$ is the gauge link in the direction $\hat{\mu}$ associated
with site $x$.
With the local tensor current defined as
$\Sigma^{a}_{\mu\nu}(x) = -\frac{1}{2}\,
\overline{\psi}(x) [\gamma_{\mu}, \gamma_{\nu}] T^a \psi(x)$,
we obtain the $\mathrm{O}(a)$-improved versions of the currents via
\begin{equation}
	\label{eq:impcv}
	J^{(\alpha),a,I}_{\mu}(x) = J^{(\alpha),a}_{\mu}(x) +
	a\cv^{(\alpha)}(g_0) \, \partial_{\nu} \Sigma^{a}_{\mu\nu}(x)
	\,,\quad \alpha=\loc,\,\cons\,.
\end{equation}
Employing the non-perturbative determination of the improvement
coefficients $\cv^{(\alpha)}(g_0)$ ensures the removal of cutoff
effects of $\mathrm{O}(a)$ in the chiral limit. 
The line of constant physics (LCP) that is
chosen in the formulation and evaluation of the relevant improvement
condition is ambiguous regarding higher-order cutoff effects. As a
consequence, matrix elements of $\mathrm{O}(a)$-improved currents that
differ in the choice of LCP approach the continuum limit with
different rates in $a^2$. 
In previous works, we have used two alternative sets of non-perturbatively
determined coefficients from \cite{Gerardin:2018kpy} and
\cite{Heitger:2020zaq} and interpreted possible deviations in the
continuum limit as systematic uncertainties of the continuum
extrapolation.

The calculation of $\cv^{(\alpha)}(g_0)$ in \cite{Gerardin:2018kpy}
was based on a preliminary determination of the improvement
coefficient $\tilde b_{\rm A}$ that enters the improvement condition.
However, the final results for $\tilde b_{\rm A}$ published in
\cite{Bali:2023sdi}
differ significantly from the preliminary ones
used in \cite{Gerardin:2018kpy}. In turn, an updated determination of
$\cv^{(\alpha)}(g_0)$ using the published values of $\tilde b_{\rm A}$ and
additional SU(3)-symmetric ensembles \cite{Harris:20XX} yield
coefficients that differ significantly from those in
\cite{Gerardin:2018kpy} while being much closer to the ones extracted
directly at vanishing quark masses in the Schrödinger functional
scheme employed in \cite{Heitger:2020zaq}. In this work, we will use
the updated values of \cite{Harris:20XX} for set~1 and employ them to
cross-check our main results which will be computed using the
published values of \cite{Heitger:2020zaq} (set~2). We note that our
earlier results for the short- and intermediate-distance windows
\cite{Ce:2022kxy, Kuberski:2024bcj} are unaffected by the change of
improvement coefficients of set~1.

The renormalization pattern of the electromagnetic current based
on Wilson quarks has been outlined in 
refs.~\cite{Gerardin:2019rua,Ce:2022eix,Ce:2022kxy},
and we use the renormalization factor and improvement coefficients
of \cite{Gerardin:2018kpy} and \cite{Heitger:2020zaq,Fritzsch:2018zym},
respectively, in combination with set~1 and set~2 of improvement 
coefficients $\cv^{(\alpha)}$.
The values of the critical hopping parameters that enter the
mass-dependent improvement via the bare subtracted quark mass 
are taken from \cite{RQCD:2022xux}.

\subsection{Noise reduction in the long-distance tail \label{s:lat_noise}}
One of the two main difficulties in the computation of the
long-distance contribution to $\ahvp$ is the exponential loss of
signal in the light-connected and disconnected correlation
functions. In this work we employ several advanced noise reduction
techniques to enhance the statistical signal, focusing on the
light-quark connected correlation function which contributes about
90\% of the total $\amu$. Specifically, we combine improved estimators
from low-mode averaging (LMA) and the spectral reconstruction of the
isovector correlation function with the widely used bounding method.

In our previous work \cite{Gerardin:2019rua}, the light-quark
connected correlation function was computed either from point sources
or from time, spin and colour-diluted time slice sources. In both
cases, assuming that multiple sources on a gauge configuration are
largely uncorrelated, the statistical uncertainty at a certain
source-sink separation scales $\propto 1 / \sqrt{N}$ with the number
of sources $N$. A brute-force reduction of the statistical error by
increasing the number of sources is infeasible in the long-distance
regime and at close-to-physical values of the light quark masses,
since the signal deteriorates exponentially fast.

In this work, we resort to an improved estimator for the
light-connected correlation function that is based on the low modes of
the Dirac operator. For small values of the light quark mass, we find
low-mode averaging (LMA)~\cite{Giusti:2004yp, DeGrand:2004qw} to yield
significantly more accurate results compared to our old setup, and
thus we employ LMA for all ensembles with pion masses smaller than
280\,MeV. We refer to appendix~\ref{a:lma} for a detailed
explanation of our setup for the LMA computation.
Our setup for computing the quark-disconnected correlation functions
via a combination of frequency-splitting techniques
\cite{Jansen:2008wv, Giusti:2019kff} and hierarchical probing
\cite{Stathopoulos:2013aci} in combination with the generalized
hopping parameter expansion \cite{Gulpers:2013uca} has been
extensively discussed in appendix~C of \cite{Ce:2022eix}.

Euclidean finite-volume two-point correlation functions at source-sink 
separation $t$ can be expressed via the spectral decomposition
\begin{align}\label{e:spectral_dec}
	\Gf{k}{l}(t) = \sum_{n=0}^{\infty} \frac{Z_n^2}{2E_n}\,\mathrm{e}^{ -E_n t} 
	\,,
\end{align}
where $Z_n$ denote the real amplitudes and $E_n$ the ordered
real and positive finite-volume energies. 
At sufficiently large time separations, only the lowest-lying states
contribute significantly, as contributions from higher-energy states 
decay faster.
In this regime, the finite-volume isovector correlation function 
$\Gf{3}{3}(t)$ is dominated by two-pion states.
These can be computed in a dedicated spectroscopy study,
as outlined in \cref{a:spec}.

In both the isovector and isoscalar channels, we make use of 
the representation in \eq{e:spectral_dec} to impose
lower and upper bounds on $\Gf{k}{l}(t)$ via
\cite{LehnerBounding2016, Borsanyi:2016lpl, Blum:2018mom, Gerardin:2019rua}
\begin{align} \label{e:bounding}
	0 
	\leq \Gf{k}{l}(t_c)\, \mathrm{e}^{-E_{\rm eff}^{*} (t - t_c)} 
	\leq \Gf{k}{l}(t) 
	\leq \Gf{k}{l}(t_c)\, \mathrm{e}^{-E_0 (t - t_c)}\,,\qquad t\geq t_c\,.
\end{align}
Here, $E_0$ is the energy level of the lowest-lying state that
contributes to the correlation function. In practice, it has to be
computed or estimated from the data. 

On those ensembles for which we performed a dedicated spectroscopy
study in the isovector channel, we employ the lowest state determined
from the generalized eigenvalue problem (GEVP).  On all other
ensembles, we use a Gounaris-Sakurai parameterization of the timelike
pion form factor, which is used to compute the finite-volume
correction (see section \ref{s:fv}), to estimate the ground-state
energy, which is always found to lie below that of two non-interacting
pions. We note that the lowest energy levels determined by the
Gounaris-Sakurai fit agree with those from the dedicated spectroscopy
study.
We use the central value minus the statistical uncertainty of this estimate
as input for $E_0$. As in \cite{Ce:2022eix}, this estimate for $E_0$ is
also used for the isoscalar correlation function, which is justified by the
fact that $m_\rho \lesssim m_\omega$, making this a conservative
choice.

The energy $E_{\rm eff}^{*}$ is determined from the effective mass of
the correlation function at some time $t_{\rm eff} < t_c$, as computed
from the logarithmic derivative of the correlation function. In this
work, we fix $t_{\rm eff}$ on each ensemble such that the effective
mass at this distance is clearly larger than in the region where the
bounding method will be applied. Empirically, we find that this
requirement is satisfied for $t_{\rm eff} = {4.5}/{E_0}$ for all
ensembles used in this work. This approach provides a strict lower
bound on the correlation function that is not affected by local
fluctuations.

Via \eq{e:bounding}, the bounds on the energy $E_0$ and $E_{\rm eff}^{*}$
are translated into bounds on the correlation function. By replacing
the measured correlation function with its upper and lower bounds for
$t > t_c$ in \eq{e:def_amu}, we obtain corresponding constraints on
$\af{k}{l}$ that depend on $t_c$. At the value of $t_c$ where the
central values of both bounds are compatible within the $1\,\sigma$
uncertainty of their respective counterpart, we average over the mean
of the two bounds in a region of $0.25\,{\rm fm}$ or at least 4 time
slices, to smooth out local fluctuations.
  
\begin{figure}[t]
	\includegraphics[width=.49\textwidth]{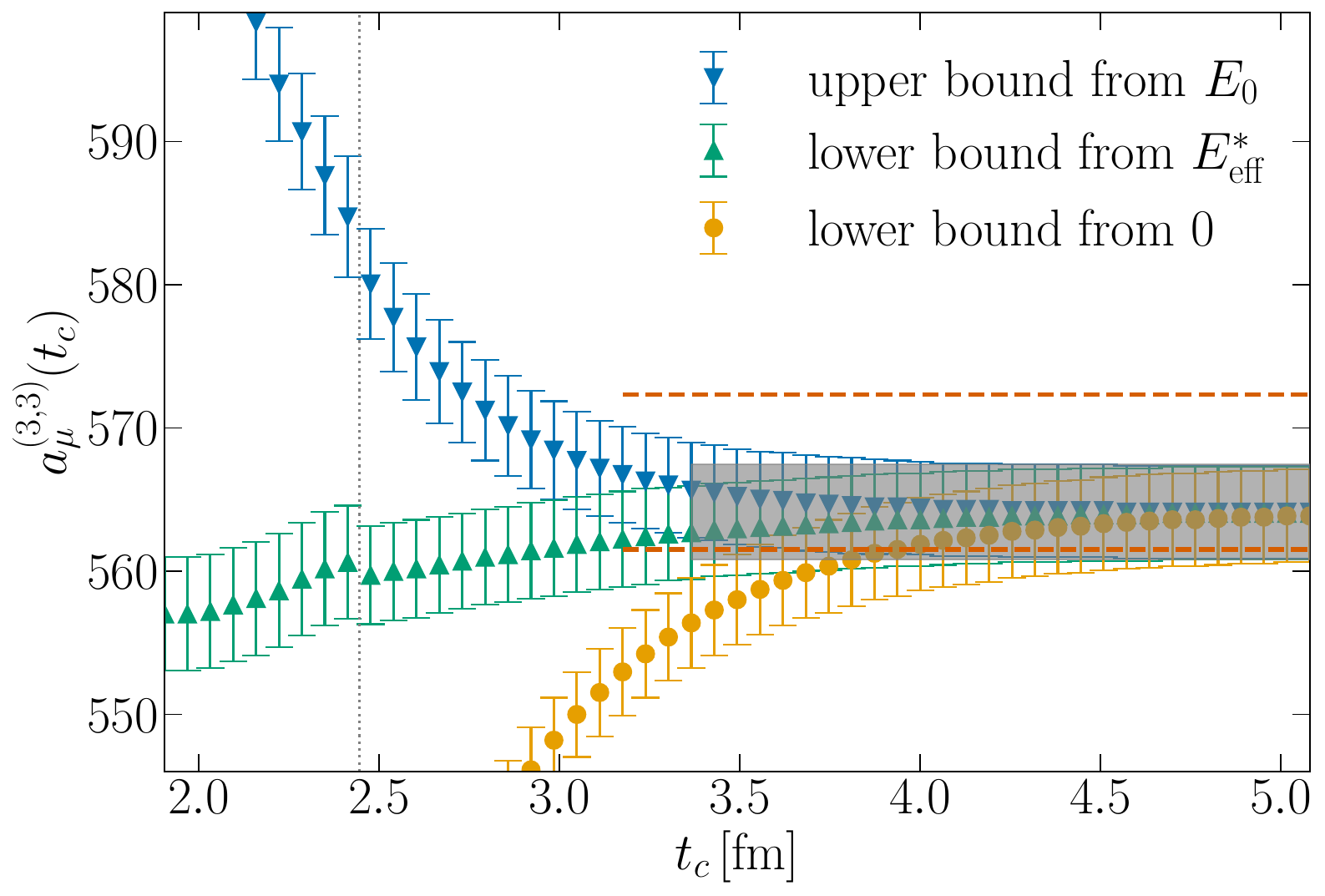}%
	\hfill
	\includegraphics[width=.49\textwidth]{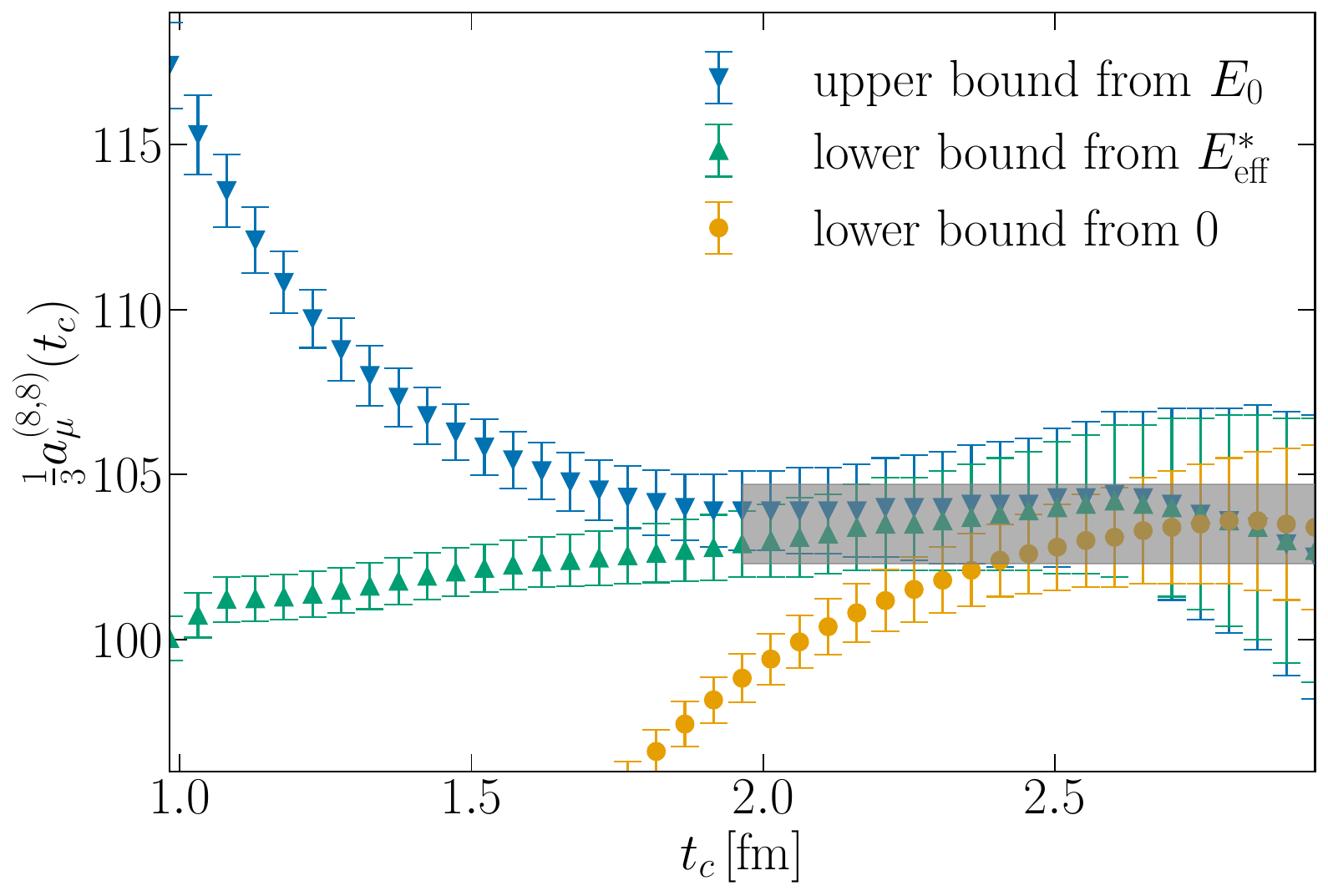}%
	\caption{\label{f:bounding}%
		Determination of $\af{k}{l}$ using the bounding method. 
		$t_c$ denotes the time where the correlator is replaced by the single state
		exponential as detailed in~\eq{e:bounding}.
		The downward triangles show the upper bound from the estimate of the ground
		state energy.
		The lower bound from the effective mass is given by the upward triangles, and
		the lower bound from setting the correlator to zero beyond $t_c$, which is just shown for 
		comparison, is represented by the circles.
		The gray area indicates the final estimate.
		\textit{Left:} In the isovector channel on ensemble E250. The dotted vertical
		line denotes the switching point between the LMA and the spectroscopy 
		correlation functions, 
		where $N_{\pi\pi} = 4$ states have been used to reconstruct the latter.
		The dashed horizontal lines indicate the starting time and the $1\,\sigma$
		uncertainty band of the estimate that would be obtained when using only 
		the LMA data set.
		\textit{Right:} In the isoscalar channel on ensemble J303.
	}
\end{figure}

On ensembles with periodic boundary conditions in the temporal direction,
we extend the above boundaries to include the contributions of wrappers
around the temporal direction of the torus, which are small in all cases that
are considered.
For the ensembles A653, A654 and B450 with $m_\pi > 335\,{\rm MeV}$
and small temporal extents, we perform a fit to a single-state in the 
region around $T/2$ and replace the correlator at large times by 
the corresponding single-exponential form \cite{DellaMorte:2017dyu}.
Note that this treatment is only possible due to the large pion mass on
these ensembles such that a single stable state dominates.

For two ensembles, D200 at $m_\pi \approx 200$\,MeV and E250 at $m_\pi
\approx 130$\,MeV, we increase the statistical precision in the
long-distance tail further by supplementing the LMA calculation with
an explicit reconstruction of the isovector correlation function in
terms of two-pion states, similarly to what was done in
\cite{Gerardin:2019rua}. To this end, we have computed the
lowest-lying energies $E_n$ and corresponding amplitudes $Z_n$ for
the vector-vector, as well as the derivative of the vector-tensor
currents, see \cite{Andersen:2018mau,Paul:2021pjz,Paul:2023ksa}. Full
details on the computation of the finite-volume energies and matrix
elements at physical pion mass are deferred to \cref{a:spec}. 

For D200 and E250, we observe that the isovector correlation
function is fully saturated by the three lowest states starting at
$t\gtrsim1.1\,$fm and by the four lowest states starting at $1.5\,$fm, 
respectively. However, since LMA still
yields smaller statistical errors for source-sink separations below
about $2.5\,$fm, we only switch to the reconstructed isovector
correlator when the latter is statistically more precise. In this way
we are able to eliminate the exponential growth of the relative
statistical noise, which is also encountered for LMA (albeit at a much
reduced level), since the signal-to-noise ratio varies only slowly for
the spectrally reconstructed correlator. We stress that the
reconstruction in terms of the four lowest-lying states occurs in a
region where all higher states have clearly decayed below any
statistical significance.

We illustrate the bounding method for the isovector channel on the
physical mass ensemble E250 and for the isoscalar channel on the finer
lattice spacing ensemble J303, as shown in \cref{f:bounding}.
The blue triangles denote the upper bound on $\af{k}{l}$ based on the
estimated ground state energy, while the green upward triangles depict
the lower bound from the effective mass.
For comparison, we also display a less strict lower bound obtained by
setting $\Gf{k}{l}$ to zero for $t > t_c$.
The gray area represents the estimate for $\af{k}{l}$, bounded by the
two limits, and begins at the time separation where we start
averaging the two.

In the left hand panel, the dotted vertical line indicates the time separation
where we switch from the LMA to the reconstructed correlation function.%
\footnote{Note that the bounding method does not have to be applied as
soon as the reconstructed correlation function is employed because the
noise is under control in this case.}
The dashed vertical lines show the uncertainty range of the estimate that
would be obtained from using only the LMA correlation function.

\subsection{Physical point extrapolation \label{s:setup_extrap}}
Our strategy to approach the physical point consists in a combined
chiral-continuum extra\-polation of our lattice data.
Whereas the continuum limit is mostly constrained by ensembles
that feature larger-than-physical pion masses, precise data
at near-physical values of the quark masses ensure that the interpolation
in $m_\pi^2$ is well controlled.
To disentangle strange from light quark mass effects, we have added
four ensembles with physical strange quark mass to our standard set of
ensembles, for which the sum of the bare quark masses is held constant
along a chiral trajectory.

As outlined in \cite{Gerardin:2019rua}, see also
\cite{Colangelo:2021moe}, a strong chiral dependence with a leading
term proportional to $1/m_\pi^2$ or $\log(m_\pi^2)$ can be expected in
the long-distance regime of $\ahvp$. Our set of ensembles (see
\cref{t:ensembles}) allows us to constrain the pion mass dependence of
the observables that are studied in this work in the full range $m_\pi
\in [131, 430]\,{\rm MeV}$. We point out that the region below
220\,MeV is especially finely sampled and that three ensembles with
approximately physical values of the quark masses enter the
interpolation, with the data set on ensemble E250 being one of the
statistically most precise ones.

For the Wilson quark action used in this work, distortions of the pion
spectrum that could affect the long-distance tail are absent, in
contrast to staggered or Wilson twisted-mass discretizations.
Data computed for six values of the lattice spacing allow us to
resolve leading and subleading cutoff effects. At our level of
precision, we have to reckon with the occurrence of
non-negligible mass-dependent cutoff effects. These can be reliably
constrained since we cover the full range of pion masses on four of
the six values of the lattice spacing that enter our result. In a
future update, a significant increase in the number of available
configurations for ensemble F300 will play a crucial role for further reducing
the systematic uncertainty of the continuum extrapolation at the
physical pion mass.

The small-$a$ behaviour of a (lattice) regularized quantum field
theory is described by Symanzik effective theory, which is expected to
work well for the observable $\aLD$. As pointed out in
\cite{Husung:2019ytz} for the case of QCD, the leading dependence on
the cutoff is modified by logarithmic corrections and can be expressed
as $a^{n_{\rm min}} \left[\alpha_{\rm s} (1/a)\right]^{\hat{\Gamma}}$,
where $\hat{\Gamma}$ is the leading anomalous dimension. 
The Wilson quark action and the currents used in our work are non-perturbatively
$\mathrm{O}(a)$-improved such that $n_{\rm min} = 2$.
As explained in ref.~\cite{Husung:2021mfl}, for our choice of action
the term with $\hat{\Gamma} = 0.76$ is expected to dominate the
description of cutoff effects in spectral quantities and those from the
sea, while $\hat{\Gamma} = 0.395$ is the lowest non-zero anomalous dimension
for quark bilinears with vector quantum numbers \cite{Husung:2024cgc}.
A potentially dangerous slowing down of the continuum extrapolation
due to large negative anomalous dimensions can thus be excluded thanks
to the existing analytic knowledge for our action. While it is not
possible to resolve one or more anomalous dimensions in the existing
range of lattice spacings, all of our continuum extrapolations include
the possibility of a non-zero anomalous dimension for the leading
term.

To convert the muon mass in the QED kernel function to lattice units
and to form dimensionless quantities that act as proxies for the quark
masses in our ensembles, we use the pion decay constant $af_\pi$.
This approach, known as $f_\pi$-rescaling, was introduced in
\cite{Gerardin:2019rua}. The key benefit of this strategy is the
mitigation of the quark mass dependence in the isovector contribution
to $\aLD$ via a partial cancellation from the corresponding dependence
of the decay constant. For each of our ensembles, we calculate
$af_\pi$ as outlined in appendix~E of \cite{Ce:2022kxy} and apply
corrections for the leading finite-size effects, following
\cite{Colangelo:2005gd}. The renormalization and improvement procedure
utilizes the results for $Z_{\rm A}$ from \cite{DallaBrida:2018tpn}
and $b_{\rm A}$, $\bar{b}_{\rm A}$ from \cite{Bali:2021qem}. To reduce
fluctuations in the decay constants, we perform a fit to the values
obtained from all large-volume ensembles. This fit is guided by the
expectations from SU(3) chiral perturbation theory
\cite{Gasser:1984gg}. The fitted results are then evaluated for the
parameter values specific to each ensemble, ensuring a stable and
consistent representation of the decay constants.

The fit proceeds by forming a dimensionless combination of 
the decay constant and the flow quantity $\sqrt{t_0}$. 
The physical value of $\sqrt{t_0}$ is irrelevant for this purpose
since we perform a local interpolation of the data.
The quark mass dependence of the pion decay constant is then
parameterized using the two variables
\begin{align}
	y = \frac{m_\pi^2}{8 \pi^2 f_\pi^2} \propto m_{\rm l}\,, \qquad
	y_{K \pi} = \frac{m_K^2 + \frac{1}{2}m_\pi^2}{8 \pi^2 f_{K \pi}^2} \propto 2 m_{\rm l} + m_{\rm s}\,,
\end{align} 
where $f_{K \pi} = \frac{2}{3} (f_K + \frac{1}{2} f_\pi)$.
The specific physical values that define our hadronic scheme
are collected in \cref{a:scheme}.
It is worth noting that only a small deviation from the physical value
of $y_{K \pi}$ is observed among the ensembles used in this work.

We refer to the pion decay constant at finite lattice spacing and physical 
values of $y$ and $y_{K \pi}$ as $a \tilde{f}_\pi$.
It is obtained according to
\begin{align}
	a \tilde{f}_\pi = 
	\left(\frac{a}{\sqrt{t_0^{\rm sym}}}\right) \cdot
	\left(\frac{\sqrt{t_0^{\rm sym}}}{\sqrt{t_0^{\rm phys}}}\right) \cdot
	\left(\sqrt{t_0} f_\pi\right)_{\rm phys}\,,
\end{align}
with the first two factors taken from \cite{RQCD:2022xux}, the second being
evaluated in the continuum.
The quantity $\left(\sqrt{t_0} f_\pi\right)_{\rm phys}$ is obtained from the fit
described above at physical
values of $y$ and $y_{K \pi}$ for each value of the lattice spacing.

For our fits to the various contributions to $\aLD$ we proceed as follows.
The proxies for the light quark mass and the sum of the sea quark masses
are defined by
\begin{align}\label{e:yzdef}
	y = \frac{m_\pi^2}{8 \pi^2 f_\pi^2} \propto m_{\rm l}\,, \qquad
	z = \frac{m_K^2 + \frac{1}{2}m_\pi^2}{8 \pi^2 \tilde{f}_{\pi}^2} \propto 2 m_{\rm l} + m_{\rm s}\,.
\end{align}
Compared to $y_{K \pi}$, we find that the simpler quark mass
dependence of $z$ helps us to disentangle the two directions in the
quark mass plane and to separate cutoff effects.  The quantity
$a/\sqrt{t_0}$ serves as a proxy for the lattice spacing.
For the muon mass entering the QED kernel used on a given ensemble, we use 
$a m_\mu = (af_\pi) \cdot (m_\mu / f_\pi)^{\rm phys}$ for the isovector 
contribution, with $af_\pi$ computed on that ensemble, while
$a m_\mu = (a\tilde{f}_\pi) \cdot (m_\mu / f_\pi)^{\rm phys}$ is used
for the other contributions. This strategy prevents the chiral
dependence of the pion decay constant from affecting the isoscalar
contribution while ensuring a consistent scale setting across all our
observables. We summarize the values of the bare lattice quantities
and quark mass proxies in \cref{tab:ps}.

We follow the general strategy of our previous 
works~\cite{Ce:2022kxy,Kuberski:2024bcj} to extrapolate to the 
physical point by performing a simultaneous fit of our data to the
quark mass and cutoff dependence.
Denoting the light quark mass proxy by $\Xpi \propto m_\mathrm{l}$, 
we describe the continuum light quark mass dependence with the
general ansatz
\begin{align} \label{e:cc_pi}
	\mathcal{O}(\Xpi) =
	\mathcal{O}(\Xpi^{\rm phys}) 
	&+ \gamma_1 \, \left( \Xpi - \Xpi^{\rm phys} \right)  \\
	&+ \gamma_2 \left( f_{\rm ch, 1}(\Xpi) -  f_{\rm ch, 1}(\Xpi^{\rm phys})\right) \nonumber\\
	&+ \gamma_3 \left( f_{\rm ch, 2}(\Xpi) -  f_{\rm ch, 2}(\Xpi^{\rm phys})\right)
	\nonumber\\
	\text{where } f_{\rm ch, 1} &\in  \{
	1 / \Xpi\,;\enspace
	\log (\Xpi)\,;\enspace
	\Xpi\log (\Xpi)\,;\enspace
	\Xpi^2\,
	\} \nonumber\,,\\
	\text{and } f_{\rm ch, 2} &\in  \{
	1 / \Xpi\,;\enspace
	\Xpi^2\,
	\}\,. \nonumber
\end{align}
Here we always include the leading term $\propto \Xpi$ and test for the
significance of the higher order terms on a case-by-case basis.
The dependence on the quark mass proxy $\XK \propto 2 m_{\rm l} + m_{\rm s}$
is always parameterized via
\begin{align}\label{e:cc_K}
	\mathcal{O}(\XK) &= 
	\mathcal{O}(\XK^{\rm phys}) 
	+ \gamma_0 \, \left( \XK - \XK^{\rm phys} \right)\,.
\end{align}
and allows us to correct for small deviations from $\XK^{\rm phys}$.

Denoting the proxy for the lattice spacing with $\Xa = a/\sqrt{t_0}$, our most general 
ansatz for the dependence on the lattice spacing in this work is
\begin{align}\label{e:cc_a}
	\mathcal{O}(\Xa) &= 
	\beta_2 \, [\alpha_{\rm s}(1 / \Xa)]^{\hat{\Gamma}} \Xa^2 
	+ \beta_3 \, \Xa^3  
	+ \delta_2 \, [\alpha_{\rm s}(1 / \Xa)]^{\hat{\Gamma}} \Xa^2 \left( \Xpi - \Xpi^{\rm phys} \right)\,,
\end{align} 
where $\hat{\Gamma} \in \{0, 0.395\}$.
We always include the leading term with the coefficient $\beta_2$ and
test for higher order cutoff effects as well as quark mass dependent
cutoff effects by including/excluding the terms proportional to
$\beta_3$ and $\delta_2$.
We perform every fit for each of the two choices for the anomalous 
dimension $\hat{\Gamma}$ that have been motivated above.
We use the five-loop running relation from \cite{Baikov:2016tgj}, using as input
$\Lambda^{(3)}_{\overline{\rm MS}}$~\cite{Bruno:2017gxd}, to evaluate
the running-coupling constant at the scale $1/a$.%
\footnote{When working in the BMW-20 scheme, we employ the quark mass proxies
$\rho_2$ and $\rho_4$ that are defined in eq.~(\ref{e:rho_variables}) and 
$a/w_0$ as proxy for the lattice spacing.}

Further variations are introduced by applying cuts to the data that
enter the fits. In addition to fitting the whole data set, we also
consider fits in which the coarsest or the two coarsest values of the
lattice spacing are excluded. To avoid overfitting when removing data
from the two coarsest lattice spacings, we do not include terms that
parameterize higher-order cutoff effects.
We also perform fits that exclude ensembles with pion masses larger
than 400\,MeV, thereby removing data at the SU(3)-symmetric
point. Further cuts in the pion mass are not considered because this
would exclude both of our ensembles at the finest lattice spacing of
$a\approx 0.039\,\mathrm{fm}$.

To quantify the systematic uncertainty from the extrapolation to the
physical point and to determine our final results, we perform a model
average over the different fit forms that are considered in this work.
As done in our previous works and following ref.~\cite{Jay:2020jkz}, 
we use the Akaike Information Criterion (AIC) \cite{Akaike1998} 
to assign a weight to each fit.
The central value and its statistical uncertainty are then obtained
from a weighted average over all analyses, whereas the systematic
uncertainty is obtained from the distribution of weighted models.
We test explicitly that using the information criterion that has been defined 
in \cite{Borsanyi:2020mff} leads to negligible differences in our final results.
The determination and propagation of statistical uncertainties is
performed using the $\Gamma$-method in the implementation of the
\texttt{pyerrors} package
\cite{Wolff:2003sm,Ramos:2018vgu,Joswig:2022qfe}. Significant
autocorrelations are present at small lattice spacing, and we reliably
take them into account for all observables considered in this work.

\subsection{Finite-volume correction \label{s:fv}}
Finite-volume effects are sizeable in the long-distance regime
of the isovector correlation function.
The origin of these effects lies in the discrete nature of the low-lying multi-pion
spectrum when the theory is formulated in finite volume.
Accordingly, it is mandatory to apply a correction for finite-size effects
lest the latter become dominant in the final error budget.

As in our previous works \cite{Ce:2022eix, Ce:2022kxy}, we employ two
methods to correct our data for finite-size effects. Based on the
electromagnetic form factor of the pion in the spacelike region, the
method by Hansen and Patella (HP) \cite{Hansen:2019rbh,Hansen:2020whp}
is expected to work especially well in the short and intermediate
distance regions. The Meyer-Lellouch-L\"uscher (MLL) formalism is
based on the timelike pion form factor \cite{Meyer:2011um} and
expected to be most successful in the long-distance region, where only
a few states contribute significantly in the spectral decomposition of
the correlation function.

We follow our strategy from \cite{Ce:2022kxy} and apply the
Hansen-Patella method to correct the isovector correlation
function for source sink separations below
$t^{*} = (m_{\pi} L/4)^2 / m_{\pi}$. 
From then on, we employ the MLL formalism.
Accordingly, the latter dominates the correction for $\aLD$ close to the
physical value of the pion mass.

Motivated by its phenomenological success and simplicity, we use the vector-meson-dominance (VMD) parameterization
of the pion form factor in the HP volume correction, $F_\pi(-Q^2) = M_{\rm VMD}^2/(Q^2+M_{\rm VMD}^2)$,
while the Gounaris-Sakurai (GS) parameterization is used in the MLL method.
In order to make the form factor consistent on the space- and time-like sides,
we proceed by matching the value of the VMD form factor with the GS one at virtuality $Q^2 = M_{\rho}^2$,
where $M_\rho$ is the $\rho$ meson mass entering the GS parameterization. The VMD and GS parameterizations
then agree to within one percent for all virtualities $0\leq Q^2\leq 0.8\,{\rm GeV}^2$.
As a result, when comparing the corrections that are predicted by the HP and MLL 
formalisms with each other, we find that they are consistent at the level of 5\% and that
any discontinuity in the finite-size correction on $\Gf{3}{3}(t)$ at the time $t=t^*$ is 
a very small effect in comparison to the correction itself.

As in \cite{Ce:2022eix,Ce:2022kxy}, we also correct for the effect
from kaons propagating in the finite box, which is relevant for
ensembles close to the SU(3) symmetric point on the chiral trajectory
where ${\rm Tr}[M_{\rm q}] = {\rm const}$. We use the HP formalism to
compute the corresponding correction.

While we previously corrected the data for the entire finite-size effect
before performing the chiral-continuum extrapolation, we now follow
a different procedure set out in~\cite{Borsanyi:2020mff}.
We first apply the finite-volume correction on all ensembles to match
a common reference value of $m_\pi L$ and then extrapolate the results
to the physical point.
By choosing a reference value of $m_\pi L$ close to the one
corresponding to our physical pion mass ensembles, we minimize the
correction that is applied to our most important data.
Furthermore, to facilitate the comparison with the results of~\cite{Borsanyi:2020mff}
without the need for a finite-volume correction with the associated
uncertainty, we define our reference target to be
\begin{align}\label{eq:mpiLref}
	(m_\pi L)^{\rm ref} = (m_{\pi^0})_{\rm phys} \cdot 6.272\,{\rm fm} \approx 4.290\,.
\end{align}
The correction from this reference value to the infinite-volume limit
is performed in the continuum.

\begin{figure}[t]
	\centering
	\includegraphics*[width=0.5\linewidth]{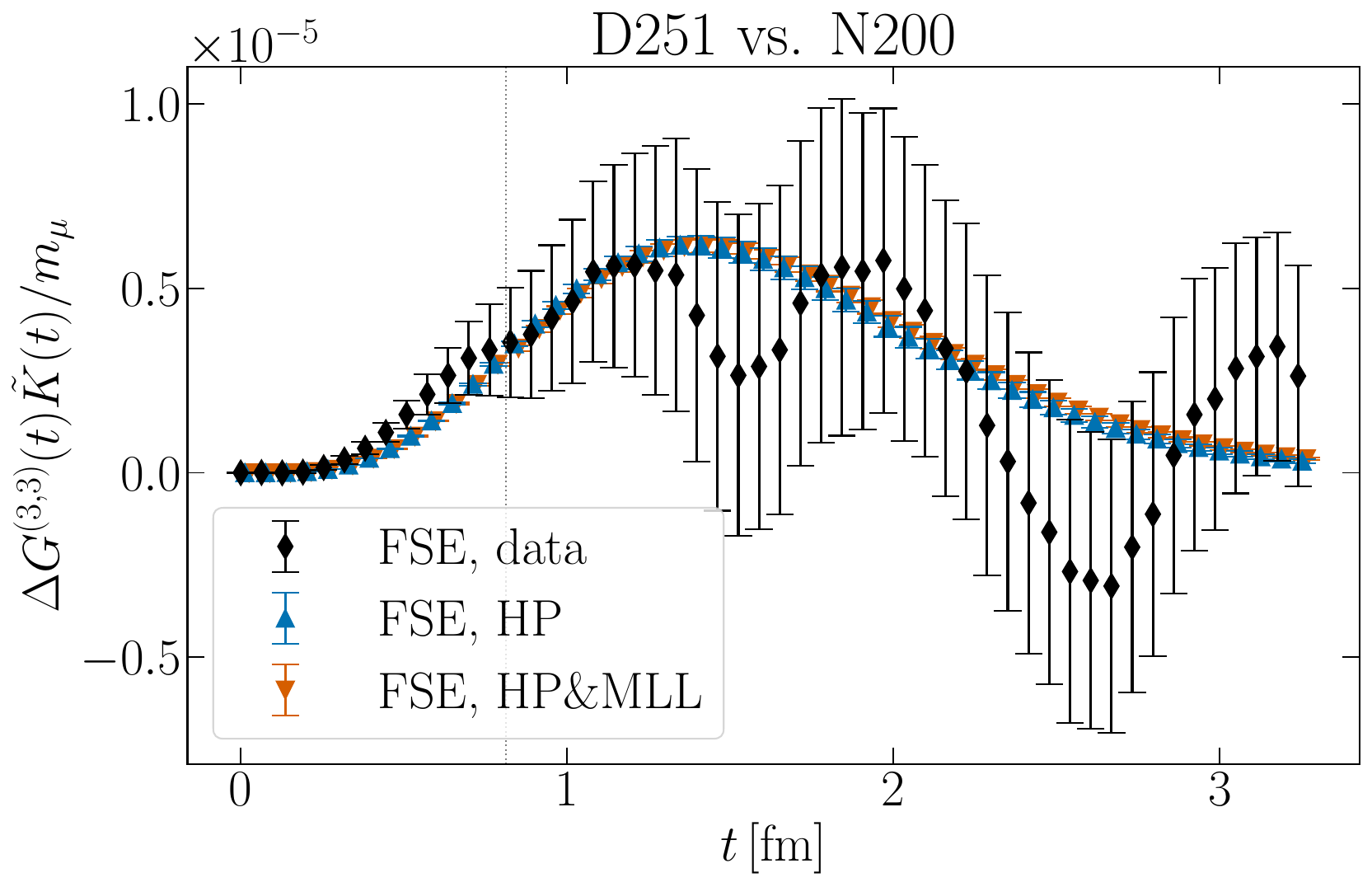}%
	\hfill
	\includegraphics*[width=0.5\linewidth]{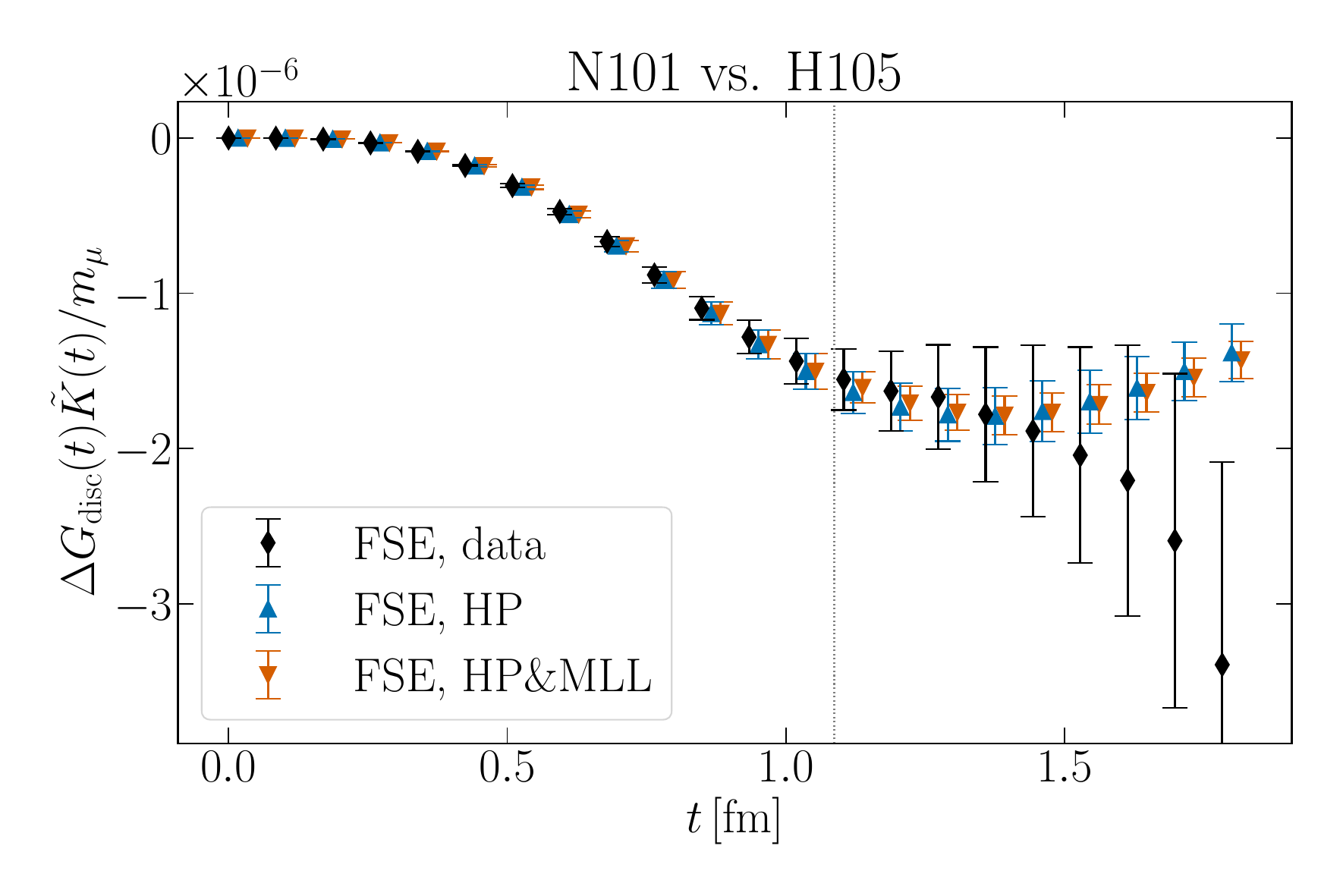}%
	\caption{%
	Illustration of finite-volume effects on the integrand for the light-connected 
	and disconnected contributions. 
	Black diamonds denote effects computed from lattice QCD in two volumes. 
	Blue upward triangles show corrections predicted by the HP method, while 
	red downward triangles represent corrections using the MLL method beyond 
	$t^\star$, indicated by the vertical dotted line for the respective smaller volume.
	\textit{Left:} Light-connected contribution for ensembles D251 and N200.
	\textit{Right:} Disconnected contribution for ensembles N101 and H105.
	}
	\label{f:fse}
\end{figure}

We find an excellent agreement of the HP\&MLL method with the effects
that we see in our data.  This is illustrated in \cref{f:fse}, where
we show in black the differences between the integrands for the
isovector contribution for $\ahvp$ as computed on the ensembles D251
and N200 with $m_\pi \approx 286\,{\rm MeV}$.  Both ensembles differ
only in their spatial extent and feature $(m_\pi L)^{\rm D251} = 5.9$
and $(m_\pi L)^{\rm N200} = 4.4$. The statistical uncertainty of this
difference is mainly driven by the smaller box N200.  Together with
the finite-volume effect from the data, we show the corresponding
finite-volume correction as predicted by our models. The blue data set
denotes the correction as given by HP, whereas the orange data uses HP
up to $t^{*}$ and MLL from then on. Both methods agree very well with
each other and also -- as far as can be judged given the statistical
uncertainties -- with the lattice data.

On the right hand side of \cref{f:fse} we show a similar comparison for
the  disconnected contribution in the $(u,d,s)$ quark sector on the 
ensembles N101 and H105 at a similar pion mass but coarser lattice spacing, 
corresponding to the effect from 
$(m_\pi L)^{\rm H105} = 3.9$ to $(m_\pi L)^{\rm N101} = 5.8$.
We use the same models, with the appropriate prefactor $-\textstyle\frac{1}{9}$ 
\cite{DellaMorte:2010aq, Francis:2013fzp}, to predict the difference.
Again, an excellent agreement between the prediction and the 
data for the finite-size correction can be observed in the region
where the statistical uncertainties of the data are still small.
We note that, while this test further increases our confidence in the correction,
it will not be applied in our analysis since we work with the isoscalar
contribution, where the leading finite-volume effects of light-connected
and disconnected contributions cancel.
As we only correct for finite-size effects from the isovector, the agreement
between the prediction in the right panel of \cref{f:fse} and the lattice data is
evidence for the smallness of these effects in the isoscalar channel.

For the final conversion of our continuum results at physical quark
masses from the reference volume specified in eq.\ (\ref{eq:mpiLref})
to infinite volume, we use the HP correction up to $t^{*}$ and MLL
from then on.  The GS parameters for the pion form factor are taken
from the dedicated calculation on ensemble E250 described in
appendix~\ref{a:spec} and a forthcoming publication~\cite{pion_ff_to_be_published}.
The VMD mass entering the HP method is based on the pion charge radius 
$r_\pi = 0.659(4)\,$fm from \cite{ParticleDataGroup:2024cfk} and we
do not find any significant deviation if we instead match to the 
GS form factor as described above.

The correction applied is
\begin{align}\label{e:result_aLDfse}
        \aLD(L = \infty) - \aLD(L_{\rm ref}) =
        16.7(1.5)        
        \,.
\end{align}
The absolute uncertainty we have assigned to it
is mainly based on the sensitivity of the correction to the values of the GS parameters
as well as to the difference to the next-to-next-to-leading order
chiral perturbation theory expression given in~\cite{Aubin:2019usy}, which we find to be on the order of 0.2.
We have also estimated the finite-size effect associated with higher channels.
Specifically, we have studied the $\pi^+\pi^-\pi^0\pi^0$ channel, approximating it
as an $\omega\pi$ channel. From here, using the cross-section measurement~\cite{SND:2023gan},
we have estimated the finite-size effect in the approximation that there are discrete, non-interacting
$\omega\pi^0$ energy levels on the torus, finding an absolute effect on $\amu$ in the range 0.2 to $0.3$.

In addition to the finite-size effects affecting the isovector
channel, we have also considered those affecting the isoscalar
channel.  One expects, on one hand, a contribution from $\bar K K$
states, which we take into account via the HP formalism. These effects
are of order $\exp(-m_K L)$.  There are however also finite-size
effects of order $\exp(-m_\pi L)$, associated mainly with the
three-pion channel. As pointed out in~\cite{Borsanyi:2020mff}, these
are heavily suppressed in the chiral power-counting compared to the
isovector channel; indeed it takes three pions and three derivatives to form an isoscalar current,
$\epsilon_{\mu\nu\alpha\beta}\partial_\nu\pi^+ \partial_\alpha\pi^-\partial_\beta\pi^0$ \cite{Wess:1971yu, Witten:1983tw}.
We expect the numerically leading effect to come
from a slight shift of the finite-volume energy level associated with
the $\omega$ meson, and possibly with the higher-lying three-pion
states.  In the former case, a lattice study~\cite{Yan:2024gwp} of
energy levels in the isoscalar channel has recently appeared for a
pion mass of 200 or 300\,MeV; no statistically significant shift of
the lowest-lying level was found between $L=2.5\,$fm and
$L=3.7\,$fm. In the case of the three-pion states above one GeV, we
approximate them as being mainly in a $\rho\pi$ configuration and
estimate the associated finite-size effect in the same way as in the
$\omega\pi$ case described in the previous paragraph; however, the
corresponding $e^+e^-\to \pi^+\pi^-\pi^0$ cross-section is a factor of about three
smaller at those energies than $e^+e^-\to\pi^+\pi^-\pi^0\pi^0$, further reducing the
importance of this effect.
Based on the above discussion, we do not include any finite-size correction for the isoscalar channel,
but assign an absolute uncertainty of 0.3 to this effect.

%% file: tables/tab_ens.tex
\begin{table}[!htbp]
\vskip 0.1in
\begin{tabular}{l@{\hskip -.15em}c@{\hskip -.1em}c@{\hskip 01em}c@{\hskip 01em}c@{\hskip 01em}c@{\hskip 01em}c@{\hskip 01em}c@{\hskip 01em}c@{\hskip 01em}c@{\hskip 01em}
	}
\hline
Id   & $\quad\beta\phantom{\Big|}\quad$   & bc & $\textstyle\big(\frac{L}{a}\big)^3\times\frac{T}{a}$   & $a\,[{\rm fm}]$   & $m_\pi\,[{\rm MeV}]$   & $m_K\,[{\rm MeV}]$   &   $m_\pi L$ &   $L\,[{\rm fm}]$ 
& MDU
\\
\hline
\textbf{A653} & 3.34 & p  & $24^3 \times 48$   & 0.097 & 430 & 430 & 5.1 & 2.3 & 20200 \\
\textbf{A654} &      & p  & $24^3 \times 48$   &       & 338 & 462 & 4.0 & 2.3 & 16000 \\
\hline
H101 & 3.4  & o  & $32^3 \times 96$   & 0.085 & 424 & 424 & 5.8 & 2.7 &  8064 \\
H102 &      & o  & $32^3 \times 96$   &       & 358 & 445 & 4.9 & 2.7 &  7832 \\
H105$^{*}$ &      & o  & $32^3 \times 96$   &       & 283 & 470 & 3.9 & 2.7 &  8260 \\
N101 &      & o  & $48^3 \times 128$  &       & 282 & 468 & 5.8 & 4.1 &  6376 \\
\textbf{C101} &      & o  & $48^3 \times 96$   &       & 222 & 478 & 4.6 & 4.1 &  8000 \\
\textbf{C102}$^{\dag}$ &      & o  & $48^3 \times 96$   &       & 224 & 506 & 4.6 & 4.1 &  6000 \\
 \textbf{D150}$^{\dag}$ &      & p  & $64^3 \times 128$  &       & 131 & 484 & 3.6 & 5.4 &  1616 \\
\hline
B450 & 3.46 & p  & $32^3 \times 64$   & 0.075 & 422 & 422 & 5.1 & 2.4 &  6448 \\
S400$^{*}$ &      & o  & $32^3 \times 128$  &       & 355 & 447 & 4.3 & 2.4 & 11492 \\
\textbf{N452} &      & p  & $48^3 \times 128$  &       & 356 & 447 & 6.5 & 3.6 &  4000 \\
\textbf{N451} &      & p  & $48^3 \times 128$  &       & 291 & 468 & 5.3 & 3.6 &  4044 \\
\textbf{D450} &      & p  & $64^3 \times 128$  &       & 219 & 483 & 5.3 & 4.8 &  2000 \\
\textbf{D451}$^{\dag}$ &      & p  & $64^3 \times 128$  &       & 219 & 509 & 5.3 & 4.8 &  3700 \\
\textbf{D452} &      & p  & $64^3 \times 128$  &       & 156 & 490 & 3.8 & 4.8 &  4000 \\
\hline
H200$^{*}$ & 3.55 & o  & $32^3 \times 96$   & 0.064 & 423 & 423 & 4.4 & 2.0 &  8000 \\
N202 &      & o  & $48^3 \times 128$  &       & 417 & 417 & 6.4 & 3.0 &  7608 \\
N203 &      & o  & $48^3 \times 128$  &       & 349 & 447 & 5.4 & 3.0 &  6172 \\
N200 &      & o  & $48^3 \times 128$  &       & 286 & 468 & 4.4 & 3.0 &  6848 \\
\textbf{D251} &      & p  & $64^3 \times 128$  &       & 286 & 467 & 5.9 & 4.1 &  5968 \\
\textbf{D200} &      & o  & $64^3 \times 128$  &       & 202 & 486 & 4.2 & 4.1 &  8004 \\
\textbf{D201}$^{\dag}$ &      & o  & $64^3 \times 128$  &       & 202 & 507 & 4.2 & 4.1 &  4312 \\
 \textbf{E250}$^{\dag}$ &      & p  & $96^3 \times 192$  &       & 131 & 495 & 4.1 & 6.1 &  4496 \\
\hline
N300$^{*}$ & 3.7  & o  & $48^3 \times 128$  & 0.049 & 425 & 425 & 5.1 & 2.4 &  8188 \\
\textbf{J307} &      & o  & $64^3 \times 192$  &       & 424 & 424 & 6.7 & 3.1 &  3200 \\
N302$^{*}$ &      & o  & $48^3 \times 128$  &       & 350 & 456 & 4.2 & 2.4 &  8804 \\
\textbf{J306} &      & o  & $64^3 \times 192$  &       & 349 & 455 & 5.6 & 3.1 &  3840 \\
\textbf{J303} &      & o  & $64^3 \times 192$  &       & 260 & 480 & 4.1 & 3.1 &  8584 \\
\textbf{J304}$^{\dag}$ &      & o  & $64^3 \times 192$  &       & 263 & 530 & 4.2 & 3.1 &  6508 \\
\textbf{E300} &      & o  & $96^3 \times 192$  &       & 177 & 498 & 4.2 & 4.7 &  7180 \\
\textbf{F300}$^{\dag}$ &      & o  & $128^3 \times 256$ &       & 136 & 496 & 4.3 & 6.3 &  1412 \\
\hline
\textbf{J500} & 3.85 & o  & $64^3 \times 192$  & 0.039 & 417 & 417 & 5.2 & 2.5 & 15000 \\
\textbf{J501} &      & o  & $64^3 \times 192$  &       & 337 & 450 & 4.2 & 2.5 & 15680 \\
\hline
 \end{tabular} 
\label{t:ensembles}
\caption{Parameters of the simulations: the bare coupling $\beta =
	6/g_0^2$, the temporal boundary conditions, open (o) or
	anti-periodic (p), the lattice dimensions, the lattice spacing $a$
	in physical units based on \cite{Strassberger:2021tsu,
		RQCD:2022xux}, the approximate pion and kaon masses, the physical size of the
	lattice and the length of the Monte Carlo chain in Molecular
	Dynamics Units (MDU). Ensembles with an asterisk are used to control finite-size
	effects, but are not included in
	the final analysis. Ensembles marked by a dagger lie on a second chiral trajectory
	where $m_{\rm s} \approx m_{\rm s}^{\rm phys}$. 
	Ensembles in bold face have either been added or the current correlator 
	has been determined with significantly improved precision with respect 
	to~\cite{Gerardin:2019rua}. 
}
\end{table}

%% file: sec_results.tex
\section{Results \label{s:res}}
In this section, we describe our calculation of $\aLD$ in isospin-symmetric
QCD. Combining the result with our earlier determinations of $\aSD$
\cite{Kuberski:2024bcj} and $\aID$ \cite{Ce:2022kxy} allows us to
present an updated result for $\ahvp$ with respect to
\cite{Gerardin:2019rua}. As in our earlier work, we prefer to perform
an isospin decomposition of the electromagnetic current and first
present the computation of the isovector and isoscalar contributions
to $\aLD$. To facilitate the comparison with other groups, we also
provide results for individual flavour components.
Based on an estimate for the leading isospin-breaking effects,
we compute an updated value of $\ahvp$ that can be directly compared
to experiment.

\subsection{The isovector contribution}
The isovector contribution dominates by far the central value and the
uncertainty of $\ahvp$ and $\aLD$. Therefore, its precise computation
is the main focus of this work. As explained in \cref{s:lat_noise}, we
use a combination of noise reduction methods to compute the isovector
correlation function to high accuracy, especially at close-to-physical
pion masses where the signal-to-noise problem is enhanced.

\begin{figure}[t]
	\centering
	\includegraphics*[width=0.8\linewidth]{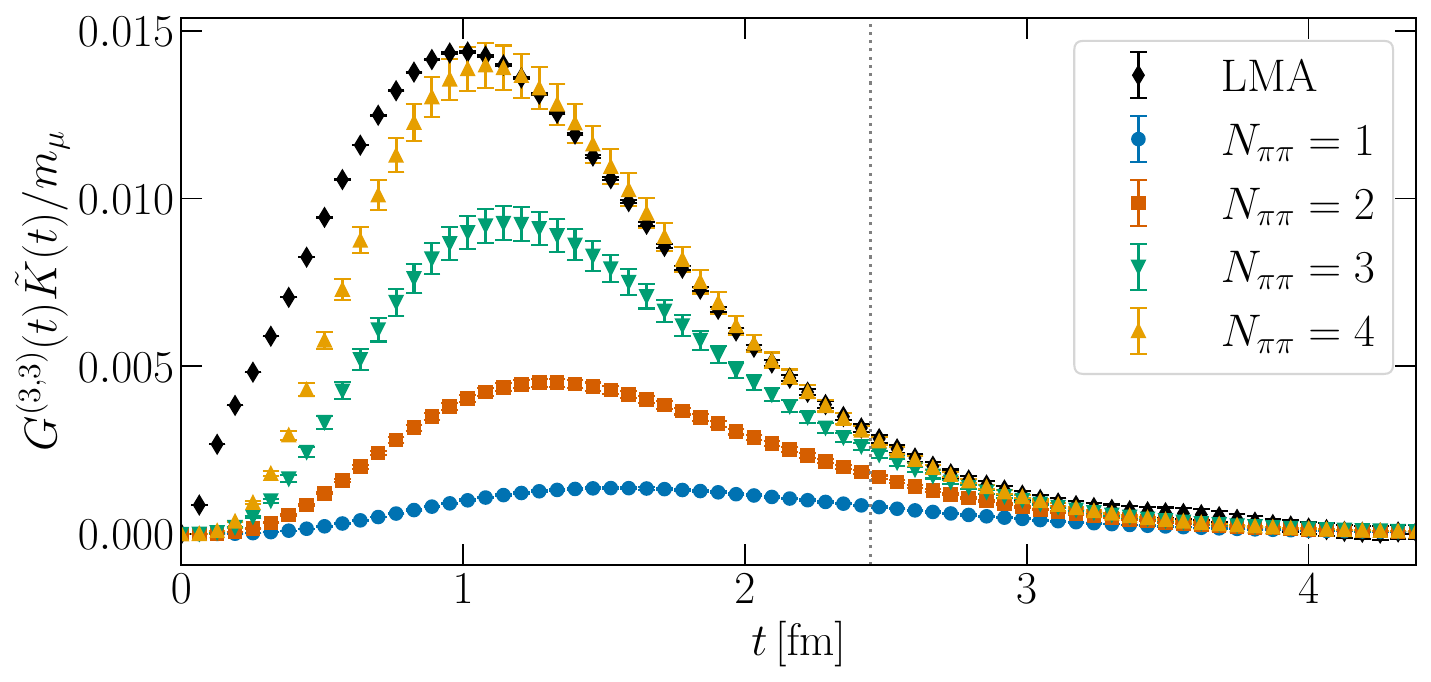}%
	\caption{%
		The integrand to compute $\ahvpf{3}{3}$ on the physical mass ensemble E250.
		The black diamonds are based on the correlation function that is computed using LMA. 
		The coloured data show the reconstructed integrand from $N_{\pi\pi}$ 
		states. 
		The vertical dotted line denotes the distance where we change from the LMA to 
		the spectroscopy data set.
	}
	\label{f:I1_E250}
\end{figure}

Here, we focus specifically on the combination of noise reduction
techniques applied on ensemble E250 at (slightly smaller than)
physical value of the pion mass. Whereas stochastic sources have been
utilized to determine the isovector correlation function on this
ensemble in \cite{Gerardin:2019rua}, we now employ LMA to maximize the
extractable information in the long-distance tail from the gauge
ensemble. The computation is described in detail in \cref{a:lma}, and
the expected dominance of low modes in the tail is highlighted in
\cref{f:LMA_contribs}. When used in combination with the bounding
method, we find that LMA alone allows us to reduce the uncertainty on
$\ahvp$ from 2.2\% in ref.~\cite{Gerardin:2019rua} to 0.8\% in this
work. In \cref{f:I1_E250} we show the integrand to compute
$\aLDf{3}{3}$ on E250, where the black diamonds denote the integrand
computed from the LMA correlation function.

Recalling that the isovector correlator in the long-distance regime is
dominated by two pions, we have also performed a dedicated study to
determine the spectrum of the lowest-lying two-pion states and their
overlap with the isovector current (see \cref{a:spec} for an in-depth
description). The coloured data points in \cref{f:I1_E250} show the
accumulated contributions of an increasing number of two-pion states
to the TMR integrand. We find that four states saturate the correlator
at a source-sink separation of about $1.5\,$fm. We note in passing
that the largest energy level that enters this reconstruction is
slightly above the mass of the $\rho$ meson. As stated
already in \cref{s:lat_noise}, the correlation function that has been
computed using LMA is more precise at this distance. However, since
the signal-to-noise ratio deteriorates exponentially in the LMA
correlation function whereas it stays basically constant for the
correlator reconstructed from two-pion states, it is clear that there
exists a distance above which the reconstruction is more precise. For
our specific calculation, this happens for $t \gtrsim 2.4\,$fm.

To combine the two sets of data, we follow one of the methods that
have been explored in our previous work \cite{Gerardin:2019rua} and
replace the directly computed correlation function with the
reconstructed one beyond a specific source-sink separation where the
reconstructed correlation function is more precise.
We note that neglected contributions from higher states are even less
significant for these larger source-sink separations. 
Four-pion states have been shown to be numerically irrelevant 
in \cite{Bruno:2019nzm}, since their overlap with the isovector
correlator is very small.
The combination of the two data sets allows us to further reduce the
relative uncertainty of $\ahvpf{3}{3}$ on this ensemble by a factor of 
two to $0.4\%$ (excluding the uncertainty of the scale setting quantity).
Since ensemble E250 has close-to-physical quark masses, this
result provides a strong constraint for the chiral-continuum fit and
has a direct influence on the attainable precision at the physical point.
The second ensemble where we employ spectroscopy data is 
D200 with $m_\pi \approx 200\,$MeV.
With respect to the previous application in \cite{Gerardin:2019rua},
we have added an LMA computation of the isovector correlation function.
The combination of both methods reduces the statistical uncertainty
by about 25\% on this ensemble, compared to pure LMA.

Across our set of 34 gauge ensembles, we reach a precision of
0.35\%$-$1.5\% for $\ahvpf{3}{3}$ and 0.55\%$-$2.4\% for $\aLDf{3}{3}$
(see \cref{tab:aLD} for an overview of results). We find that
high-precision results on ensembles with close-to-physical pion masses
are crucial to constrain our global fit in the relevant region of the
parameter space. For most of the ensembles with lattice spacings
$a\leq 0.05\,$fm, autocorrelations limit the attainable precision such
that longer Monte Carlo chains are needed to reduce the uncertainties.

As outlined in \cref{s:setup_extrap}, we scan over a variety of fit forms
and data selections to determine our final result at the physical point
from a model average.
For the chiral dependence, see eq.~(\ref{e:cc_pi}), we find that fits without
a chirally divergent term do not lead to acceptable fit quality, which is why
we exclude them from the model average. 
The term $1 / \Xpi$ in $f_{\rm ch, 2}$ is only used in conjunction with
$\log(\Xpi)$ in  $f_{\rm ch, 1}$, following an observation in 
\cite{Colangelo:2021moe} that this combination could be favoured for 
pion masses below the physical point.
This leads to five different ansätze for the chiral behaviour that are combined
with eight ansätze for the continuum extrapolations and four subsets of 
the data.

Upon inspecting the different classes of fits with their respective model 
weights, we make the following observations. 
Fits that only include a single term to parameterize the lattice spacing
dependence generally have good quality and are thus preferred over 
fits that include higher-order lattice artifacts, which however have a
non-negligible model weight.
Fits that include mass-dependent cutoff effects favour slightly larger 
values of $\aLDf{3}{3}$ at the physical point than fits without this extra
term.
Varying the anomalous dimension $\hat{\Gamma}$ does not lead to
significant changes in the fit quality or the result in the continuum
limit. However, fits with a non-zero anomalous dimension prefer
slightly smaller values of $\aLDf{3}{3}$.

The chiral behaviour is tightly constrained by the precise data point of the
E250 ensemble at physical pion mass. 
Two parameters are sufficient to describe the chiral behaviour with good
fit quality.
The inclusion of a third parameter to parameterize the dependence on the 
squared pion mass leads to an insignificant shift towards larger values
of $\aIDf{3}{3}$.

\begin{figure}[t]
	\includegraphics[width=.51\textwidth]{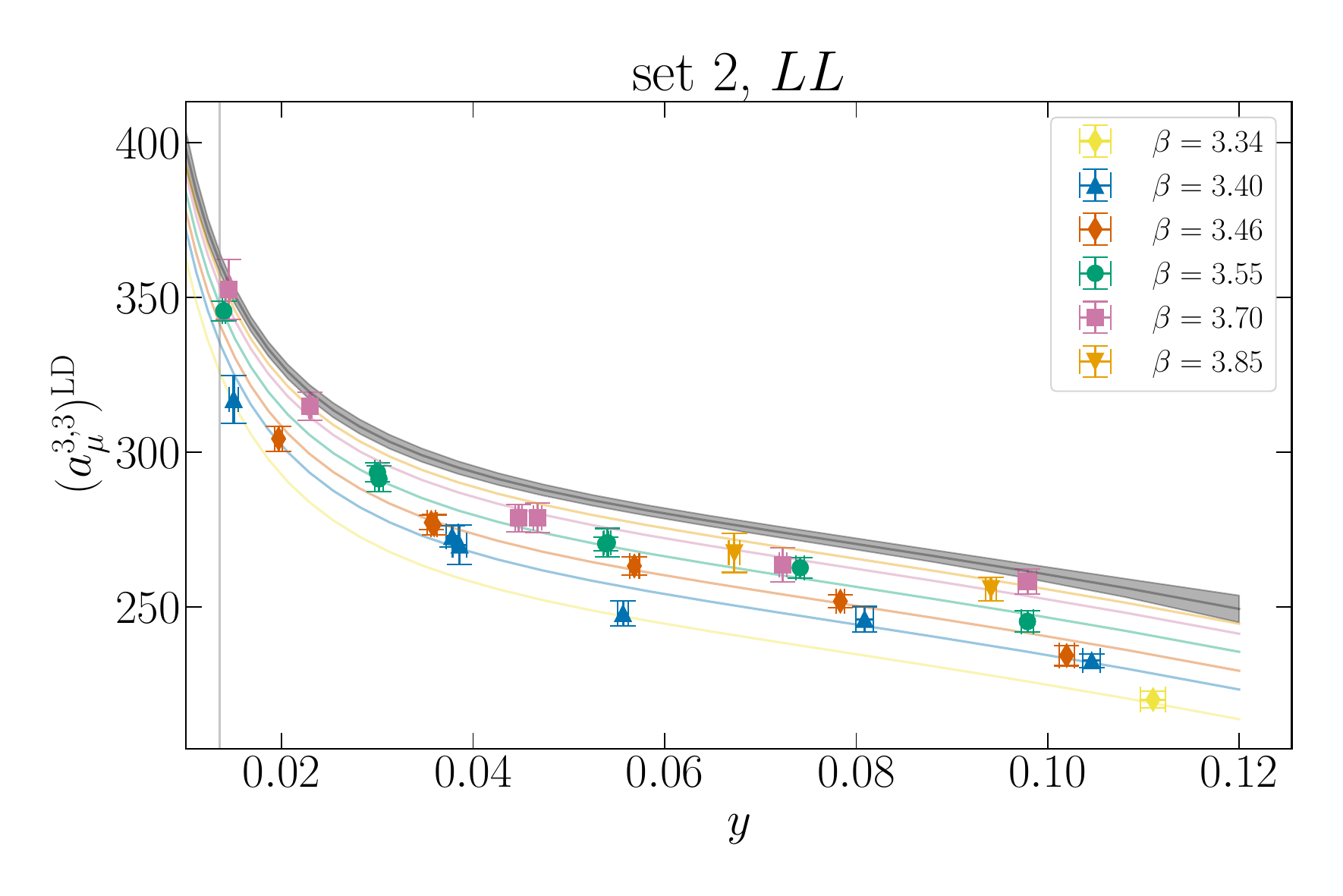}%
	\includegraphics[width=.48\textwidth]{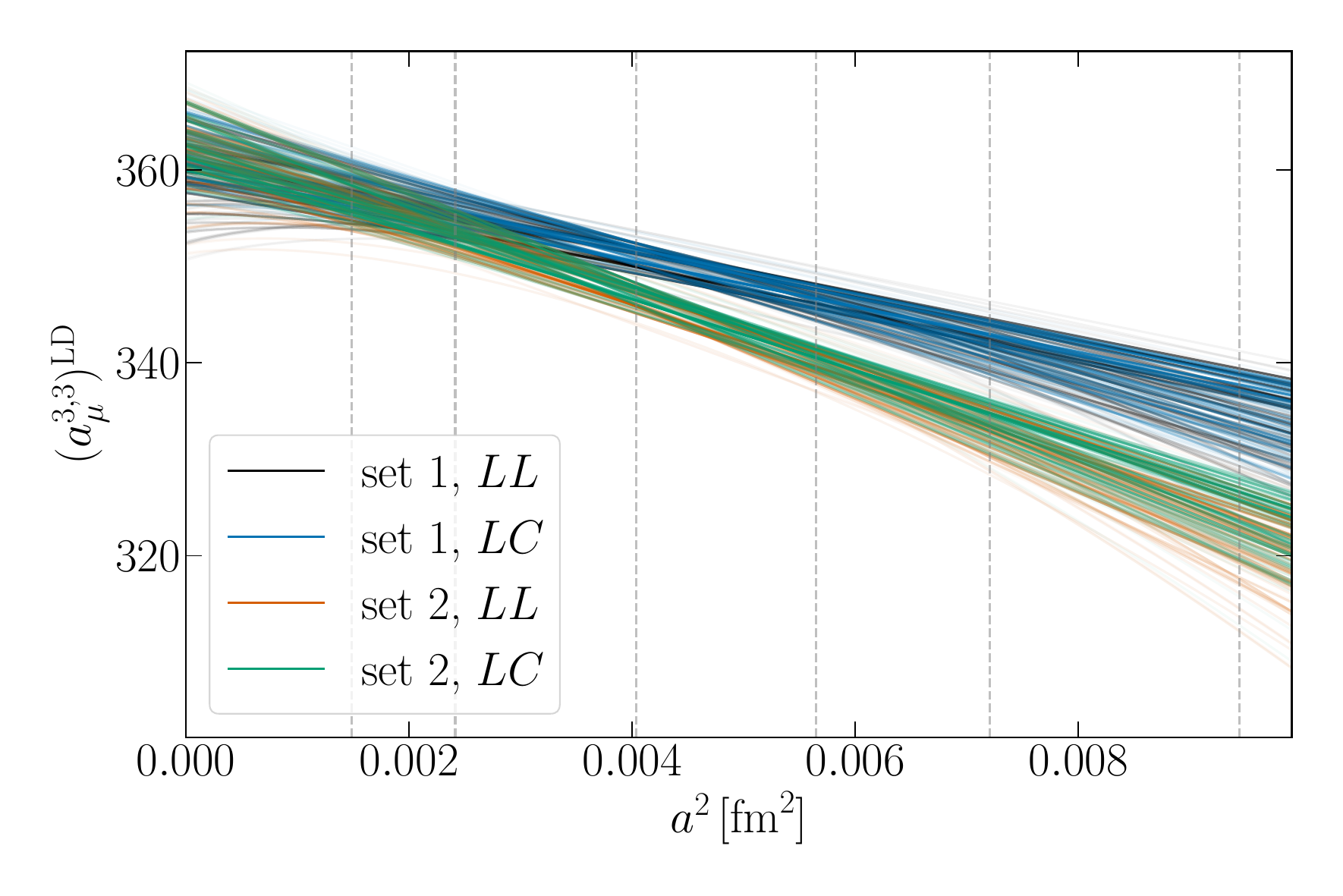}%
	\caption{\label{f:fit_I1}%
		Chiral-continuum extrapolation of the isovector contribution to $\aLD$.
		\textit{Left:} Illustration of the best fit according to the AIC to the data 
		based on the improvement scheme set 2 and the $\loc\loc$ discretisation. 
		The data points denote the result for each ensemble, 
		corrected for a deviation from the physical value of $z$. 
		The black line denotes the chiral dependence in the continuum limit and 
		the grey area the statistical uncertainty. The coloured lines correspond 
		to the chiral dependence at non-zero lattice spacing.
		The dotted vertical line denotes the physical value of $y$.
		\textit{Right:} Approaches to the continuum limit for four sets of data 
		based on the improvement schemes of set~1 and 2 and the $\loc\loc$ 
		and $\loc\cons$ discretizations of the current based on a scan over fit models.
		Each line shows the result from one single fit and the opacity of
		the lines corresponds to the weight of the fit in the model average.
		Dashed vertical lines indicate the lattice spacings used in this work.
		The conversion to fm has been performed for illustrative purposes only.
	}
\end{figure}

The left-hand side of \cref{f:fit_I1} depicts the chiral-continuum
extrapolation with the highest model weight for the $\loc\loc$
discretization of the vector current, using set~2 of the improvement
and renormalization coefficients.
No cuts in the data have been applied in this instance.
The data are adjusted for deviations from $z^{\rm phys}$ and presented
alongside the evaluation of the chiral behaviour at finite lattice spacing
(shown by the coloured lines) and in the continuum limit (represented
by the grey error band).
As can be seen from the figure, the chiral dependence is well constrained
over the full range of pion masses.
This includes the region  $m_\pi < 230\,$MeV or $y < 0.04$, respectively, 
which is sampled by eleven gauge ensembles and exhibits a strong curvature.

The panel on the right-hand side of \cref{f:fit_I1} illustrates the
continuum extrapolation at physical quark masses for each of the fits
included in the model average across four data sets.
Each fit is represented by a line whose opacity corresponds to the weight
in the model average for the respective data set.
The local and conserved discretizations of the vector current show only
marginal differences.
Comparing sets~1 and~2, a small difference is observed at finite lattice
spacing, which, as anticipated, disappears in the continuum limit where all
extrapolations are in close agreement.

For the coarsest lattice spacing, the relative size of the cutoff
effects is about 10\%, while the extrapolation from the finest lattice
spacing is very small. We note that replacing the scale setting
quantity by, for instance, $\sqrt{t_0}$ or $w_0$, has a significant
impact on the magnitude of the cutoff effects. From this observation,
we infer that the relative cutoff effects between $\aLD$ and the scale
setting quantity dominate over the intrinsic cutoff effects of $\aLD$
itself.  More precise data points at the two finest lattice spacings
will help to further constrain the continuum extrapolation.

Our final result in the reference volume, based on the $\loc\loc$
discretization and set~2 is
\begin{align}\label{e:result_aLD33fv}
	\aLDf{3}{3}(L_{\rm ref}) =
	362.0(3.7)_{\rm stat}(2.7)_{\rm syst}[4.6]
	\,.
\end{align}
We note that statistical uncertainties dominate over the systematic
uncertainties from the variation of the fit models.
The final uncertainty of the long-distance, isovector contribution 
in finite volume, reported in square brackets and obtained by adding statistical 
and systematic uncertainties in quadrature, is at the level of $1.3\%$.
Combining the result of \eq{e:result_aLD33fv} with the finite-volume effects
that have been computed in \eq{e:result_aLDfse} for the isovector channel, 
we obtain
\begin{align}\label{e:result_aLD33}
	\aLDf{3}{3} =
	378.7(3.7)_{\rm stat}(3.1)_{\rm syst}[4.8]
	\,.
\end{align}

\subsection{The isoscalar contribution}
At the SU(3) symmetric point, where light and strange quark masses are equal,
the quark-disconnected contribution from light and strange quarks vanishes,
and the isoscalar contribution is trivially related to the isovector one.
As one approaches the physical values of quark masses, a strong
signal-to-noise problem is observed, since the absolute error of the
quark-disconnected correlation function remains constant as a function
of the source-sink separation. As described in \cref{s:lat_noise}, we
employ the bounding method to obtain reliable estimates for
$\aLDf{8}{8}$.

In contrast to the isovector case, where some of our most precise data
points are at small pion masses, we find that statistical
uncertainties in the isoscalar channel grow towards physical quark
masses. However, the chiral dependence in the isoscalar channel is
much more benign, as the singular behaviour of light-connected and
disconnected contributions in the isoscalar channel cancels, as do the
leading finite-size effects. We perform a small finite-size
correction of the strange-connected contribution, which is relevant
only at or near the SU(3)-symmetric point, where the kaon mass is
relatively small ($m_K L \geq 5.1$ across all of the ensembles
entering the final fits).

To describe the chiral dependence of the isoscalar contribution, we only
include terms that do not diverge in the chiral limit and find the
variation of the results for different ansätze to be mild.
Fits that include mass-dependent cutoff effects are strongly favoured 
by the AIC and lead to smaller results at the physical point,
compared to fits with pure $a^2$ cutoff effects.
Models with non-zero anomalous dimension have very similar fit quality
compared to fits with $\hat{\Gamma}=0$ and lead to slightly larger results
in the continuum limit.

\begin{figure}[t]
	\includegraphics[width=.51\textwidth]{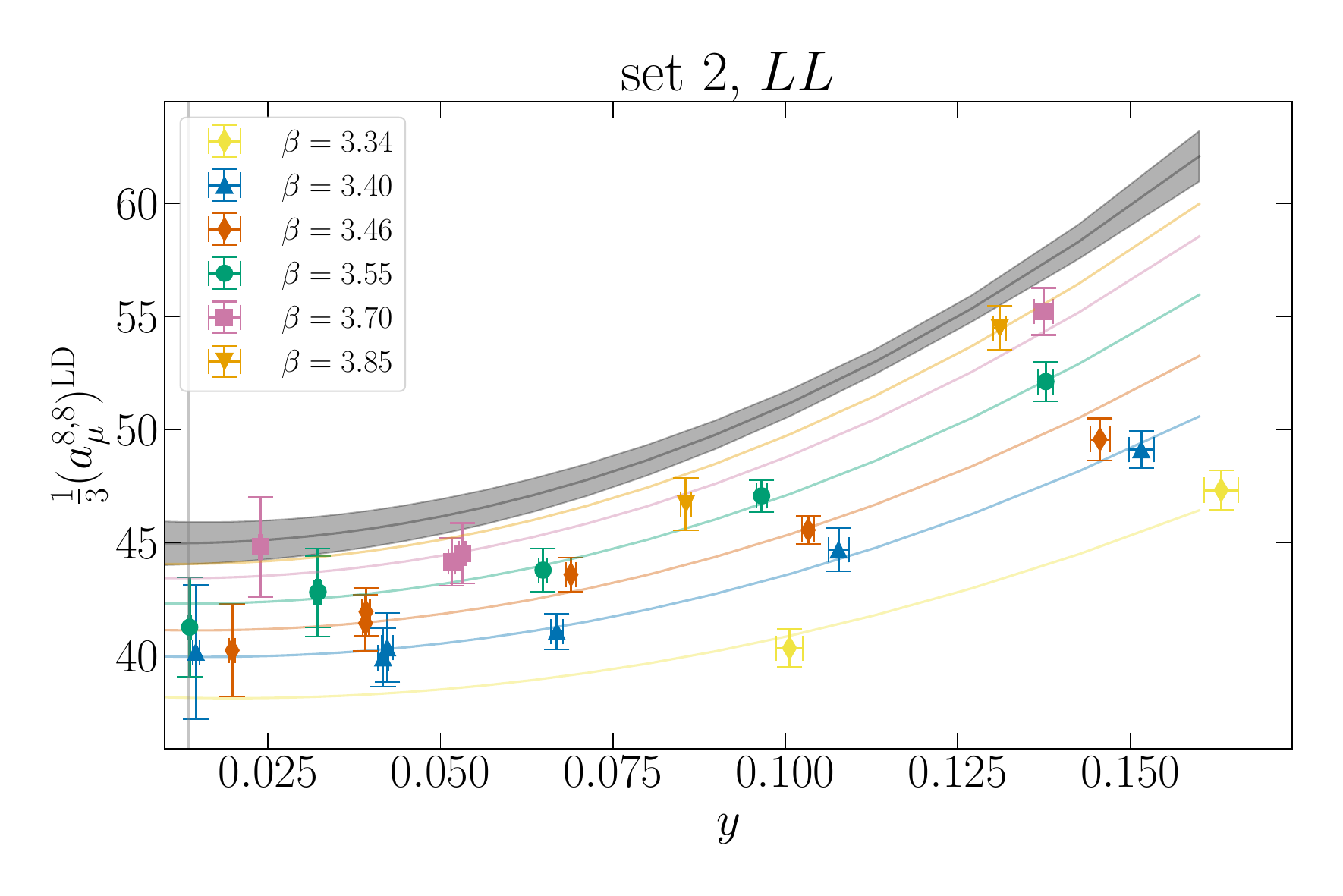}%
	\includegraphics[width=.48\textwidth]{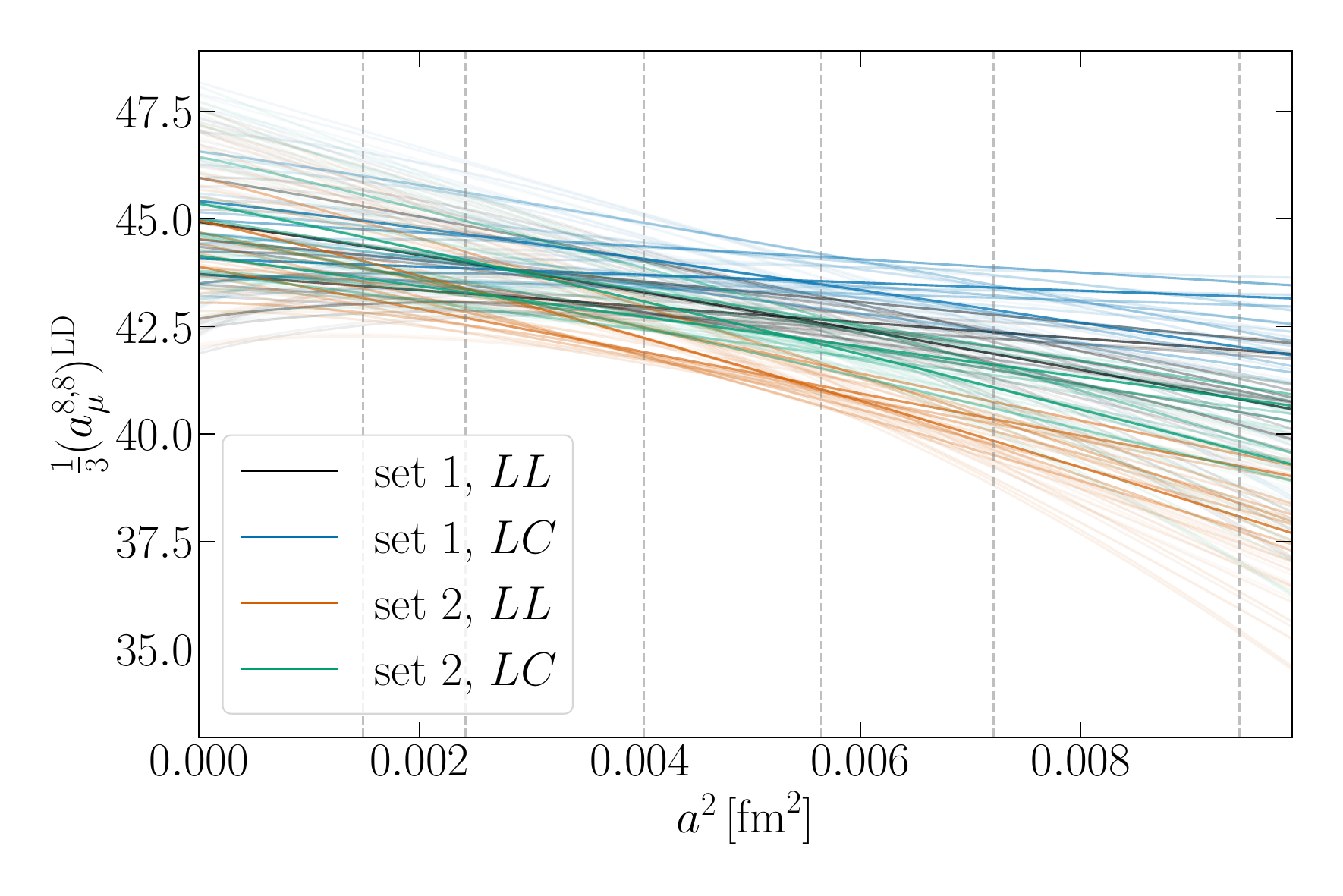}%
	\caption{\label{f:fit_I0}%
          Same as \cref{f:fit_I0} for the isoscalar contribution.
	}
\end{figure}

The left-hand side of \cref{f:fit_I0} shows our best fit for set~2 and
the $\loc\loc$ discretization, which employs two terms to describe the
chiral dependence of the data.
The dependence on the variable $z$ (see \eq{e:yzdef}), which
is expected due to the strange-quark mass content of the isoscalar
contribution, is well described by the linear term in our fit ansatz and
constrained by the four ensembles on the chiral trajectory where the
strange quark mass is kept near its physical value.

The scan over the different ansätze to perform the continuum extrapolation
is depicted on the right hand side.
Again, the variation between the different discretization prescriptions is
negligible in the continuum.
Based on the model average, we find
\begin{align}\label{e:result_aLD88}
	\txts\frac{1}{3}\aLDf{8}{8} =
	44.5(1.2)_{\rm stat}(1.1)_{\rm syst}(0.3)_{\rm FV}[1.6]
	\,,
\end{align}
for the isoscalar contribution, including the numerically irrelevant estimate 
of $0(0.3)$ for the finite-volume correction, see \cref{s:fv}.
Statistical and systematic uncertainties have a similar size and the combined
uncertainty is at the level of $3.6\%$, mainly driven by the
statistical noise encountered for the quark-disconnected contribution.

\subsection{Further contributions}
The charm-connected contribution at long distances is very small since
the correlator falls off quickly and, as described below, we find that
its contribution to the total $\aLD$ is smaller than the overall
error. For our evaluation of the long-distance contribution, we employ
the same data set as the one that has been used in \cite{Ce:2022kxy,
  Kuberski:2024bcj}. The charm quark is partially quenched in the sea
of light and strange quarks, and we tune the hopping parameter to
reproduce the $D_{\rm s}$ meson mass at finite lattice spacing (see
\cref{a:scheme}). Statistical noise is irrelevant and we simply sum
the integrand in the long-distance region. In line
with~\cite{Ce:2022kxy, Kuberski:2024bcj} we use the $\loc\cons$
correlation function to compute the final result since it exhibits
more benign cutoff effects. Our result
\begin{align}\label{e:result_aLDcc}
	\txts\frac{4}{9}\aLDf{c}{c} =
	0.01409(35)_{\rm stat}(60)_{\rm syst}[69]
	\,,
\end{align}
is negligibly small with respect to the full $\aLD$ and even the
charm-connected contribution to $\ahvp$, which arises predominantly
from the short- and intermediate-distance windows.

Effects from charm-disconnected contributions have been shown to be
numerically irrelevant already for the short-distance contribution in
\cite{Kuberski:2024bcj} such that we do not consider them here.  We
have estimated the effect from missing charm loops in our computation
in appendix~C of \cite{Ce:2022kxy} and section~H of
\cite{Kuberski:2024bcj} and expect them to be irrelevant in the
long-distance regime compared with the uncertainties reported here.
Note also that the RBC/UKQCD collaboration has investigated the effect
of charm quenching on $\aID$ and did not find a numerically relevant
contribution~\cite{RBC:2023pvn}.
Furthermore, no evidence for a charm quenching effect on the quantity $r_1$ defined
from the static potential is seen at the 1\% level when comparing the
results of~\cite{MILC:2010hzw} and ~\cite{Dowdall:2013rya}  (the scale being defined by $f_\pi$ in both cases; see the discussion in~\cite{FlavourLatticeAveragingGroupFLAG:2021npn}).
Dedicated studies of charm quenching effects on generic low-energy
observables in \cite{Bruno:2014ufa,Knechtli:2017xgy} find an
effect at the level of $0.2\%$.
We include this effect as additional uncertainty in our final estimate for $\aLD$.

\subsection{Flavour decomposition}
To allow for cross-checks of the various contributions to $\aLD$ among
different lattice calculations, we also perform an analysis of the
strange-connected contribution that enters our final result via the
isoscalar contribution in \eq{e:result_aLD88}. After performing the
model average, we find
\begin{align}\label{e:result_aLDss}
	\txts\frac{1}{9}\aLDf{s}{s} =
	17.73(17)_{\rm stat}(13)_{\rm syst}[21]
	\,.
\end{align}
By combining this result with 
\eq{e:result_aLD33fv}, \eq{e:result_aLD88} and \eq{e:result_aLDfse}
we determine the disconnected contribution in the infinite-volume limit as
\begin{align}\label{e:result_aLDdisc}
	\aLD_{\rm disc} =
	-15.3(1.2)_{\rm stat}(1.2)_{\rm syst}[1.6]
	\,.
\end{align}
To complete the set of results according to their decomposition in
terms of quark flavours, we note that the light-quark connected
contribution is obtained by multiplying the isovector contribution of
\eq{e:result_aLD33} by 10/9.

\subsection{The long-distance contribution}
\input{./tables/tab_scales_LD}
We combine our results for isovector, isoscalar and charm-connected
contributions in 
eqs.~(\ref{e:result_aLD33}, \ref{e:result_aLD88}, \ref{e:result_aLDcc})
in the infinite-volume limit to obtain
\begin{align}\label{e:result_aLD}
	\aLD =
	423.2(4.2)_{\rm stat}(3.3)_{\rm syst}(0.8)_Q[5.4]
	\,,
\end{align}
for the long-distance contribution to $\ahvp$ in isospin-symmetric
QCD, where we include an additional uncertainty due to the quenching
of the charm quark, denoted by the subscript~$Q$. Our hadronic scheme
is defined by
\begin{align}
  f_\pi&=130.56\,\mathrm{MeV}, \qquad f_\mathrm{K}=157.2\,\mathrm{MeV}, \nonumber\\
  m_\pi&=134.9768\,\mathrm{MeV},\quad m_K = 495.011\,\mathrm{MeV},
  \quad m_{D_{\rm s}} = 1968.47\,\mathrm{MeV}\,. 
\end{align}
More details can be found in \cref{a:scheme}.
The conversion to other schemes can be easily performed using the
information collected in table~\ref{t:scaledep_LD}, where we list the
dimensionless scale dependencies
\begin{align}\label{e:scaledep}
  \frac{S}{O}\frac{\partial O}{\partial S}\,,
\end{align}
for $O=\aLDf{3}{3},\,\aLDf{8}{8},\,\aLDf{\rm c}{\rm c},\,\aLD$ and
each quantity $S$ that is used to define the scheme.
As anticipated, the dependence on $f_\mathrm{K}$ is strongly
suppressed in the dominant contributions with respect
to the dependence on $f_\pi$. 
For the numerically irrelevant charm-connected contribution, the
scale dependence is dominated by the tuning of the valence charm
quark mass.
Small changes in the scheme can be performed a posteriori given the
information provided in the table. To convert our results to a scheme
that employs a different quantity to set the scale, such as the $\Omega$
baryon mass, the derivative of $f_\pi$ and $f_\mathrm{K}$ with respect
to this quantity must be determined.

Currently, there is only one other result~\cite{Blum:2024drk} for the isovector 
contribution to $\aLD$, while no further results currently exist for the isoscalar contribution.%
\footnote{After the completion of this manuscript, the Fermilab/HPQCD/MILC 
collaborations released a preprint reporting a result for 
$\aLDf{3}{3}$~\cite{Bazavov:2024eou}. 
Their analysis, based on rooted staggered fermions, tested two different 
scale-setting choices, $f_\pi$ and $m_\Omega$, and in both cases obtained 
a value smaller than the one presented here.  
The ETM collaboration has also recently computed the strange- 
and charm-connected contributions to the long-distance window~\cite{ExtendedTwistedMassCollaborationETMC:2024xdf}, 
setting the scale with $f_\pi$. 
Their results are compatible with ours.}
Before proceeding to comparisons, we comment on the dependence of the 
results on the chosen hadronic scheme that defines isoQCD.
First, we remark that, on CLS ensembles, determinations of the flow-scale $t_0$ via
the physical quantities $(f_K + \frac{1}{2}f_\pi)$~($1.0\%$ precision, \cite{Strassberger:2021tsu}),
$m_{\Xi}$~($0.6\%$,~\cite{RQCD:2022xux}), $m_\Omega$~(0.22\%, \cite{Hudspith:2024kzk}) and $m_N$ ($0.6\%$,~\cite{Lingscheid2024})
yield consistent results within their respective uncertainties.
Among the determinations by different collaborations of the flow scales $t_0$ and $w_0$
in terms of various input quantities ($m_\Omega$, $f_\pi$, \dots),
however, somewhat more variation is observed.
In particular, the determinations of $t_0$ from~\cite{Dowdall:2013rya,MILC:2015tqx} are significantly lower
than that of~\cite{ExtendedTwistedMass:2021qui}, even though all three use $N_f=2+1+1$
simulations and rely on $f_\pi$ as input quantity.
There is mild evidence that flow scales determined with Wilson-type fermions~\cite{Strassberger:2021tsu,RQCD:2022xux,ExtendedTwistedMass:2021qui}
are systematically larger than those obtained with staggered quarks,
including the result in~\cite{Borsanyi:2020mff} for $w_0$ that defines the BMW20 scheme.
At present,
this makes it difficult to quantitatively address the question of the hadronic scheme dependence
in the physical values of $t_0$ and $w_0$.
However, the dependence of $\aLD$ on the hadronic scheme 
could be relevant due to its high precision, its enhanced sensitivity to the scale setting 
and relatively large contributions from isospin-breaking corrections.
The size of the latter is not yet precisely known, and care is needed when 
combining results from different isoQCD schemes.

\begin{figure}[t]
	\centering
	\includegraphics[width=.65\textwidth]{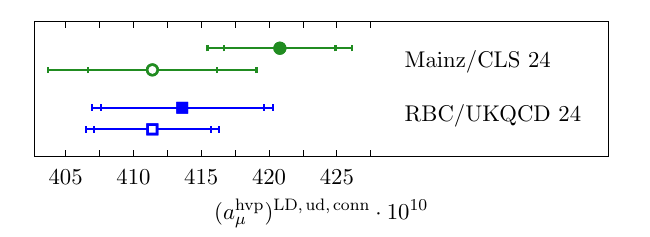}%
	\caption{\label{f:comparison_aLD33}%
	Overview of results for the light-connected contribution to $\aLD$. 
	Open symbols denote the results in the BMW20 scheme whereas 
	results that are shown by filled symbols have been computed in the
	Mainz and RBC/UKQCD worlds, respectively. 
	}
\end{figure}

In \cref{f:comparison_aLD33} we compare our results for $\aLDf{3}{3}$
in our preferred scheme~\eq{e:result_aLD33} and in the BMW20
scheme~\eq{e:result_aLD33_BMW20} with the recent determination by
RBC/UKQCD \cite{Blum:2024drk} in their RBC/UKQCD18 scheme and in
the BMW20 scheme~\cite{Borsanyi:2020mff}. We find excellent agreement between
the two calculation when the same scheme is employed.\footnote{In this
comparison, the contribution of the scale uncertainty to the error is
not included.} This is a reassuring indication of universality
between two different lattice actions in the pure long-distance regime
of $\ahvp$ and strengthens our confidence in the reliability of 
lattice QCD results for this quantity.
However, the results differ noticeably when a different scheme is
employed. While this is not unexpected in isospin-symmetric QCD, it is
clear that any scheme dependence would have to be compensated upon
properly including  isospin-breaking effects.

We stress that we observe sizeable higher-order cutoff effect when
$w_0$ is used to set the scale, leading to larger overall
uncertainties in the continuum limit. This is why we have chosen
$f_\pi$ and $f_\mathrm{K}$ in $2+1$-flavour QCD as scale-setting
quantities, as outlined in appendix~\ref{a:scheme}.
By contrast, the BMW20 scheme is based on the $\Omega$-baryon mass
computed in $2+1+1$-flavour QCD+QED, which is used to determine the
value of $w_0$ at the physical point.
When the latter is used as input in our calculation in order to connect to the BMW20 scheme,
we observe a shift in the central value of our result. We cannot presently resolve
whether this shift is entirely explained by the different choice of scale.

\subsection{Full hadronic vacuum polarization contribution}
Having computed $\aLD$ in isoQCD, we can combine it with our
results from \cite{Ce:2022kxy} and \cite{Kuberski:2024bcj} which,
without the inclusion of isospin-breaking effects, read
\begin{align}\label{e:result_aSDaID}
	\aSD &=
	68.76(21)_{\rm stat}(38)_{\rm syst}[44]
	\,,\\
	\aID &=
	236.60(79)_{\rm stat}(1.13)_{\rm syst}[1.38]
	\,.
\end{align}
We take the small correlation between the three observables into
account when summing them.
It is worth noting that we have used the intermediate scale setting
quantity $\sqrt{t_0}$ from \cite{Strassberger:2021tsu} in our 
computation of $\aSD$.
Since it has been determined from $f_{K \pi}$ using the exact same
values for $f_\pi$ and $f_K$ as the ones that were used for $\aID$ and
$\aLD$, see \cref{a:scheme}, we can consistently combine the three
windows.
%
\input{./tables/tab_contribs}

As our final result for $\ahvp$ in isospin-symmetric QCD as defined in
\cref{a:scheme}, we quote
\begin{align}\label{e:result_ahvp}
	(\ahvp)^{\rm isoQCD} 
	&= \aSD + \aID + \aLD  \nonumber\\
	&=
	728.6(4.3)_{\rm stat}(3.6)_{\rm syst}(0.8)_Q[5.6]
	\,.
\end{align}
Similarly, the light-quark connected contribution, which dominates in
the final result, is obtained by summing the results for the isovector
contribution listed in the second row of Table~\ref{t:contribs} and
multiplying by 10/9:
\begin{align}\label{e:result_lqconn}
	(\amu)^{ud,\,\rm conn} &=
	675.7(4.1)_{\rm stat}(3.7)_{\rm syst}[5.5]
	\,.
\end{align}
The electromagnetic and strong isospin-breaking corrections to these results are discussed in the next subsection.

We quote the results in eqs.~(\ref{e:result_ahvp}) and~(\ref{e:result_ahvp}) 
only in the $f_\pi$ scheme because we did not determine
the short and intermediate-distance window observables in the scheme
of~\cite{Borsanyi:2020mff}.
In their recent work \cite{RBC:2023pvn} the RBC/UKQCD collaboration
found only slight variations of these quantities between the BMW20 
and their own scheme which, however, also employs $m_\Omega$ to set the scale.

For all quark-connected flavour contributions we find excellent agreement 
with our previous work~\cite{Gerardin:2019rua}, albeit with significantly
reduced uncertainties.
In the case of the quark-disconnected contribution, we observe an upward shift that
can be understood from the fact that only a small fraction of the current data 
set was available in~\cite{Gerardin:2019rua} and an extrapolation to physical quark
masses had to be performed.
This increase in the quark-disconnected contribution is the main reason for the 
shift in the central value of $\ahvp$ between~\cite{Gerardin:2019rua} and this work that,
however, is entirely within the uncertainty of~\cite{Gerardin:2019rua}.

\begin{figure}[t]
\centering
\includegraphics[width=.75\textwidth]{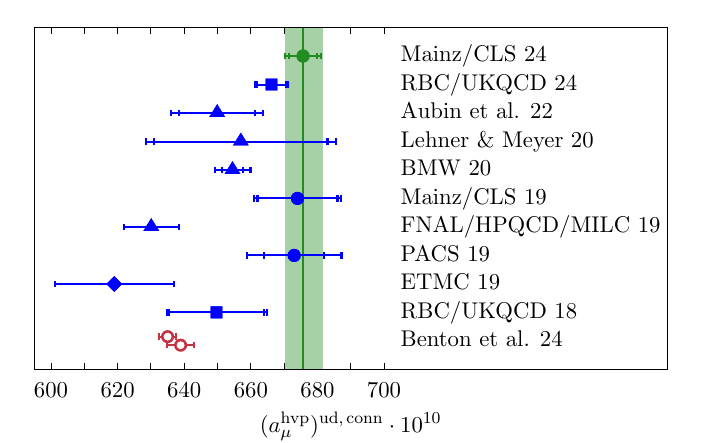}%
\caption{\label{f:comparison_ud}Comparison of our result for the
  light-quark connected contribution $\amul$ with other lattice
  calculations. Note that the numbers plotted refer to the particular
  scheme defining isospin-symmetric QCD adopted in each
  calculation. Our result quoted in \eq{e:result_lqconn} is
  represented by the green circle and vertical band. Different
  discretizations of the quark action are denoted by circles (Wilson
  fermions) \cite{Shintani:2019wai, Gerardin:2019rua}, triangles
  (staggered fermions) \cite{Davies:2019efs, Borsanyi:2020mff,
    Lehner:2020crt, Aubin:2022hgm}, squares (domain wall fermions)
  \cite{Blum:2018mom,Blum:2024drk} and diamonds (twisted-mass Wilson
  fermions) \cite{Giusti:2019xct}. The data-driven evaluations of
  Ref. \cite{Benton:2024kwp} based on the KNT
  \cite{Keshavarzi:2019abf} and DHMZ \cite{Davier:2019can} data sets
  are represented as open red circles.  }
\end{figure}

In~\cref{f:comparison_ud} we compare our results for the light-quark
connected contribution to other recent calculations. We stress that
the data in the figure have not been shifted to a common reference
scheme defining isospin-symmetric QCD. Our result for $\amul$ in our
preferred scheme is compatible with the recent high-precision result
of RBC/UKQCD~\cite{Blum:2024drk}. There is a clear difference with the
2021~result of the BMW
collaboration~\cite{Borsanyi:2020mff}. Unfortunately, BMW did not
provide an updated value for this contribution in their most recent
publication~\cite{Boccaletti:2024guq}. Assuming that the shift between
their two results for $\ahvp$ is mainly due to the light-quark
connected contribution, the difference would be reduced accordingly.
We note that our result is in clear tension with the evaluation from
the data-driven dispersive approach in~\cite{Benton:2024kwp},
regardless of whether the exclusive channel analysis of either
ref.~\cite{Keshavarzi:2019abf} or \cite{Davier:2019can} is used for
the latter.

\subsection{Electromagnetic and strong isospin-breaking effects}

In the following, we present the status of our calculations of the
electromagnetic and strong isospin-breaking effects. While not
complete, these calculations of some of the dominant diagrams already
allow us to estimate the full correction without the associated
uncertainty dominating the final error budget.

\begin{figure}[t]
	\centering
	\includegraphics[width=.75\textwidth]{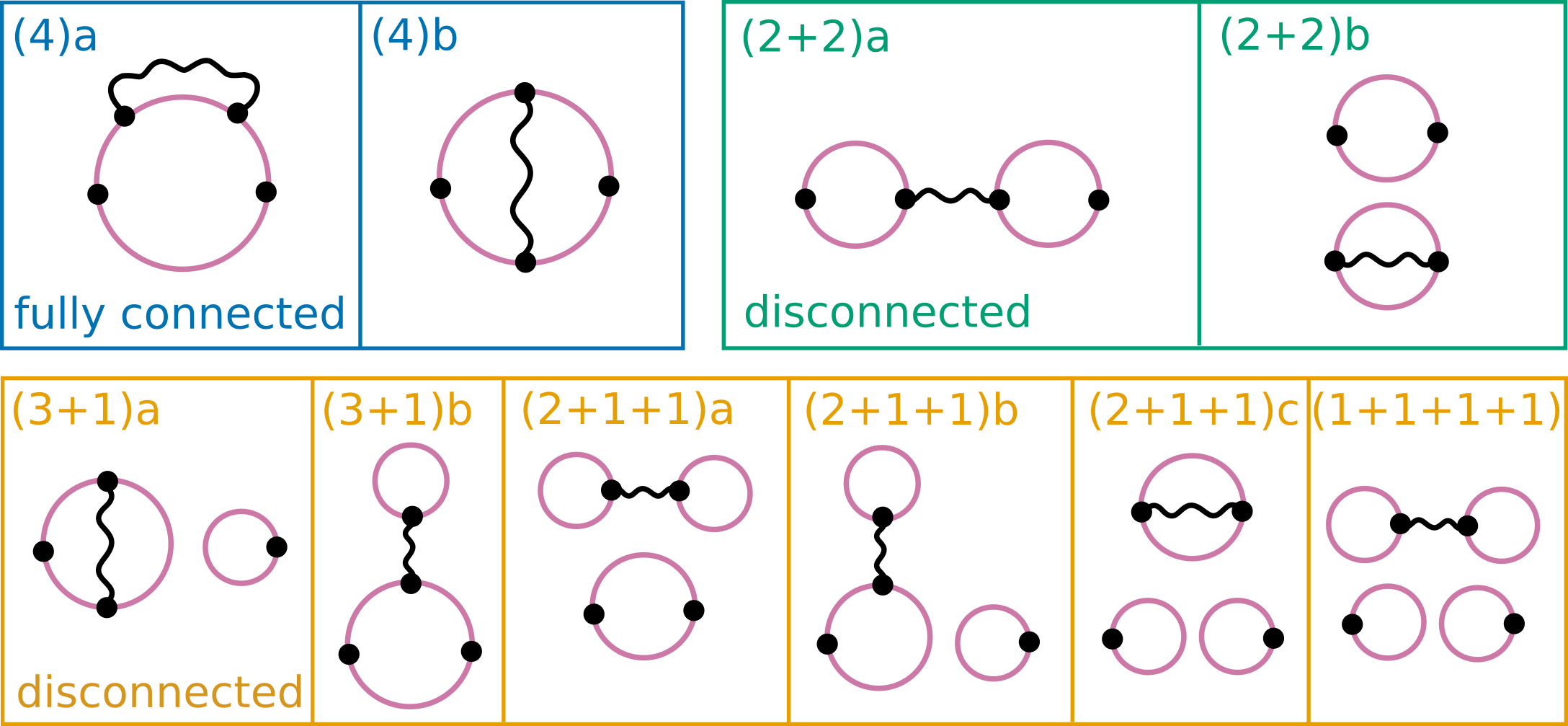}
	\caption{\label{f:ib_feyn}%
		Overview of diagrams relevant for QED corrections.
		} 
\end{figure}

An overview of the diagrams involving internal photons is shown in \cref{f:ib_feyn}.
They are classified by the number and type of quark loops involved,  
with five classes identified: fully connected (4), and the classes  
(2+2), (3+1), (2+1+1), and (1+1+1+1).
We have computed the connected part of the quark-mass insertion, as
well as the connected QED diagrams (4)a and (4)b on the lattice, with the photon
propagator evaluated in lattice regularization and infrared-regulated
by removing the spatial zero mode on each
timeslice, a method commonly referred to as ${\rm QED}_L$~\cite{Hayakawa:2008an, deDivitiis:2013xla, Risch:2021hty,Risch:2021nfs, Ce:2022kxy, Kuberski:2024bcj}.
These diagrams exclusively involve single-quark loops and form a UV-finite set.

In addition, we have computed the (2+2)a diagram~\cite{Chao:2023lxw},
consisting of two (valence) quark loops connected by an internal
photon, down to physical quark masses~\cite{Parrino:2025afq}. 
This diagram is UV-finite and has been computed with a photon propagator  
evaluated in the continuum and infinite volume  
using the coordinate-space approach from~\cite{Biloshytskyi:2022ets}. 
At small pion masses, it is found to be dominated by the charged pion loop.
The observation that the $\pi\pi\gamma$ vertex only involves the isovector part of the electromagnetic current
leads to two non-trivial homogeneous relations between the charged pion loop
contributions to the various classes of diagrams.
Neglecting the diagrams of the classes (2+1+1) and (1+1+1+1),
one then arrives at the following partition of the charged pion loop among
the diagrams~\cite{Chao:2021tvp},
\ba
a_\mu^{\pi\;{\rm loop},(4)} &=& \phantom{-}\frac{34}{81} a_\mu^{\pi\;{\rm loop}},
\\
a_\mu^{\pi\;{\rm loop},(2+2)} &=& \phantom{-}\frac{75}{81} a_\mu^{\pi\;{\rm loop}},
\\
a_\mu^{\pi\;{\rm loop},(3+1)} &=& -\frac{28}{81} a_\mu^{\pi\;{\rm loop};}.
\ea
The quantity $a_\mu^{\pi\;{\rm loop}}$ refers to the pion loop contribution,
which, at the simplest level (i.e.\ without a pion form factor), could be estimated
using scalar QED \cite{Parrino:2025afq}.
The three coefficients multiplying $a_\mu^{\pi\;{\rm loop}}$ sum to unity.
This partition is used below.

\begin{figure}[t]
	\centering
	\includegraphics[width=.8\textwidth]{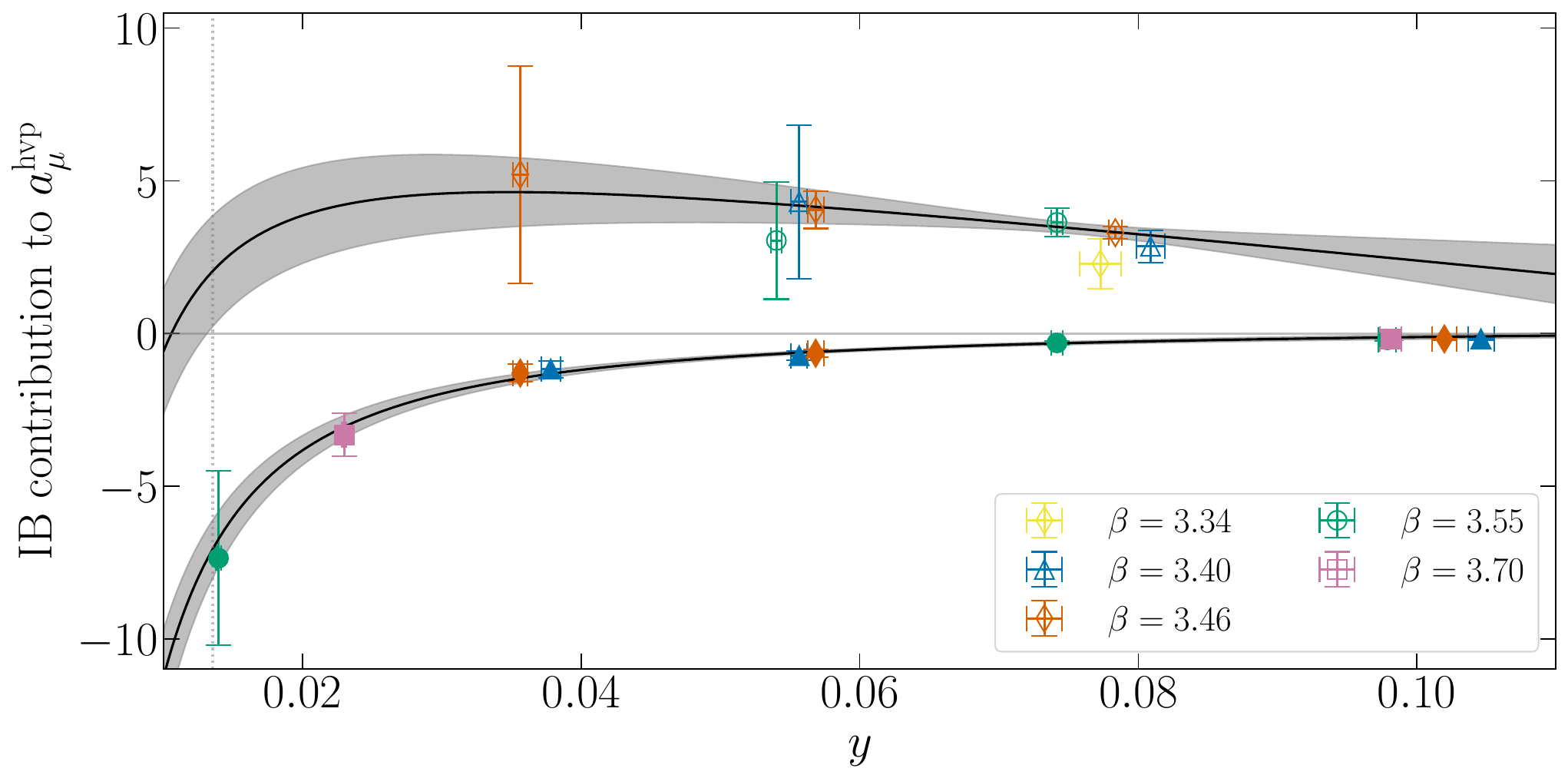}%
	\caption{\label{f:ib_fit}%
		Chiral extrapolation of the isospin-breaking
		corrections to the physical pion mass based on
		eqs.\,(\ref{e:IBconn}) and~(\ref{e:IB2p2a}).
		Open symbols denote the fully connected (4) plus disconnected scalar-insertion
		diagrams and filled symbols the (2+2)a diagrams. 
		The dotted vertical line denotes the physical value of $y$.
		} 
\end{figure}

We derive an estimate for the full QED correction by
extrapolating simultaneously the single-quark-loop diagrams and the diagram (2+2)a;
the remaining QED diagrams, consisting of at most two quark loops, are estimated
based on the charged pion loop.
Similarly, we add to the lattice results for the purely connected diagrams a chiral perturbation theory based estimate of the contribution
of the disconnected scalar-insertion diagram that is generated by the derivative $(m_u-m_d)\partial a_\mu^{\rm hvp}/\partial(m_u-m_d)$.

Based on the observation that the pion mass is insensitive to $(m_u-m_d)$ at linear order,
this diagram has been shown~\cite{Lehner:2020crt} to largely cancel the corresponding connected scalar-insertion diagram; individually however,
these two diagrams have a steep pion mass dependence. We determine the value of the prefactor $(m_u-m_d)$ from the kaon mass splitting,
specifically from the difference (amounting to about 6.0\,MeV) of the elastic part of the electromagnetic splitting and the physical $m_{K^+}-m_{K^0}$ splitting.

We remark that the size of the disconnected scalar-insertion diagram is small compared to the connected diagrams
at the pion masses for which we have computed the latter, however it becomes non-negligible at the physical pion mass.
Thus, the quantity notated $a_\mu^{\rm hvp1\gamma^*,(4)}$ below is defined to contain, in addition to all fully connected diagrams,
the disconnected scalar-insertion diagram estimated as described above.
The QED diagrams containing three or more quark loops are neglected.
Indeed, they are $1/N_c^2$ suppressed compared to the fully connected diagrams. 
Both the BMW\,2020 calculation~\cite{Borsanyi:2020mff} 
and the calculation of the light-by-light scattering contribution~\cite{Chao:2021tvp} found them to be small.%
\footnote{For a recent phenomenological estimate of isospin-breaking effects in $\ahvp$ see ref.~\cite{Hoferichter:2023sli}.}

We use the superscript $1\gamma^*$ to denote contributions that contain
one internal photon, including the required counterterms.
These contributions are one-photon irreducible, i.e.\ they are part of the
leading-order HVP contribution to $a_\mu$ in the standard nomenclature~\cite{Aoyama:2020ynm}.
The charged pion loop, computed with a pion form factor of the vector-dominance
form $M_V^2/(Q^2+M_V^2)$, has been found to behave like $1/m_\pi^3$ in the
mass range of 135 to 300\,MeV in a continuum calculation in scalar QED~\cite{Parrino:2025afq}.
Based on this observation, our ansatz for a combined fit reads
\ba\label{e:IBconn}
a_\mu^{\rm hvp1\gamma^*,(4)} &=& \frac{34}{81} \frac{A}{m_\pi^3}  + bm_\pi^2 + c + 0.22\,\log\frac{m_V^2}{m_\pi^2},
\\
a_\mu^{\rm hvp1\gamma^*,(2+2)a} &=& \frac{50}{81} \frac{A}{m_\pi^3}  + d\,,\label{e:IB2p2a}
\ea
where $A,\,b,\,c$ and~$d$ are fit parameters.
The logarithmic term corresponds to the neutral pion exchange contribution~\cite{Biloshytskyi:2022ets}, including its $34/9$ enhancement factor in
the connected part~\cite{Bijnens:2016hgx}.\footnote{In principle, the same contribution with a coefficient $-25/34 \cdot 0.22 = -0.16$
should be added to the (2+2)a diagram, however here we know that this contribution is largely compensated
by the $\eta$ and $\eta'$ contribution~\cite{ParrinoLAT24,Parrino:2025afq}.}
The importance of this term is marginal.
Using the fitted coefficients of the combined ansatz (\ref{e:IBconn}--\ref{e:IB2p2a}),
we obtain our estimate for the total correction to $a_\mu^{\rm hvp}$ in isoQCD as follows,
\be
a_\mu^{\rm hvp1\gamma^*} = \frac{A}{m_\pi^3}  + bm_\pi^2 + c + d + 0.22\,\log\frac{m_V^2}{m_\pi^2}.
\ee
This expression amounts to summing $a_\mu^{\rm hvp1\gamma^*,(4)}$ and $a_\mu^{\rm hvp1\gamma^*,(2+2)a}$, as well as to
including estimates of the missing electromagnetic diagrams via the pion loop, while diagrams containing three or more quark loops
are neglected altogether.

To account for cutoff, finite-volume, and higher-order quark mass effects,  
we explore multiple fit models and combine them in a model average.  
Our fits extend the ansatz in eqs.~(\ref{e:IBconn} -- \ref{e:IB2p2a})  
by incorporating terms for cutoff effects and additional components  
to parameterize the pion mass dependence in the  
$a_\mu^{\rm hvp1\gamma^*,(2+2)a}$ contribution~\cite{Parrino:2025afq}.  
We also apply cuts on lattice spacing, pion mass, and $Lm_\pi$,  
and average over both local-local and local-conserved discretizations  
of the current for the fully connected contribution.

Figure~\ref{f:ib_fit} shows the chiral dependence of each of the two  
contributions in the continuum according to the best fit in the model average,  
represented by the black line and the gray uncertainty band.  
The open symbols represent data from the combination of fully connected 
diagrams with an estimate for the disconnected scalar-insertion diagram,
while the filled symbols denote the (2+2)a contribution.  
The chiral dependence of the connected contribution near the physical pion mass,  
marked by the vertical line, is highly constrained by the curvature of the (2+2)a
contribution.
Our final estimate is given by\footnote{Ignoring the log term from the outset  
	would have yielded an irrelevant shift of $-0.2$.}  
\be\la{eq:QEDest}
a_\mu^{\rm hvp1\gamma^*} = -4.1(2.4)(0.9)(3.5)[4.4] 
\ee  
The first two contributions to the total uncertainty are the statistical  
and systematic uncertainties as obtained from the model average.
The third quoted error corresponds to half the absolute size
of the disconnected diagram $a_\mu^{\rm hvp1\gamma^*,(2+2)a}$.
We assign this uncertainty to our result to account for missing contributions 
from electrically charged sea quarks, as well as potential systematic effects 
from our parameterization based on the pion loop.
Indeed, due to the observed cancellations between diagrams,  
in particular between the connected and the (2+2)a diagram,
we estimate half the size of the latter to provide a conservative estimate of the total uncertainty.

\begin{figure}[t]
\centering
\includegraphics[width=.75\textwidth]{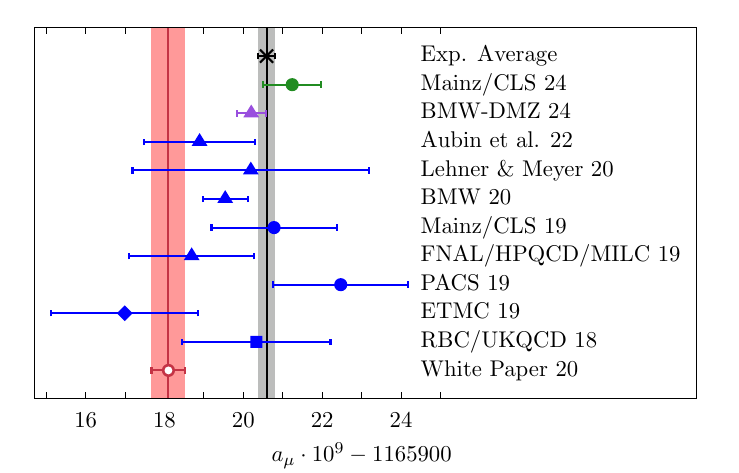}%
\caption{\label{f:comparison_full} 
  Compilation of results for the leading-order hadronic vacuum polarization
  contribution, including isospin-breaking effects, combined with the remaining
  contributions to $a_\mu$ as summarized in the 2020 White Paper \cite{Aoyama:2020ynm}.
  Our result quoted in
  \eq{e:result_ahvp_full} is shown in green. Different discretizations
  of the quark action are denoted by circles (Wilson fermions)
  \cite{Shintani:2019wai, Gerardin:2019rua}, triangles (staggered
  fermions) \cite{Davies:2019efs, Borsanyi:2020mff, Lehner:2020crt,
    Aubin:2022hgm}, squares (domain wall fermions) \cite{Blum:2018mom}
  and diamonds (twisted-mass Wilson fermions)
  \cite{Giusti:2019xct}. The data-driven estimate from the 2020 White
  Paper \cite{Aoyama:2020ynm} and the current experimental average
  \cite{Bennett:2006fi, Muong-2:2021ojo, Muong-2:2023cdq} are
  represented by the red and grey vertical bands, respectively. The
  recent estimate by BMW-DMZ \cite{Boccaletti:2024guq} is based on a
  combination of lattice and data-driven evaluations.
}
\end{figure}

Combining our evaluation of $\ahvp$ in isoQCD from \eq{e:result_ahvp} with
\eq{eq:QEDest}, we obtain
\begin{align}\label{e:result_ahvp_full}
	\ahvp = 724.5(4.9)_{\rm stat}(5.2)_{\rm syst}[7.1]\,,
\end{align}
for the full leading-order hadronic vacuum polarization contribution
to $a_\mu$. Our result is in tension with the data-driven evaluation
of the 2020 White Paper at the level of $3.9\,\sigma$ but yields a SM
prediction for the entire $a_\mu$ that agrees with the current
experimental average, as can be inferred from \cref{f:comparison_full}.

%% file: tables/tab_scales_LD.tex
\begin{table}[!t]
	\renewcommand*\arraystretch{1.2}
	\centering
	\begin{tabular}{c|*{5}{c}}
		\toprule
		& \multicolumn{5}{c}{$S$} \\
		\cmidrule(lr){1-6}
		$O$ & $f_{\pi}$    & $f_K$       & $m_{\pi}$    & $m_K$       & $m_{Ds}$  \\
		\midrule
		$\aLDf{3}{3}$        
		& $ -1.8982$ & $ -0.0277$ & $ -0.5737$ & $ -0.2816$   & --        \\
		$\aLDf{8}{8}$        
		& $ -2.0891$ & $ +0.1691$ & $ +0.0948$ & $ -1.6577$     & --        \\
		$\aLDf{\rm c}{\rm c}$  
		& $ +0.6830$ & $ +1.6446$ & $ -0.0695$ & $ +0.0675$ & $-10.5355$   \\
		$\aLD$   
		& $ -1.9190$ & $ -0.0061$ & $ -0.5005$ & $ -0.4322$ & $ -0.0004$  \\
		\bottomrule
	\end{tabular}
	\caption{\label{t:scaledep_LD}%
		Dimensionless scheme dependencies of observable $O$ with respect to
		the quantity $S$ according to \eq{e:scaledep}.
		}
\end{table}

%% file: tables/tab_contribs.tex
\begin{table}[!t]
	\renewcommand*\arraystretch{1.2}
	\centering
	\begin{tabular}{c
			|
			*{3}{r}
			l
			}
		\toprule
		$O$ 
		& \multicolumn{1}{c}{$\aSD$} 
		& \multicolumn{1}{c}{$\aID$} 
		& \multicolumn{1}{c}{$\aLD$} 
		& \multicolumn{1}{c}{$\ahvp$}
		\\
		\midrule
		$\ahvp$   
		&  68.76(0.21)(0.38)%
		& 236.60(0.79)(1.13)%
		& 423.2(4.2)(3.3)%
		& 728.6(4.3)(3.6)[5.5]
		\\ \midrule
		$\ahvpf{3}{3}$        
		&  43.06(0.05)(0.21)%
		& 186.30(0.75)(1.08)%
		& 378.7(3.7)(3.1)%
		& 608.1(3.7)(3.3)[5.0]
		\\
		$\frac{1}{3}\ahvpf{8}{8}$        
		&  13.86(0.16)(0.78)%
		&  47.41(0.23)(0.29)%
		&  44.5(1.2)(1.1)%
		& 105.8(1.3)(1.4)[1.9]
		\\
		$\frac{4}{9}\ahvpf{\rm c}{\rm c}$  
		&  11.53(0.17)(0.23)%
		&   2.89(0.13)(0.03)%
		&  0.0141(4)(6)
		&   \hspace{.38em}14.4(0.2)(0.2)[0.3]
		\\ \midrule
		$\frac{1}{9}\ahvpf{\rm s}{\rm s}$  
		&   9.07(0.01)(0.06)%
		&  27.68(0.18)(0.22)%
		&  17.73(0.17)(0.13)%
		&  \hspace{.38em}54.5(0.3)(0.3)[0.4]
		\\
		$a_\mu^\mathrm{disc}$
		&  1.3(2.6)(4.1)$\cdot 10^{-3}$%
		&  $-$0.81(0.04)(0.08)%
		& $-$15.3(1.2)(1.2)%
		& \hspace{-.42em}$-$16.1(1.2)(1.2)[1.6]
		\\
		\bottomrule
	\end{tabular}
	\caption{\label{t:contribs}
		Contributions to $\ahvp$ in units of $10^{-10}$ in the infinite volume limit 
		and isospin symmetric QCD as computed in
		\cite{Ce:2022kxy, Kuberski:2024bcj} and in this work.
		Note that the light-connected contribution that is conventionally quoted can be
		obtained from $\frac{10}{9}\ahvpf{3}{3}$.
		}
\end{table}

%% file: sec_conclusion.tex
\section{Conclusion \label{s:concl}}
\begin{figure}[t]
	\centering
	\includegraphics[width=.50\textwidth]{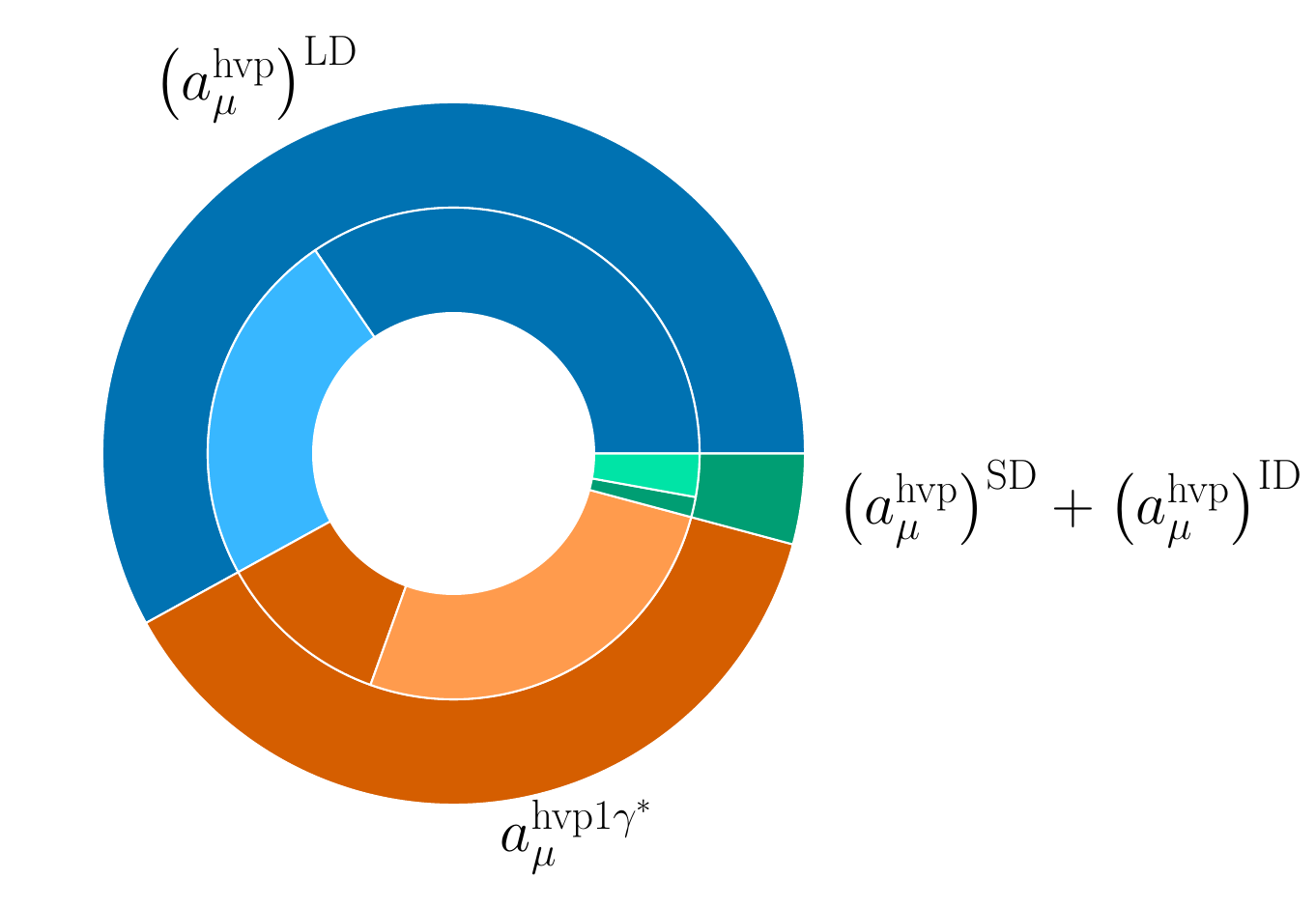}%
	\caption{\label{f:pie}%
          Squared uncertainty of our final estimate for $\ahvp$ in
          eq.~(\ref{e:result_ahvp_full}).  Each of the three
          contributions can be divided into statistical (dark colours)
          and systematic uncertainties (light colours) that are
          displayed in the inner circle.
        }  
\end{figure}

We have performed a fully blinded, high-precision determination of the
long-distance contribution, $\aLD$, to the leading-order hadronic
vacuum polarization contribution of the muon $g-2$. After combining
the result with our previous calculations of the short- and
intermediate-distance window
observables~\cite{Ce:2022kxy,Kuberski:2024bcj}, we have obtained the
entire HVP contribution in isospin-symmetric QCD with a total precision of
0.77\% and a good balance between statistical and systematic
uncertainties.

Compared to ref.~\cite{Gerardin:2019rua}, we have improved the
precision of our estimate for $\ahvp$ in isospin-symmetric QCD by a
factor~2.6. The key ingredients that allowed us to reach that level of
precision were the addition of several high-statistics gauge ensembles
at fine lattice spacing and close-to-physical quark mass, as well as
the application of state-of-the-art noise reduction techniques to
mitigate the exponential loss of signal in the long-distance regime.

Furthermore, to quote a result for $\ahvp$ that can be
straightforwardly compared with the data-driven result, we have
determined electromagnetic and strong isospin-breaking
corrections. The resulting estimate, shown in
eq.~(\ref{e:result_ahvp_full}), has a relative precision of just under
1\% and corroborates the strong tension observed between lattice
calculations and data-driven evaluations derived from $e^+e^-$
hadronic cross sections published prior to the result by CMD-3.

For our current result, the pie chart in figure~\ref{f:pie} shows the
squared uncertainties associated with the short, intermediate and
long-distance window observables along with the isospin-breaking
corrections. It is obvious that our efforts must focus on improving
the precision for both the long-distance contribution and the
isospin-breaking corrections. The latter are small in absolute
terms but make a sizeable contribution to the error.

While we still have a long way to go to reach our long-term goal of
reducing the overall error to the level of about 0.2\%, there is room
for improvement: We are currently extending the set of gauge ensembles
at fine lattice spacings, with a special focus on the ensemble F300 at
physical value of the pion mass. This will allow us to further
constrain mass-dependent and mass-independent cutoff effects in future
analyses, which is crucial given that higher-order cutoff effects or
modifications of the leading-order effects by non-zero anomalous
dimensions cannot be excluded with our current data set. We also aim
for improving the precision of our estimates for the isospin-breaking
corrections by extending our lattice calculations beyond the
electroquenched approximation. This includes the effect of isospin-breaking 
on scale setting, which is the subject of current
investigations~\cite{Segner:2023igh}.

%% file: app_scheme.tex
\section{The hadronic scheme \label{a:scheme}}
Care has to be taken when working in isospin-symmetric QCD
with respect to the definition of the physical point of the theory, 
since this definition is ambiguous. 
As soon as isospin-breaking effects are incorporated,
this ambiguity is lifted. 
The exact definition of the scheme thus has an impact on the 
size of the isospin-breaking corrections.
For a meaningful comparison of multiple independent calculations
in isoQCD the exact definition of the physical point has to match, 
if the precision is of the order of these corrections.

In line with our calculations of the short and intermediate distance
contributions to the HVP, we define our scheme for 
isoQCD via the conditions
\begin{eqnarray}
	m_\pi = (m_{\pi^0})_{\rm phys}, \qquad 
	2 m^2_K - m^2_{\pi} = (m^2_{K^+} + m^2_{K^0}  - m^2_{\pi^+})_{\rm phys},
\end{eqnarray}
corresponding to
\begin{eqnarray}
	m_\pi = 134.9768(5)~\MeV \,, \quad m_K = 495.011(10)~\MeV \,,
\end{eqnarray}
together with the pion decay constant in the isospin-symmetric 
theory~\cite{Tanabashi:2018oca,FlavourLatticeAveragingGroupFLAG:2021npn}
\begin{eqnarray}
	f_\pi = 130.56(14)~\MeV \,.
\end{eqnarray}
As outlined in section~\ref{s:setup_extrap}, we employ the combination
$f_{K \pi} $ to correct for  small deviations from the chiral trajectory
on the CLS ensembles in our data set. 
Here, we employ the value
$f_K = 157.2(5)~\MeV$~\cite{Tanabashi:2018oca,FlavourLatticeAveragingGroupFLAG:2021npn}
to define the physical point. 
It implies a ratio $f_K/f_\pi$ that is consistent with the latest lattice determinations~\cite{Bazavov:2017lyh,Miller:2020xhy,ExtendedTwistedMass:2021qui}.
As can be inferred from table~\ref{t:scaledep_LD}, the dependence on
$f_K$ is largely suppressed with respect to the dependence on $f_\pi$.
The charm quark, included in the partially quenched approximation,
is fixed via the condition
\begin{eqnarray}
	m_{D_{\rm s}} = 1968.47\,{\rm MeV}\,.
\end{eqnarray}
We parameterize the sea quark mass dependence of observables via
the dimensionless combinations of eq.~(\ref{e:yzdef}).

To be able to compare our result in isoQCD with that of \cite{Borsanyi:2020mff},
we also evaluate all observables in this work in the BMW20 scheme, defined by
\begin{eqnarray} \label{e:scheme_BMW20}
	m_\pi = 134.9768(5)~\MeV \,, 
	\quad M_{\rm ss} = 689.89(49)\,{\rm MeV}\,,
	\quad w_0 = 0.17236(70)\,{\rm fm}\,,
\end{eqnarray}
where $M_{\rm ss}$ is the meson with two mass-degenerate quarks with
the mass of the strange quark and $w_0$ is computed from the gradient
flowed gauge field \cite{Borsanyi:2012zs}.
As quark mass proxies in this scheme, we use the variables
\begin{align}\label{e:rho_variables}
	\rho_2 = w_0^2 m_\pi^2 \propto m_{\rm l},\,,
	\qquad
	\rho_4 = w_0^2 (m_\pi^2 + \frac{1}{2} M_{\rm ss} ^2)\propto 2m_{\rm l} + m_{\rm s}\,,
\end{align}
and parameterize the lattice spacing with $w_0/a$ as measured 
on our ensembles.

\subsection{Results in the alternative scheme}
\begin{figure}[t]
	\includegraphics[width=.50\textwidth]{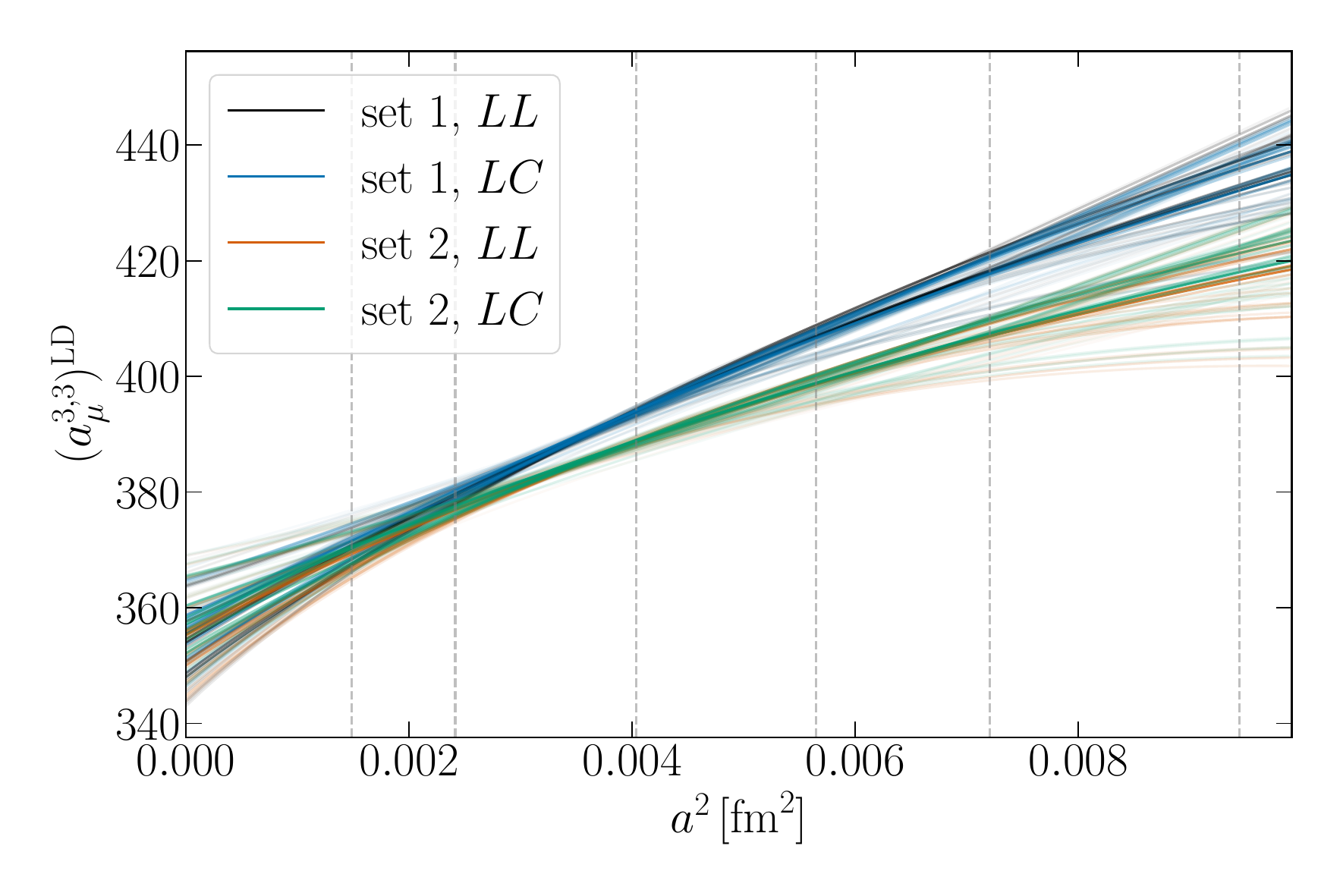}%
	\includegraphics[width=.50\textwidth]{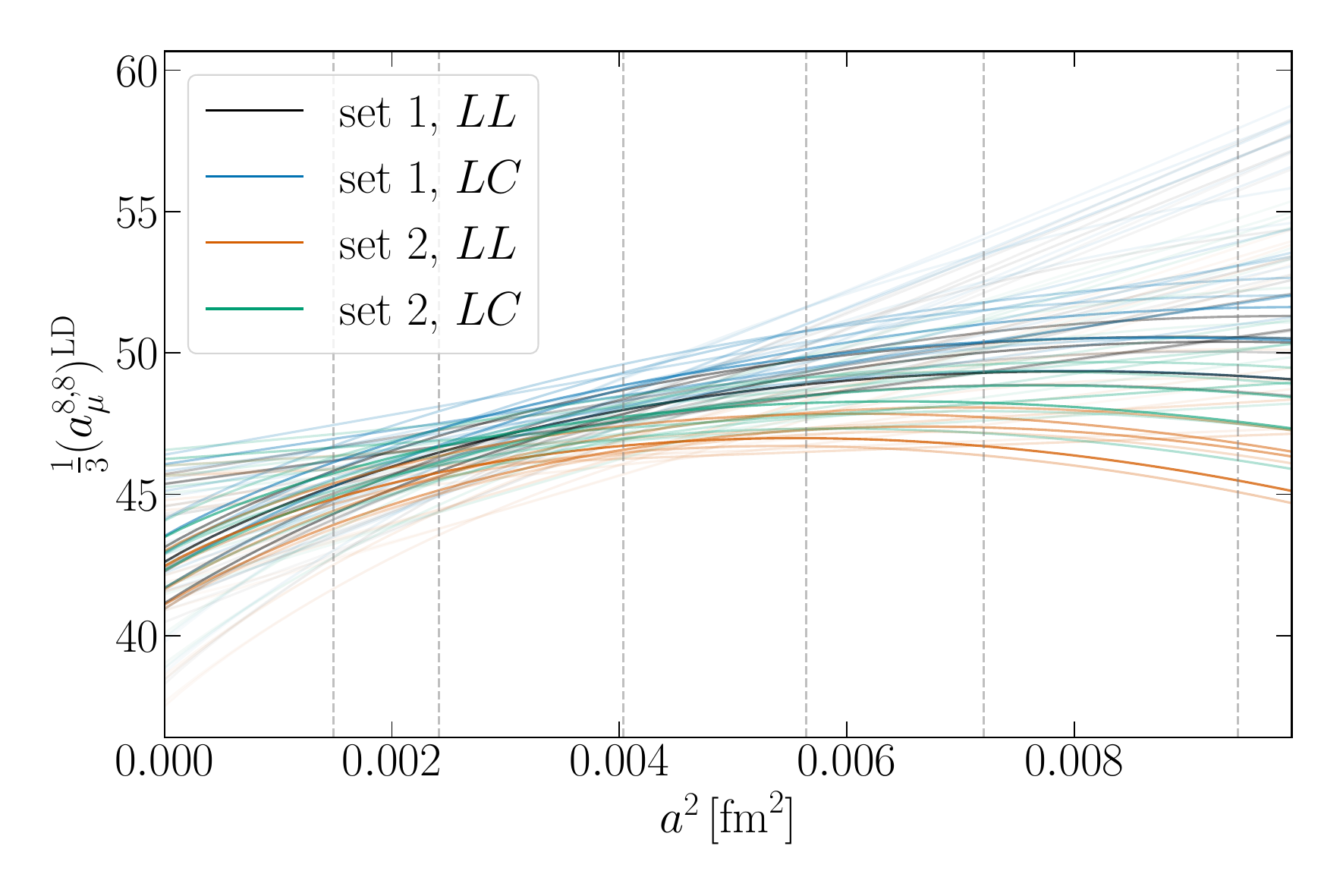}%
	\caption{\label{f:fit_w0}%
		Continuum extrapolations using $w_0/a$ to set the scale.
		\textit{Right:} Approaches to the continuum limit for four sets of data based on the improvement schemes of set~1 and 2 and the $\loc\loc$ and $\loc\cons$ discretizations of the current based on a scan over fit models.
		Each line shows the result from one single fit and the opacity of
		the lines corresponds to the weight of the fit in the model average.
		Dashed vertical lines indicate the lattice spacings used in this work.
	}
\end{figure}

To allow for comparisons in isoQCD without scheme ambiguity, we follow the 
approach of~\cite{RBC:2023pvn} and perform a full second analysis using 
the scheme of eq.~(\ref{e:scheme_BMW20}).
Compared to our preferred scheme, we note that the size of the cutoff effects 
in the contributions to $\aLD$ is significantly larger and that these have a different
sign compared to the case where we use $f_\pi$ to make the muon mass in 
the QED kernel dimensionless.
When performing the continuum extrapolations, fits that parameterize higher
order lattice artefacts as well as mass-dependent cutoff effects are preferred 
over the other variations.
The variation in continuum extrapolations within this scheme, as obtained from 
the model averages, is shown in figure~\ref{f:fit_w0}.

Due to the larger cutoff effects and the need to include terms that parameterize
higher orders in the Symanzik expansion, we observe larger systematic and 
statistical uncertainties, when using $w_0$ to set the scale, compared to 
the $f_\pi$ scheme.
Our results in the BMW20 scheme are
\begin{align}
	\aLDf{3}{3}_\mathrm{BMW20}(L_\mathrm{ref}) 
	&= 353.6(4.3)_\mathrm{stat}(5.2)_\mathrm{syst}(3.0)_\mathrm{scale}[7.3]
	\,,\label{e:result_aLD33_BMW20}\\
	\txts\frac{1}{3}\aLDf{8}{8}_\mathrm{BMW20}(L_\mathrm{ref})
	&= 42.5(1.8)_\mathrm{stat}(1.5)_\mathrm{syst}(0.4)_\mathrm{scale}[2.4]
	\,,\label{e:result_aLD88_BMW20}\\
	\txts\frac{1}{9}\aLDf{s}{s}_\mathrm{BMW20}(L_\mathrm{ref})
	&= 16.81(0.14)_\mathrm{stat}(0.23)_\mathrm{syst}(0.13)_\mathrm{scale}[0.29]
	\,,\label{e:result_aLDss_BMW20}\\
	\aLD_\mathrm{disc, BMW20}(L_\mathrm{ref})
	&= -13.6(1.7)_\mathrm{stat}(1.6)_\mathrm{syst}(0.1)_\mathrm{scale}[2.4]
	\,.\label{e:result_aLDdisc_BMW20}
\end{align}
Note that the contribution of the scale to the final error should not enter when
comparing two results in the same scheme.
The finite-volume correction in eq.~(\ref{e:result_aLDfse}) has been evaluated
in the continuum limit and may be applied to correct the isovector and 
disconnected (with the appropriate scaling factor of $-1/9$) contributions.

%% file: app_blinding.tex
\section{The blinding strategy \label{a:blinding}}

To describe our blinding strategy, it is useful to recall the master
formula for the time-momentum representation of $\amu$ (see
\eq{eq:TMRdef})
\be \amu = \Big(\frac{\alpha}{\pi}\Big)^2
\int_0^\infty dt\; G(t)\; \widetilde K(t;m_\mu),
\ee
where the
kernel function $\widetilde K(t;m_\mu)$ is obtained by convoluting
the momentum-space kernel $K(Q^2;m_\mu)$ defined in \cite{Blum:2002ii}
with a time-dependent function \cite{Bernecker:2011gh}
\be \widetilde
K(t;m_\mu) = 4\pi^2 \int_0^\infty dQ^2\, K(Q^2;m_\mu)\,\Big[t^2 -
  \frac{4}{Q^2} \sin^2(Qt/2)\Big].  
\ee
A simplified form of $K(Q^2;m_\mu)$ is given by
\be
K(Q^2;m_\mu) = \frac{1}{m_\mu^2} \, \frac{\left(\sqrt{\frac{4m_\mu^2}{Q^2}+1}-1\right)^3}{4 \left(\sqrt{\frac{4m_\mu^2}{Q^2}+1}+1\right)
   \sqrt{\frac{4m_\mu^2}{Q^2}+1}}\,.
\ee
Our approach to blinding is based on suitable modifications of the TMR
kernel, which, when convoluted with (unmodified) numerical data for
the current correlator, converge to the same result at the physical
point, up to a multiplicative factor.

\subsection{Modified kernel}

One set of modified TMR kernels at a fixed value of $t$ is defined
via the function
\be \widetilde K_{\rm bld}(t;a;m;B;\vec c) = 4\pi^2B
\int_0^{(\pi/a)^2} dQ^2\, K(Q_{\rm lat}^2(Q,a); m)\,\Big[t^2 -
  \frac{4}{Q_{\rm lat}^2(Q,a)} \sin^2(Qt/2)\Big], 
\ee
where
\be
Q_{\rm lat}^2(Q,a) = c_2 \Big( c_1 \frac{1}{2a} \sin(2Qa) + (1-c_1)
\frac{1}{a} \sin(Qa)\Big)^2 + (1-c_2)\Big(\frac{2}{a}
\sin(Qa/2)\Big)^2.
\ee
We restrict the value of $c_1$ and $c_2$ to
\be
   0\leq c_1\leq 1, \qquad 0<c_2<0.7.  
\ee
Indeed, $c_2$ should not be chosen too large, to ensure that $Q_{\rm lat}$ remains large as $Q\to \pi/a$.

Concretely, we take the following steps: One ``kernel set'' is defined
by a choice for the values of $c_1$, $c_2$, $\sigma$ and $\ell$. Then
the quantities to be analyzed are
\be
B a_\mu^{\rm HVP,bld}(+1,\ell,\sigma,\vec c) = \Big(\frac{\alpha}{\pi}\Big)^2 \lim_{a\to 0}  \int_0^\infty dt\; G(t,a)\;
\widetilde K_{\rm bld}\Big(t;a \tanh\Big(\frac{t}{\ell}\Big);m_\mu (1+\sigma a^2);B;\vec c\Big).
\ee
Reasonable values of the parameters are
\be
0.75\lesssim \ell\lesssim 1.0\,{\rm fm}, \qquad -4\lesssim\sigma [{\rm fm}^{-2}]\lesssim 4 .
\ee
We also consider
\ba 
&& (2-B) a_\mu^{\rm HVP,bld}(-1,\ell,\sigma,\vec c)
\\ && = \Big(\frac{\alpha}{\pi}\Big)^2 \lim_{a\to 0}  \int_0^\infty dt\; G(t,a)\;
\Big(2 \widetilde K(t;m_\mu) - \widetilde K_{\rm bld}\Big(t;a \tanh\Big(\frac{t}{\ell}\Big);m_\mu (1+\sigma a^2);B;\vec c\Big)\Big),
\nonumber
\ea
which reverses the sign of the deviation of $\widetilde K_{\rm bld}$ from $\widetilde K$ at a given $t$.
The test is based on the expectation that
\be
a_\mu^{\rm hvp,bld}(s,\ell,\sigma,\vec c) = \amu, \qquad \forall (s=\pm1, \ell,\sigma,\vec c)
\ee
at the physical point. The parameters of the five modified kernels
used in the analysis of the lattice QCD data computed on the CLS
ensembles are listed in table~\ref{tab:paramsBlindTrue}.

\begin{table}
\centerline{
  \begin{tabular}{|l|l|l|l|l|l|l|}
    \hline
      Set & $B$ & $s$ & $\ell [{\rm fm}]$ & $\sigma \,[{\rm fm}^{-2}]$ & $c_1$ & $c_2$ \\
      \hline
      I   & 1.03628 & + & 0.80 & $-2.54$ & 0.35 & 0.48 \\ 
      II  & 0.94348 & $-$ & 0.73 & +4.11 & 0.14 & 0.38 \\
      III & 1.00971 & $-$ & 0.84 & +1.41 & 0.62 & 0.23 \\  
      IV  & 0.96732 & + & 0.94 & $-0.92$ & 0.45 & 0.55 \\
      V   & 1.02756 & + & 0.78 & +1.82 & 0.26 & 0.69 \\
      \hline
  \end{tabular}}
\caption{Parameters of the modified kernels I through V used for analyzing the lattice QCD data.\la{tab:paramsBlindTrue}}
\end{table}

For the purpose of testing our blinding procedure, we generated five
additional kernels (VI--X) that were used together with synthetic data
for the vector correlator $G(t)$. For the latter we used the
phenomenological model of \cite{Bernecker:2011gh} supplemented by an
artificial pion mass and lattice spacing dependence. Indeed we were
able to verify that the results obtained for kernels VI--X agreed with each
other and the input in the continuum limit and at the physical pion mass.

\subsection{Use of the modified kernels in a blinded analysis}
The procedure followed to perform a blinded analysis involved one of
us generating and providing the five modified kernels to two
co-authors in charge of the data analysis. Both data analysts reported
on their progress and mutual cross-checks during weekly meetings to
the entire group. Throughout the period in which the data analysis
took place, the parameters of the modified kernels were not known to
any member of the group, and the true TMR kernel was never used.  

Once a single analysis procedure was found that was deemed satisfactory 
for each of the five blinded kernels,
the relative unblinding of the kernels took place during a
collaboration meeting: the factors to bring kernels II through V
to the same overall normalisation as kernel I were looked up. Since at that
point the continuum results of the five kernels were in agreement,
we proceeded to the absolute unblinding step by looking up the blinding factor of kernel I.
At that point, the analysis procedure from above was applied without modifications to the true TMR kernel,
thus yielding the results presented in the main part of this article.

\begin{figure}[t]
	\includegraphics[width=.50\textwidth]{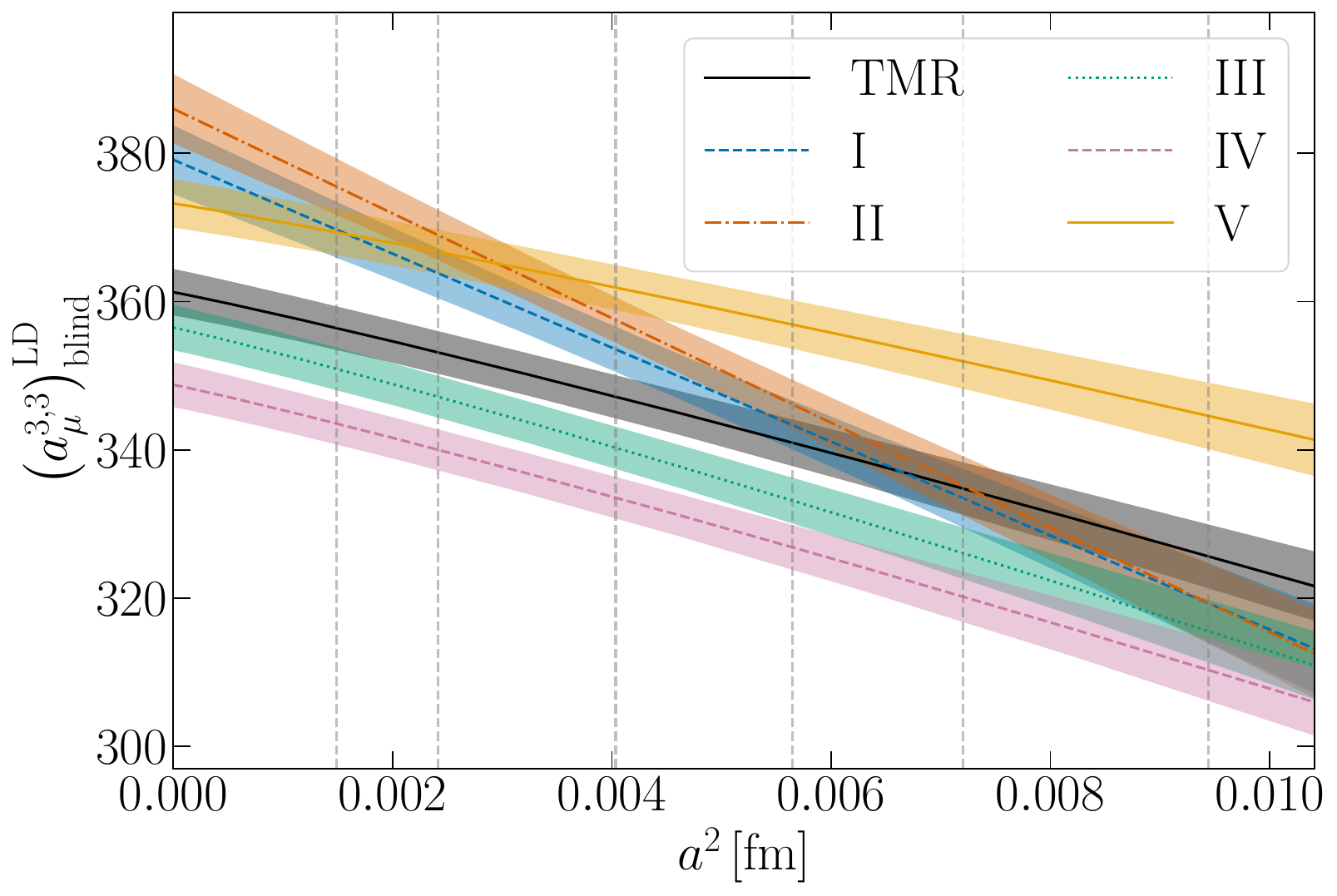}%
	\includegraphics[width=.50\textwidth]{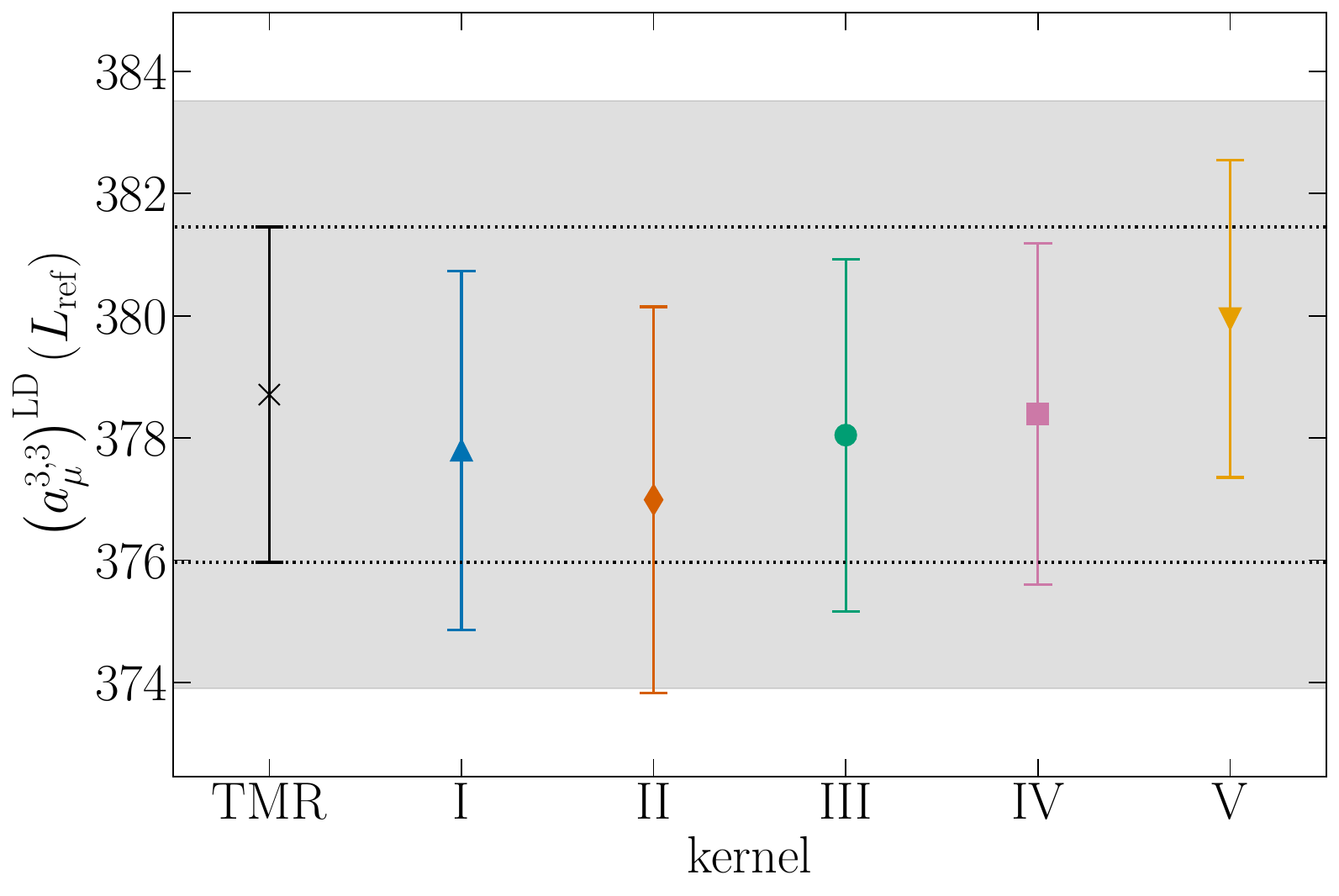}%
	\caption{\label{f:blinding}%
		\textit{Left:} Continuum extrapolation of the long-distance isovector contribution at the physical point based on the fit with the largest model weight for each of the five blinded kernels (coloured curves) as well as for the true TMR kernel (black curve).
		The dotted vertical lines denote the six values of the lattice spacing employed in this work.
		\textit{Right:} Results for the isovector contribution, based on model averages for each of the five kernels and the true kernel, after full unblinding.
		The error bars denote the systematic uncertainties only. 
		The dotted lines show the systematic uncertainty based on a model average with the true TMR kernel and the gray error band denotes the full uncertainty..
	}
\end{figure}

The left panel of figure~\ref{f:blinding} illustrates the differences in the approach 
to the continuum limit for the five blinded kernels and the true TMR kernel.
The curves and the corresponding uncertainty bands represent the continuum 
extrapolation at physical quark masses, based on the fit with the 
largest weight for each of the six model averages, before any unblinding.
The right panel displays the results for the six kernels, obtained from the
model averages, after the absolute unblinding in the continuum limit.
Because of the strong statistical correlation, we only show the systematic
uncertainty of each data point.
The gray error band denotes the full statistical and systematic uncertainty of
the final result in eq.~(\ref{e:result_aLD33}).

%% file: app_lma.tex
\section{The vector correlator from low-mode averaging \label{a:lma}}
The computation of an all-to-all estimator for the vector-vector correlation
function, taking into account all possible pairs of source and sink,
is prohibitively expensive in a large-scale lattice QCD computation.
Low-mode averaging as introduced in 
ref.~\cite{Giusti:2004yp, DeGrand:2004qw}
is based on the computation of the low eigenmodes of the
Dirac operator to allow for an all-to-all sampling of the low mode 
contribution to the correlation function.
If this contribution has a dominant weight in the long-distance 
tail, where the signal to noise problem hinders the reliable
extraction of the correlator, it allows to significantly increase
the available statistics in the most important region.

\subsection{Low modes of the Dirac operator}
We work with $\mathrm{O}(a)$ improved Wilson fermions,
see ref.~\cite{Bruno:2014jqa} for the exact definition of the Dirac
operator $D$, and focus on the hermitian operator
\begin{align}
	Q = \gamma_5 D\,,
\end{align}
where we suppress the flavour index of the massive Dirac operator
and assume to work with light quarks in the following.
Its inverse $Q^{-1}$ can be expressed via the eigenmodes of $Q$,  
denoted by $v_i$, via 
\begin{align} 
	Q^{-1}  = \sum_{i=0}^{N} \frac{1}{\lambda}_i\, {v}_i \cdot {v}_i^\dagger \,,
\end{align}
where $N$ is the dimension of the operator and ${\lambda}_i$
are the real eigenvalues.
The eigenmodes with the \(N_{\rm L}\) smallest (in magnitude) 
eigenvalues are referred to as the ``low modes''.
We define the projectors
\begin{align}
	\mathbf{{P}}_\mathrm{L} \equiv \sum_{i=0}^{N_\mathrm{L}} {v}_i \cdot {v}_i^\dagger\,,
	\qquad 
	\mathbf{{P}}_\mathrm{H} \equiv \mathbf{1} - \mathbf{{P}}_\mathrm{L}\,.
\end{align}
on the space of the low modes and the corresponding orthogonal
space.
These allow to express ${D}^{-1}$ via 
\begin{align} \label{e:Dinv_LMA}
	{D}^{-1} 
	= {Q}^{-1}(\mathbf{{P}}_\mathrm{L} + \mathbf{{P}}_\mathrm{H}) \gamma_5 
	= \sum_{i=0}^{N} \frac{1}{{\lambda}_i}\, {v}_i \cdot {v}_i^\dagger \gamma_5
	+ {Q}^{-1} \mathbf{{P}}_\mathrm{H} \gamma_5\,,
\end{align}
and to split it into low and high mode contributions of $Q$.

\subsection{Mesonic correlation functions \label{a:LMA_cf}}
The computational challenge that is addressed in this appendix 
is the precise computation of a quark-connected, zero-momentum
two-point function. After integrating out the fermions,
we write
\begin{align}
	C_{\Gamma_A \Gamma_B}(x_0, y_0) = - \sum_{\mathbf{x}, \mathbf{y}} \langle \mathrm{tr} \left[ \Gamma_A S(x, y) \Gamma_B S(y, x) \right] \rangle^\mathrm{gauge}\,, \label{e:2pt_cf}
\end{align}
with the source and sink positions $y$ and $x$, respectively, 
the trace tr that acts in colour and spin space and the gamma matrices
$\Gamma_A$ and $\Gamma_B$.
In this work, these matrices are equal to $\gamma_i$ or 
$\gamma_0 \gamma_i$, were the latter combination is needed for
the $\mathrm{O}(a)$ improvement of the current.
Furthermore, we include the conserved (point-split) vector current
in this work, but refrain from extending the notation for the sake
of clarity in this appendix.
The quark propagators $S$ are defined via
\begin{align}
	\sum_y D(x, y) S (y, z) = \mathbf{1} \delta_{x, z}\,.
\end{align}
Based on eq.~(\ref{e:Dinv_LMA}), each of the two propagators in 
eq.~(\ref{e:2pt_cf}) can be exactly split into a low and a high mode
contribution,
\begin{align} \label{e:prop_LMA}
	S(x, y)
	= \sum_{i=0}^{N} \frac{1}{{\lambda}_i}\, {v}_i(x) \cdot {v}_i^\dagger(y) \gamma_5
	+ S_{\rm H} (x, y)\,,
\end{align}
where $S_{\rm H} (x, y)$ is the propagator in the high mode space.
Correspondingly, the correlation function can be
decomposed into four terms which we denote as,
\begin{align}
	C_{\Gamma_A \Gamma_B}(x_0, y_0) = 
	C_{\Gamma_A \Gamma_B}^{\rm (ee)}(x_0, y_0) + 
	C_{\Gamma_A \Gamma_B}^{\rm (re)}(x_0, y_0) + 
	C_{\Gamma_A \Gamma_B}^{\rm (er)} (x_0, y_0)+ 
	C_{\Gamma_A \Gamma_B}^{\rm (rr)}(x_0, y_0)\,.
\end{align}

The ``eigen-eigen'' contribution is purely built from the low modes
of the Dirac operator such that both propagators can be expressed
in terms of the low modes and the corresponding eigenvalues,
\begin{align} \label{e:LMA_ee}
	C^{(\mathrm{ee})}_{\Gamma_A \Gamma_B}(x_0, y_0) =
	- \sum_{i, j}^{N_\mathrm{L}}
	\sum_{\mathbf{x}, \mathbf{y}}
	\frac{1}{\lambda_i \lambda_j}
	\left\langle
	[v_j^\dagger \gamma_5 \Gamma_A v_i ](x)
	[v_i^\dagger \gamma_5 \Gamma_B v_j ](y)
	\right\rangle\,.
\end{align}
Since the eigenmodes are lattice wide objects, this contribution
can be computed in an all-to-all fashion without any further inversion
of the Dirac operator. With the cost being purely due to contractions,
it is possible to average over all source and sink positions.

The ``rest-rest'' contribution is defined only in the orthogonal
subspace of the low mode space. It can be written as
\begin{align}\label{e:LMA_rr}
	C^{(\mathrm{rr})}_{{\Gamma_A}{\Gamma_B}}(x_0, y_0) =
	- \sum_{\mathbf{x}, \mathbf{y}} \langle \mathrm{tr} \left[ \Gamma_A S_{\rm H}(x, y) \Gamma_B S_{\rm H}(y, x) \right] \rangle\,, 
\end{align}
where the only difference with respect to eq.~(\ref{e:2pt_cf}) is the occurrence
of the high mode propagator $S_{\rm H}$. From a computational perspective,
compared to a standard evaluation of the correlation function in 
eq.~(\ref{e:2pt_cf}), the operator $\gamma_5 P_{\rm H} \gamma_5$ is 
applied to the source before each inversion of the Dirac operator.
The correlation function can be sampled with standard methods.

The ``rest-eigen'' and ``eigen-rest'' contributions each contain a low and
a high mode propagator and thus connect the two spaces. 
We can write
\begin{align}\label{e:LMA_re}
	C^{(\mathrm{re})}_{{\Gamma_A}{\Gamma_B}}(x_0, y_0) =
	- \sum_{i}^{N_\mathrm{L}}
	\sum_{\vec{x}, \vec{y}}
	\frac{1}{\lambda_i}
	\left\langle
	v_i^\dagger(x)\, \gamma_5 \Gamma_A\, S_{\rm H}(x , y)  \Gamma_B \,v_i(y)
	\right\rangle \,,
	\\ \label{e:LMA_er}
	C^{(\mathrm{er})}_{{\Gamma_A}{\Gamma_B}}(x_0, y_0) =
	- \sum_{i}^{N_\mathrm{L}}
	\sum_{\vec{x}, \vec{y}}
	\frac{1}{\lambda_i}
	\left\langle
	v_i^\dagger(y)\, \gamma_5 \Gamma_B\, S_{\rm H}(y , x)  \Gamma_A \,v_i(x)
	\right\rangle \,,
\end{align}
and notice that for $\Gamma_A = \Gamma_B$, the two functions are
trivially related. 
The explicit inversion of the Dirac operator is performed in the
high-mode space.
We note that in some works, this contribution is not explicitly computed
but instead estimated as bias correction, see e.g. \cite{Aubin:2019usy}.
As we will point out below, we find that the dedicated computation 
is vital for precision in our case.

\subsection{Even-odd preconditioning}
The dimension of the eigenproblem and with it the memory
requirement of the computation can be reduced by a factor
of two when considering the even-odd preconditioned Dirac
operator $\hat{D}$, as pointed out for Wilson quarks in 
ref.~\cite{Blossier:2010vz}.
We define the Schur complement of the asymmetric
even-odd preconditioning of the hermitian Dirac operator
\cite{DeGrand:1988vx},
\begin{align}
	\hat{Q} = Q_{\rm ee}  - Q_{\rm eo}Q_{\rm oo}^{-1}Q_{\rm oe} 
	\quad \text{with} \quad
	\hat{Q}= \gamma_5 \hat{D}\,,
\end{align}
and work with its eigenmodes, with support only on the even points of
the lattice, to define the projectors
\begin{align}
	\mathbf{\hat{P}}_\mathrm{L} \equiv \sum_{i=0}^{N_\mathrm{L}} \hat{v}_i \cdot \hat{v}_i^\dagger\,,
	\qquad 
	\mathbf{\hat{P}}_\mathrm{H} \equiv \mathbf{1} - \mathbf{\hat{P}}_\mathrm{L}\,,
\end{align}
such that the even-odd preconditioned Dirac operator can be expressed as
\begin{align}
	\hat{D}^{-1} 
	= \hat{Q}^{-1}(\mathbf{\hat{P}}_\mathrm{L} + \mathbf{\hat{P}}_\mathrm{H}) \gamma_5 
	= \sum_{i=0}^{N} \frac{1}{\hat{\lambda}_i}\, \hat{v}_i \cdot \hat{v}_i^\dagger \gamma_5
	+ \hat{Q}^{-1} \mathbf{\hat{P}}_\mathrm{H} \gamma_5\,.
\end{align}
For computing the correlation function of \cref{a:LMA_cf}, the 
eigenmodes need to be projected back onto the space of the full 
Dirac operator. When even-odd preconditioning is used for the 
inversion of the Dirac operator, the 
projection operator can be inserted after projecting to the 
even lattice sites and before performing the inversion.

\subsection{Computational details}
\input{./tables/tab_lma}
Four tasks contribute dominantly to the effort of computing 
correlation functions with our implementation of LMA. 
These are the cost to compute a sufficiently large number of
eigenmodes, the contraction for the ``eigen-eigen'' contribution,
the inversion of the Dirac operator and the preceding projection
to the high-mode space.
In this subsection, we point out the specific setup that we have 
used in our computation after an extensive tuning towards optimal
performance for the problem at hand.

Since the precise computation of the long-distance tail of the
vector-vector correlation function is hindered by the signal-to-noise
problem, this is the region where we want to make use of the
all-to-all sampling of the ``eigen-eigen'' contribution.
We have optimized the setup such that in this region, starting
at a source-sink separation of about $1.5\,$fm, the central
value and the variance of the full correlation function are 
dominated by the contribution of $C^{(\mathrm{ee})}$.
This choice ensures that all information of the gauge fields is
used to sample the long-distance tail and all noise stems from
the fluctuations of the gauge field configurations.
It has a direct impact on the cost of the calculation because a
sufficiently large number of eigenmodes has to be computed and
the remaining correlation function, especially the mixed contributions,
have to be known precisely enough not to spoil the signal.
We note that a similar strategy has been chosen in 
ref.~\cite{Borsanyi:2020mff}.

\paragraph{Solving the eigensystem.}
A large number of eigenmodes has to be computed to achieve
low mode dominance in the long distance tail. 
This number varies significantly across the ensembles that have been
included in this study.
On the one hand, the number of modes with an eigenvalue below 
some fixed threshold scales with the  lattice volume \cite{Giusti:2008vb}.
On the other hand, the dominance of the low modes is enhanced
when the quark mass is lowered towards the chiral limit.
In this work, these are competing effects since the volumes of the 
ensembles are increased as the pion mass is lowered.

One of the questions that determine whether LMA can be implemented
cost-effectively, is if a sufficiently large number of eigenmodes
can be computed with reasonable cost.
We have observed that a first estimate for the number of eigenmodes
can be found by requiring that the modulus of the largest eigenvalue
of the low modes is of the order of the strange quark mass 
(or half of it when even-odd preconditioning is used).
For the largest lattices in this work, at physical value of the pion mass,
this amounts to computing 800 eigenmodes. 
Table~\ref{tab:lma} collects the number of eigenmodes that has been 
used for each of the ensembles where LMA has been applied.

For the solution of the hermitian eigenproblem, we utilize the 
Krylov-Schur algorithm in the implementation of the
\texttt{SLEPc} package \cite{slepc-toms,slepc-manual} which relies on
\texttt{PETSc} \cite{petsc-user-ref,petsc-efficient}.
When used on its own, we observe that a large number of iterations
is needed to solve the eigensystem, resulting in a prohibitively 
large cost.
The key ingredient for the efficient computation of the eigenmodes in this
work is the use of a shift-and-invert spectral transformation:
Instead of solving the equation
\begin{align}
	\hat{Q} \hat{v} = \hat{\lambda}\, \hat{v}\,,
\end{align}
we solve for 
\begin{align}
	\hat{Q}^{-1} \hat{v} = \theta\, \hat{v} \quad \text{where}\quad
	\theta = 1 / \hat{\lambda}\,.
\end{align}
This transformation has the effect of dramatically enhancing the 
convergence properties of the solver such that only a small number
of iterations is necessary, between four and eight in our setup, 
with more than half of the modes converging in the first iteration.
In turn, the Dirac operator has to be inverted for each of the vectors
in the search space. 
We thus shift the work from the eigensolver of \texttt{SLEPc} 
to the deflated solver of the
\texttt{openQCD} package \cite{Luscher:2007se, openQCD}
and are able to profit from the physics informed optimizations
of the solver. 
With a sufficiently well tuned setup, about 2 inversions have to be
performed to compute one eigenmode. 
We have found this cost to scale linearly in the region of up to 
1000 eigenmodes that we have explored in the context of this work.

\paragraph{Computing the eigen-eigen contribution.}
We have to compute the local-local and local-conserved 
vector-vector and vector-tensor currents for the full set of
correlation functions that is used in this work. 
Efficient contraction routines are needed in order to keep the 
computational effort at a reasonable level and symmetries
in the correlation function of eq.~(\ref{e:LMA_ee}) can be utilize
to reduce the number of contractions. 
On ensembles with antiperiodic boundary conditions, a full four-volume
average can be performed. 
In contrast, on ensembles with open boundary conditions in the
time direction, all pairs of source and sink where one of the two
is in the boundary region has to be discarded from the average.
The determination of the boundary region is performed at the stage
of the analysis, based on the data that has been obtained for all
source positions.

\paragraph{Computing the rest-rest contribution.}
The rest-rest contribution of eq.~(\ref{e:LMA_rr}) can be computed 
with standard methods and we choose spin diluted stochastic 
time slice sources \cite{ETM:2008zte} for the computation.
Since this contribution dominates at short distances only,
it can be easily computed to the desired precision.
To reduce the computational effort, we use the truncated solver 
method \cite{Bali:2009hu}.
We perform low-precision solves on $\mathrm{O}(100)$ stochastic
sources per configuration and correct for the small bias with a handful 
of high-precision solves. 
We note that the setup has been chosen such that the bias is completely 
negligible with respect to the statistical uncertainty for all relevant 
source-sink separations. 
A projection onto the high-mode space has to be performed before 
each inversion. 
On ensemble E250, the cost for one projection is about half of the 
cost of one truncated solve. 
Therefore, there is a limit to the computer time that can be gained with 
the truncated solves. 
It could be expected that the solves on the deflated 
sources are significantly faster than standard solves, given the large 
number of eigenmodes that is projected out. 
However, we find the improvement to be marginal when the 
\texttt{openQCD} solver, which is based on inexact deflation \cite{Luscher:2007se}, is used to solve the Dirac equation.

\paragraph{Computing the rest-eigen contribution.}
The rest-eigen and eigen-rest contributions of 
eqs.~(\ref{e:LMA_re}--\,\ref{e:LMA_er}) provide a computational challenge,
because a significant effort has to be made to compute it 
precisely enough such that its statistical uncertainty is small
compared to that of the ``eigen-eigen'' contribution. 
As for the ``rest-eigen'' contribution, stochastic sources
may be used for the computation.
When inserted at the appropriate place, all Dirac structures can
be computed from a single source, at the cost of contractions
with all low modes. A sufficiently large number of
sources has to be employed to reduce the stochastic noise.
Due to the projection and contraction cost, the truncated solver 
method cannot be applied as efficiently in this case.

We have found the ansatz that has already been used in 
ref.~\cite{Giusti:2004yp} to be most effective for our purpose.
It amounts to projecting an eigenmode to a specific source
time slice before multiplying it with the appropriate Dirac matrix,
projecting out the eigenmodes and inverting. 
The solution is then contracted with the eigenmode. 
This operation has to be performed for each Dirac matrix and 
eigenmode, leading to a very large number of inversions
that makes up the largest fraction of the computational cost.
To reduce the computational burden, we follow the approach of 
ref.~\cite{Borsanyi:2020mff} and perform truncated solves
\cite{Bali:2009hu,Blum:2012uh}. 
The small bias, again negligible with respect to the statistical
uncertainty, is corrected by computing high-precision solves
on a small number of eigenmodes that are selected 
via Monte Carlo sampling.

This approach to compute the ``rest-eigen'' and ``eigen-rest''
contributions (which are related to each other) takes into account 
some of the all-to-all information of the eigenmodes. 
If performed for each source time slice, the
result would indeed be an exact all-to-all estimator.

\paragraph{Synthesis}
\begin{figure}
	\includegraphics[width=.49\textwidth]{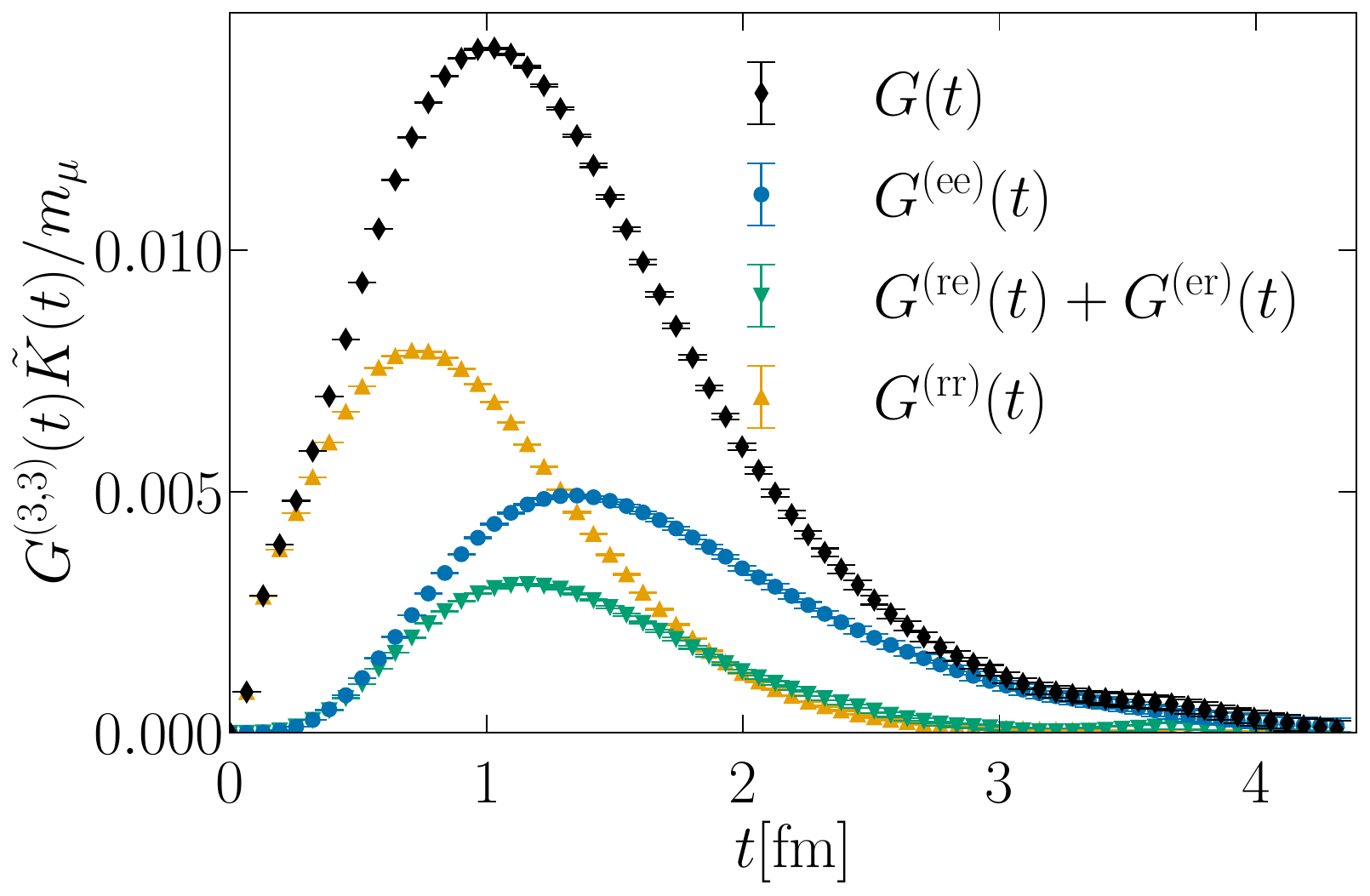}%
	\hfill
	\includegraphics[width=.465\textwidth]{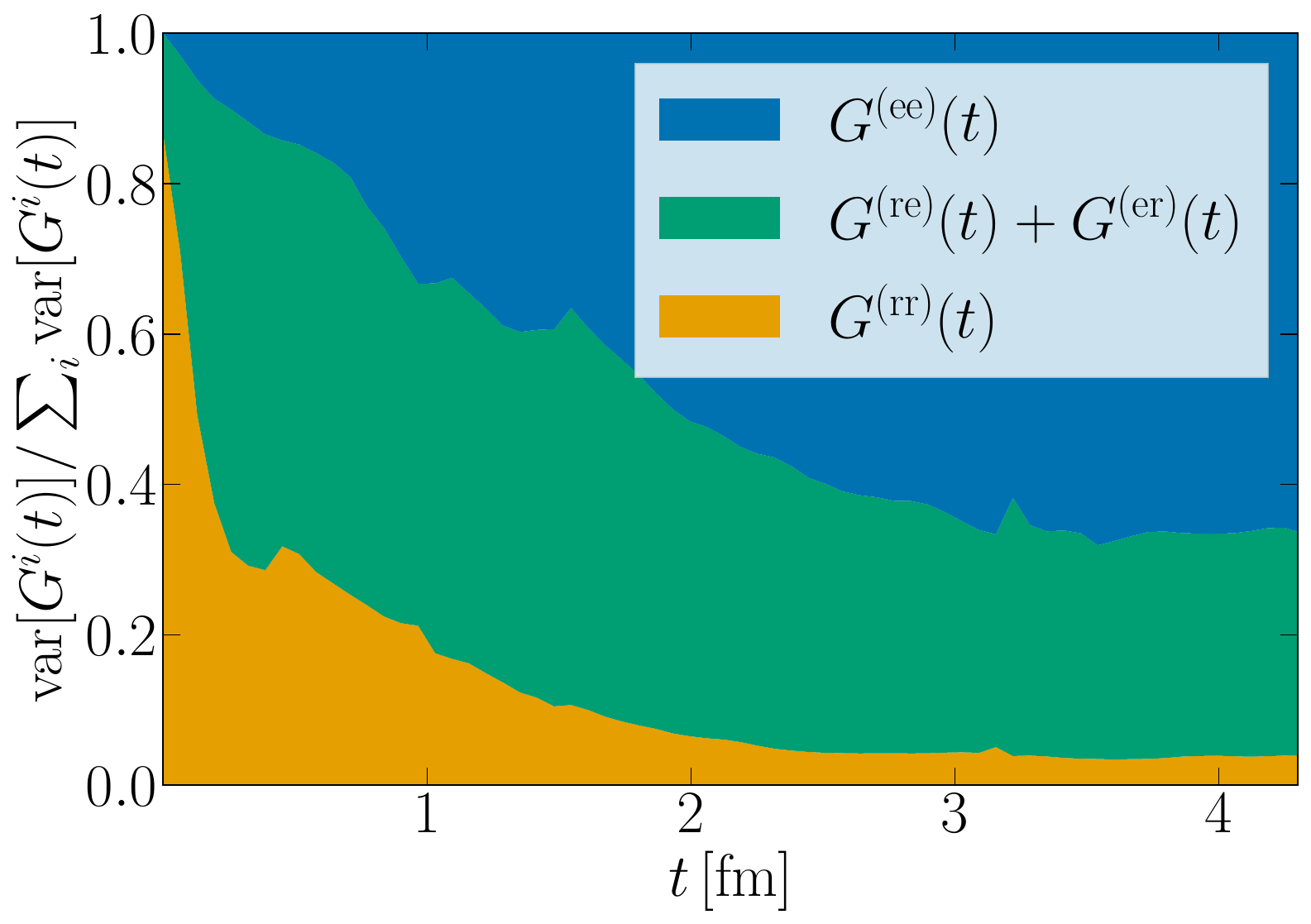}%
	\caption{\label{f:LMA_contribs}%
		LMA computation on ensemble E250 at physical quark masses.
		\textit{Left:} Integrand of $a_\mu^{(3, 3)}$ (black) and the three contributions
		of the LMA computation.
		\textit{Right:} Variances of the three contributions normalized by 
		the variance of the full correlation function.
	}
\end{figure}
Despite the significant computational effort, we have found 
low-mode averaging to be more efficient in computing the
long-distance tail of the vector-vector correlation function
than stochastic sampling, if the quark mass is small enough.%
\footnote{%
For our ensembles with pion masses above 300\,MeV, where
the correlation function can be precisely computed with 
stochastic methods, LMA is not more efficient. }
A bit surprisingly, after optimizing the solution of the eigensystem
and the contractions, the computation of the mixed contribution
of low and high modes turned out to be the most costly part
of our computation.

In \cref{f:LMA_contribs} we show on the left hand side the integrand
to compute the light-connected contribution to $\ahvp$ on ensemble
E250. The total, denoted by the black diamonds, is composed of the 
sum of the three coloured data sets. 
It is apparent that the low-mode contribution dominates for source-sink
separations $t>1.4\,{\rm fm}$.
On the right hand panel of \cref{f:LMA_contribs} we show the time dependent
variance of each contribution, normalized by the sum of the three.
Whereas the variance of the eigen-eigen contribution dominates the 
total in the long-distance regime, the variance of the rest-eigen contribution
is non-negligible although the contribution to the isovector correlation
function is small.

%% file: tables/tab_lma.tex
\begin{table}[t]
\centering
\vskip 0.1in
\renewcommand{\arraystretch}{1.1}
\begin{tabular}{lrrcc}
\hline
 Id   &   $T$\,[fm] &   $L$\,[fm] & $m_\pi\,[{\rm MeV}]$   &   $N$ \\
\hline
 C101 &         8.1 &         4.1 & 222(2)                 &   384 \\
 D150 &        10.9 &         5.4 & 131(3)                 &   608 \\
 D450 &         9.6 &         4.8 & 219(2)                 &   608 \\
 D451 &         9.6 &         4.8 & 219(1)                 &   608 \\
 D452 &         9.6 &         4.8 & 156(2)                 &   640 \\
 D251 &         8.1 &         4.1 & 286(1)                 &   480 \\
 D200 &         8.1 &         4.1 & 202(1)                 &   480 \\
 D201 &         8.1 &         4.1 & 202(2)                 &   480 \\
 E250 &        12.2 &         6.1 & 131(1)                 &   800 \\
 J303 &         9.4 &         3.1 & 260(1)                 &   288 \\
 J304 &         9.4 &         3.1 & 263(1)                 &   288 \\
 E300 &         9.4 &         4.7 & 177(1)                 &   704 \\
 F300 &        12.6 &         6.3 & 136(1)                 &   800 \\
\hline
\end{tabular}
\caption{\label{tab:lma}%
	Overview of ensembles where low-mode averaging has been used to compute
	the isovector correlation function. 
	The volume is given by $T\times L^3$, and $m_\pi$ is the pion mass.
	$N$ denotes the number of eigenmodes of $\hat{Q}$ that have been used
	in the computation.
}
\end{table}

%% file: app_spec.tex
\section{\texorpdfstring{$I=1$ $\pi\pi$}{I=1 pipi} scattering at physical pion mass \label{a:spec}}

At relatively late times, we can further improve on the LMA correlator by replacing it with the spectral reconstruction of the isovector component. The LMA correlator, despite its precision, nevertheless suffers from an exponential loss in signal-to-noise; the reconstruction, in contrast, benefits from a constant signal-to-noise ratio and is therefore guaranteed to beat the LMA correlator eventually. For the E250 ensemble, the improvement from switching to the reconstructed current correlator occurs around $t \approx 2.5$ fm.

\subsection{Measuring the finite-volume energies and matrix elements}

To reconstruct the isovector current correlator $\langle J(t) J^\dagger(0) \rangle$, we employ the correlation functions \cite{Andersen:2018mau}
\begin{align}
\langle [\pi\pi](t)\, [\pi\pi]^\dagger(0) \rangle &= Z_{\pi\pi}^* \,Z_{\pi\pi} e^{-E^{(0)} t}  + \cdots \,,
\\
\langle J(t) \, [\pi\pi]^\dagger(0) \rangle &= Z_J^*  \,\,\, Z_{\pi\pi}e^{-E^{(0)} t}  +  \cdots \,,
\end{align}
using a set of $N$ different $\pi\pi$ interpolators $\{[\pi\pi]^{(1)}, \cdots ,[\pi\pi]^{(N)}\}$. The set of correlation functions formed from any two of these $N$ interpolators forms an $N \times N$ correlation matrix which should, in principle, describe the lowest lying $N$ states. 

Given a set of interpolators describing a state of interest (in this case, two pions), we can form an optimized set of interpolators which approximately project onto particular excitations of that state by solving the associated generalized eigenvalue problem~\cite{Michael:1982gb}
\begin{align}
    C(t) v_n(t, t_0) = \lambda_n(t, t_0) C(t_0) v_n(t, t_0)\,,
\end{align}
for the eigenvalues $\lambda_n(t, t_0)$ and eigenvectors $v_n(t, t_0)$, where $C(t)$ is a matrix of correlation functions formed by the outer product of a set of interpolators with itself. 

If we consider the case of $\pi\pi$ scattering, then given a set of $N$ interpolators $[\pi\pi]^{(n)}$, the solution to the generalized eigenvalue problem allows us to form a set of $N$ optimized operators $[\Pi\,\Pi]^{(n)}(t; t_0) \equiv ([\pi\pi](t), v^{(n)}(t,t_0))$ describing $N$ energy levels. Typically one then determines the energy levels either by fitting the principal correlators (the eigenvalues of the GEVP) to 
\begin{equation}
    \lambda^{(n)}(t, t_0) = e^{-E^{(n)} (t-t_0)} + \text{ h.o.}\,,
    \label{eqn:principal_corr_lo}
\end{equation}
or by fitting the rotated correlators (formed from the eigenvectors of the GEVP) to

\begin{align}
    \langle [\Pi\,\Pi]^{(n)}(t; t_d, t_0) \, [\Pi\,\Pi]^{(n)\, \dagger}(0; t_d, t_0)  \rangle
    &= |Z^{(n)}|^2 e^{-E^{(n)} t}  + \text{ h.o.}\,,
\end{align}
where $t_d$ indicates that we have reused the eigenvalues from the solution to the GEVP at $t=t_d$ for all times. 

The crucial part to either approach, however, is estimating what those higher-order corrections should be. Naively we expect that the corrections should not be worse than $\mathcal O (e^{-t \delta E})$ for an arbitrary choice of $t_0$, with $\delta E = \min_{n\ne m} | E_n - E_m |$ the smallest gap between energy levels in the spectrum~\cite{Luscher:1990ck}. However, the work of \cite{Blossier:2009kd, Blossier:2010mk} show that it is possible to do better than this provided one is clever about the asymptotic behaviour or the choice of GEVP parameters (for instance, by imposing the restriction $t_0/2>t$).

As an example, let us consider the fits to the principal correlators. A result from \cite{Blossier:2009kd} is that the higher-order corrections to~\eqref{eqn:principal_corr_lo} should be parameterized by
\begin{equation}
\epsilon_\lambda^{(n)}(t, t_0) \sim \mathcal O \left(e^{-(E^{(N)} - E^{(n)}) t_0}e^{-E^{(n)}(t-t_0)} \right)  + \mathcal O \left(e^{-E^{(N)} t} e^{+E^{(n)} t_0} \right) \,.
\end{equation}
Therefore, for a fixed choice of $t_0$, a better choice of fit function for the principal correlators is given by 
\begin{equation}
    \label{eqn:principal_corr_nlo}
    \lambda^{(n)}(t, t_0) \approx e^{-E^{(n)}(t - t_0)} \left[
    1 
    + A
    + B e^{-\Delta E^{(n)} t}
    \right]\,,
\end{equation}
where $\Delta E^{(n)} \equiv E^{(N)} - E^{(n)} \geq \delta E^{(n)}$. 
This dependence on $\Delta E^{(n)}$ rather than $\delta E^{(n)}$ has two advantages: (1) since $\Delta E^{(n)}$ is larger, the correction is smaller; and (2) it is simpler to implement the constraint on $\Delta E^{(n)}$ in a simultaneous fit to all levels than the constraint on $\delta E^{(n)}$.
    
In this work we advocate the use of ``sliding-pivot'' fits to the effective masses and overlaps as motivated by the insights of \cite{Blossier:2009kd, Blossier:2010mk}, in which the authors showed that the corrections to the effective energies and overlaps are described by
\begin{align}
        \epsilon^{(n)}_E(t, t_0) &= \mathcal O \left(e^{- \Delta E^{(n)} t} \middle) + 
        \mathcal O \middle(e^{-2(\Delta E^{(n)} - \delta E^{(n)}) t_0} e^{-\delta E^{(n)} t} \right)
        \label{eqn:blossier_correction_en}\\
        &= \mathcal O \left(e^{- \Delta E^{(n)} t} \right) & \text{when } t/2 \leq t_0 < t \, ,
        \nonumber \\ \nonumber \\
        \epsilon^{(n)}_Z(t, t_0) &= \mathcal O \left(e^{-\Delta E^{(n)}  t_0} \middle)  + \mathcal O \middle(e^{-(\Delta E^{(n)} -\delta E^{(n)}) t_0} e^{-\delta E^{(n)} t} \right) \label{eqn:blossier_correction_pn}
        \\
        &= \mathcal O \left(e^{- \Delta E^{(n)} t_0} \right) & \text{when } t/2 \leq t_0 < t \, .
        \nonumber
        \end{align}
On the second line, we have shown the correction after restricting the choice of $t_0$ to the interval shown. Although both reduced expressions are similar, we note that the correction to the effective masses depends on $t$ while the corrections to the effective overlaps depend on $t_0$.

Here we take this restriction on $t_0$ seriously: rather than fixing $t_0$ as is often done, we allow the parameter to vary with $t$, choosing the value of $t_0$ closest to (but greater than) $t/2$. We then construct/fit the effective masses per/to the following expressions:
\begin{align}
    E^{(n)}_\text{eff}(t) &= \left\{ \log\left(\frac{\lambda^{(n)}(t-1, \lceil t/2\rceil)}{\lambda^{(n)}(t, \lceil t/2\rceil)}\right) \quad\middle|\quad 
    t = 4, 5, 6, \dots
    \right\} \, ,
    \\
    E_\mathrm{eff}^{(n)}(t) &\approx E^{(n)} \left(1 + A^{(n)}_E e^{- \Delta E^{(n)} t} \right) \, .
    \label{eqn:effective_mass_fitfcn}
\end{align}
We simultaneously fit all $N$ energy levels in order to better constrain the shared parameter $E^{(N)}$.

\begin{figure}[t]
	\centering
	\includegraphics[width=0.49\textwidth]{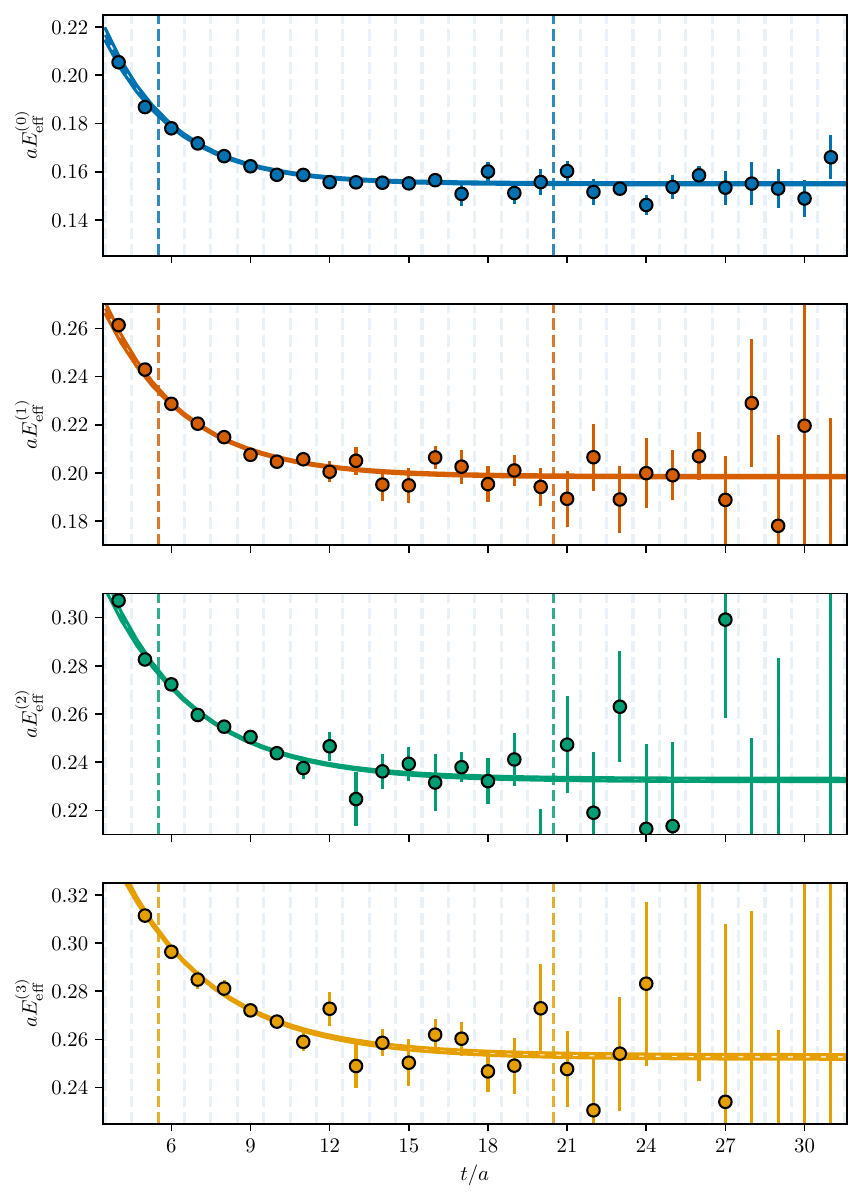}
	\includegraphics[width=0.49\textwidth]{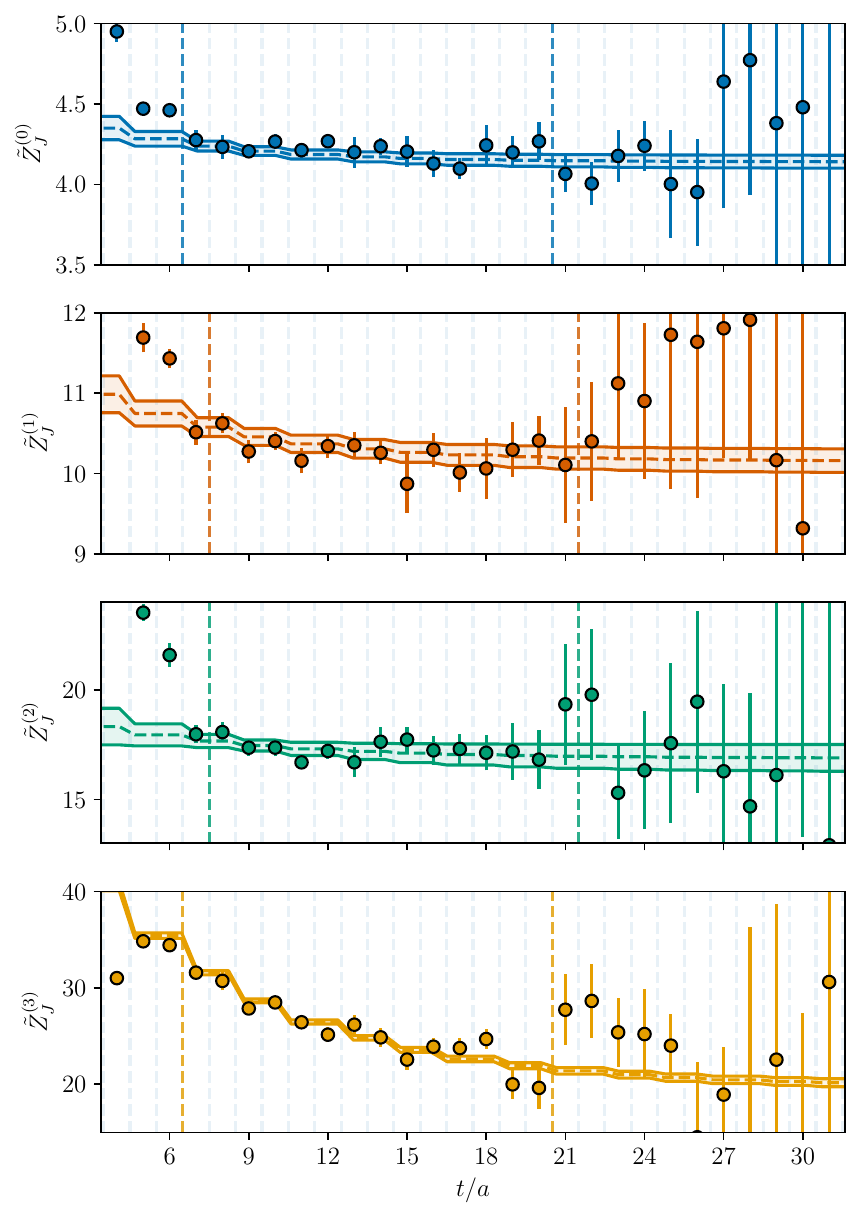}
	\caption{\label{fig:blossier_effective_quantities} A representative fit to the first four effective masses (left) and effective matrix elements (right) on E250. The vertical bands denote the fit windows. For the effective matrix elements, we only fit every other data point, starting with the point right of the left-most dashed, vertical line.}
\end{figure}

Similarly, the finite-volume matrix elements are determined by constructing the following effective quantities from the optimized mixed-current correlator $\langle J(t) \,  [\Pi\,\Pi]^{(n) \dagger}(0)\rangle$ and two pion correlator $\langle [\Pi\,\Pi]^{(n)}(t)\, [\Pi\,\Pi]^{(n) \dagger}(0) \rangle$ and fitting them using the higher-order term described in~\eqref{eqn:blossier_correction_pn},
\begin{align}
    \tilde Z^{(n)}_J(t) &= \frac{\langle J(t)\, [\Pi\,\Pi]^{(n)\, \dagger}(0; \lceil t/2 \rceil) \rangle}{\sqrt{\langle [\Pi\,\Pi]^{(n)}(t; \lceil t/2 \rceil) \, [\Pi\,\Pi]^{(n)\, \dagger}(0; \lceil t/2 \rceil) \rangle}} \left(\frac{\lambda^{(n)}(\lceil t/2 \rceil + 1, \lceil t/2 \rceil)}{\lambda^{(n)}(\lceil t/2 \rceil + 2, \lceil t/2 \rceil)}\right)^{t/2} \, ,
    \label{eqn:effective_pn}
    \\
    \tilde Z^{(n)}_J(t) &\approx Z^{(n)}_J \left( 1 + A^{(n)}_Z e^{- \Delta E^{(n)}\lceil t/2\rceil} \right) \, .
    \label{eqn:effective_pn_fitfcn}
\end{align}
Again, we emphasize that the corrections to the effective matrix elements depend on $t_0$, not $t$. 

Although the fits to the effective matrix elements contain the energy levels as parameters, the fits to the effective masses are significantly more efficient at distinguishing these energy levels. Therefore, rather than simultaneously fit the effective masses and matrix elements, we first fit the energy levels using~\eqref{eqn:effective_mass_fitfcn} before passing the posterior as a prior into the fit to the matrix elements using~\eqref{eqn:effective_pn_fitfcn}. Representative fits to the energy levels and matrix elements are shown in figure~\ref{fig:blossier_effective_quantities}, respectively.

To minimize the systematic bias from our choice of fit windows $(t_\mathrm{min}, t_\mathrm{max})$ when fitting the effective masses and matrix elements, we vary the windows and calculate the posterior under a model-averaging framework using the Bayesian Akaike information criterion for the model weights~\cite{Jay:2020jkz, Neil:2022joj}. We find the model space for the spectrum fits to be strongly peaked around the representative fit shown in figure~\ref{fig:blossier_effective_quantities}. In contrast, there is some noticeable spread among the fits to the effective matrix elements.

\subsection{Transition Point and Gounaris-Sakurai Parameters}

\begin{figure}
	\centering
	\includegraphics[width=\textwidth]{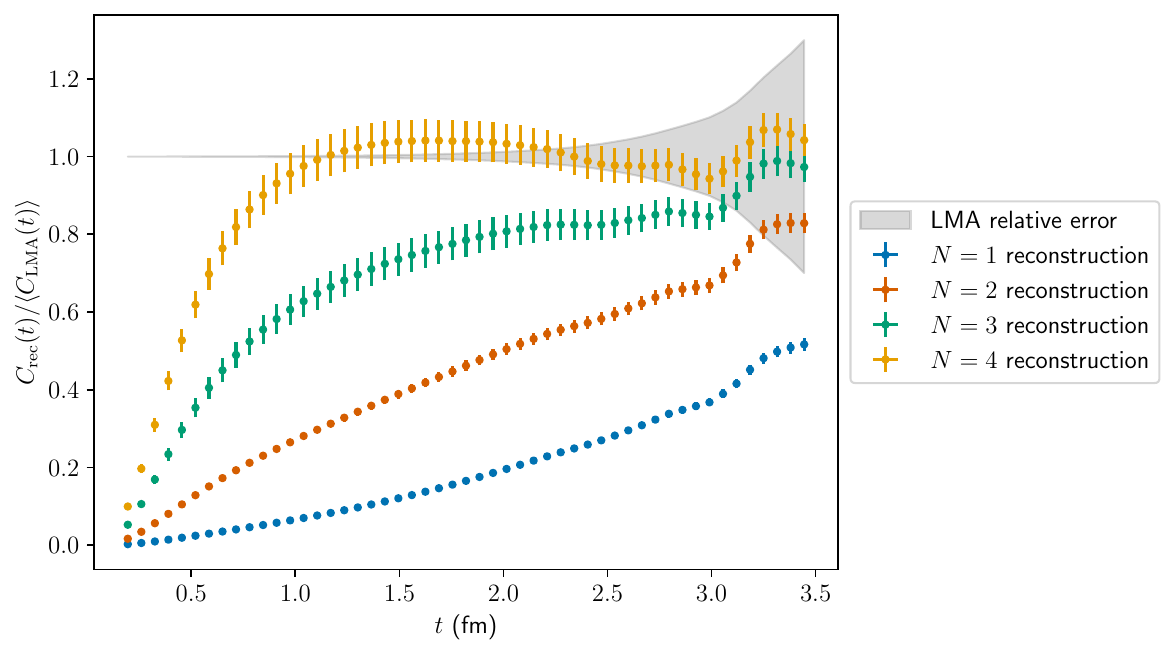}
	\caption{\label{fig:jellyfish} Saturation of the LMA correlator by the reconstructed correlator. The data points show the relative error of the reconstructed correlator if one normalizes by the LMA correlator instead of the reconstructed correlator. From the plot, one sees that the 4-state reconstruction saturates the LMA correlator around $1.2$ fm, with the LMA correlator becoming less precise after $2.5$ fm or so. }
\end{figure}

Rather than apply the bounding method~\cite{Borsanyi:2016lpl, Blum:2018mom}, we choose to replace the LMA correlator with the reconstructed correlator after some Euclidean distance. To identify the transition point, we first verify that the reconstructed correlator saturates the LMA correlator and then find the point for which the error for the reconstructed correlator is smaller than the LMA correlator (see figure~\ref{fig:jellyfish}).

We find that including four states is sufficient to saturate the LMA correlator. However, we note that including the states above the third excited state causes the reconstructed correlator to slightly overshoot the LMA correlator at the $1\sigma$ level. We therefore avoid including these states from the reconstruction for a few reasons: (1) after the third excited state, there is a pronounced decline in data quality, with the fourth excited state no longer exhibiting an exponential decay in the matrix element at the earliest times; (2) \emph{a priori} we do not consider this overshooting to be physical but rather a systematic stemming from the difficulty of constraining higher-level states; and (3) including the fourth (or higher) states in our reconstruction has no bearing on the final result, as the contribution from these states has decayed-off before we reach the transition point near $2.5$ fm.

Through this dedicated spectroscopy study, we are able to reduce the uncertainty on the physical pion mass ensemble E250 by a factor of two. An example application of the bounding method is shown in figure~\ref{f:bounding}.

To compute the Gounaris-Sakurai parameters, we follow the procedure outlined in \cite{Andersen:2018mau, Erben:2019nmx}; the only notable deviation is the manner in which we calculate the finite-volume energy levels and matrix elements. After fitting the phase shifts, we find $g_{\rho\pi\pi} = 6.02(30)$ and $m_\rho / m_\pi = 5.76(9)$.

\FloatBarrier

%% file: app_tables.tex
\section{Tables \label{a:tab}}
This appendix contains tables \ref{tab:ps} to \ref{tab:fv_ref} with detailed
results for individual gauge ensembles.
\newpage
\input{./tables/tab_ps}
\input{./tables/tab_aLD}

\input{./tables/tab_fv_ref}
\FloatBarrier

%% file: tables/tab_ps.tex
\begin{table}[!ph]
	\vskip 0.1in
	\renewcommand{\arraystretch}{1.1}  
	\begin{tabular}{
			l
			|
			S[table-format=1.8]
			S[table-format=1.8]
			S[table-format=1.8]
			|
			S[table-format=2.6]
			S[table-format=1.7]
			S[table-format=1.7]
		}
		\hline
		id      &       $a m_{\pi}$    &       $a m_{K}$               &       $af_\pi$      &        {$t_0/a^2$ }     &       $y$ &       $z$ \\
		\hline
		  A653 & 0.21184(105) & 0.21184(105) & 0.07144(25) & 2.173(07)  & 0.1110(13) & 0.2448(37) \\
		 A654 & 0.16633(131) & 0.22727(112) & 0.06725(25) & 2.194(10)  & 0.0773(13) & 0.2381(38) \\
		 \hline
		 H101 & 0.18250(71)  & 0.18250(71)  & 0.06364(30) & 2.847(06)  & 0.1046(09) & 0.2274(27) \\
		 H102 & 0.15383(80)  & 0.19135(71)  & 0.06044(31) & 2.882(12)  & 0.0809(09) & 0.2205(27) \\
		 H105 & 0.12155(115) & 0.20223(85)  & 0.05781(83) & 2.886(09)  & 0.0559(11) & 0.2198(29) \\
		 N101 & 0.12120(56)  & 0.20146(35)  & 0.05773(35) & 2.892(03)  & 0.0556(06) & 0.2182(22) \\
		 C101 & 0.09570(78)  & 0.20584(44)  & 0.05511(36) & 2.913(05)  & 0.0378(06) & 0.2137(22) \\
		 C102 & 0.09640(87)  & 0.21766(50)  & 0.05507(46) & 2.870(06)  & 0.0386(07) & 0.2368(25) \\
		 D150 & 0.05654(94)  & 0.20835(35)  & 0.05222(30) & 2.944(04)  & 0.0150(05) & 0.2049(20) \\
		 \hline
		 B450 & 0.16081(50)  & 0.16081(50)  & 0.05685(20) & 3.663(13)  & 0.1020(08) & 0.2184(21) \\
		 S400 & 0.13503(46)  & 0.17022(41)  & 0.05399(34) & 3.692(08)  & 0.0781(06) & 0.2145(20) \\
		 N452 & 0.13546(30)  & 0.17031(26)  & 0.05462(15) & 3.673(04)  & 0.0784(04) & 0.2150(18) \\
		 N451 & 0.11064(45)  & 0.17822(26)  & 0.05229(15) & 3.682(07)  & 0.0568(05) & 0.2133(18) \\
		 D450 & 0.08346(51)  & 0.18393(26)  & 0.04977(14) & 3.697(06)  & 0.0356(05) & 0.2101(18) \\
		 D451 & 0.08338(35)  & 0.19382(16)  & 0.05000(24) & 3.665(03)  & 0.0359(03) & 0.2311(18) \\
		 D452 & 0.05932(59)  & 0.18645(18)  & 0.04758(15) & 3.727(04)  & 0.0197(04) & 0.2056(16) \\
		 \hline
		 H200 & 0.13625(64)  & 0.13625(64)  & 0.04775(34) & 5.151(33)  & 0.1000(11) & 0.2125(24) \\
		 N202 & 0.13436(32)  & 0.13436(32)  & 0.04846(13) & 5.153(17)  & 0.0979(06) & 0.2066(16) \\
		 N203 & 0.11249(27)  & 0.14395(23)  & 0.04643(16) & 5.147(07)  & 0.0742(04) & 0.2064(15) \\
		 N200 & 0.09221(29)  & 0.15065(24)  & 0.04420(18) & 5.163(07)  & 0.0540(04) & 0.2056(15) \\
		 D251 & 0.09203(16)  & 0.15041(12)  & 0.04461(10) & 5.164(05)  & 0.0538(03) & 0.2050(14) \\
		 D200 & 0.06502(28)  & 0.15630(17)  & 0.04237(20) & 5.179(06)  & 0.0300(03) & 0.2026(14) \\
		 D201 & 0.06498(43)  & 0.16308(24)  & 0.04263(25) & 5.137(08)  & 0.0302(04) & 0.2191(16) \\
		 E250 & 0.04232(23)  & 0.15936(08)  & 0.04018(12) & 5.202(04)  & 0.0140(02) & 0.2006(13) \\
		 \hline
		 N300 & 0.10574(30)  & 0.10574(30)  & 0.03817(16) & 8.560(32)  & 0.0981(07) & 0.2072(17) \\
		 J307 & 0.10547(42)  & 0.10547(42)  & 0.03785(17) & 8.597(31)  & 0.0979(09) & 0.2062(20) \\
		 N302 & 0.08707(54)  & 0.11363(46)  & 0.03658(21) & 8.526(25)  & 0.0721(09) & 0.2064(22) \\
		 J306 & 0.08690(19)  & 0.11335(19)  & 0.03653(13) & 8.585(17)  & 0.0723(04) & 0.2054(14) \\
		 J303 & 0.06467(22)  & 0.11963(19)  & 0.03439(15) & 8.618(14)  & 0.0447(04) & 0.2027(14) \\
		 J304 & 0.06561(20)  & 0.13187(17)  & 0.03418(12) & 8.500(14)  & 0.0467(04) & 0.2415(16) \\
		 E300 & 0.04399(12)  & 0.12402(09)  & 0.03264(12) & 8.614(05)  & 0.0230(02) & 0.2020(12) \\
		 F300 & 0.03381(23)  & 0.12358(17)  & 0.03168(23) & 8.656(05)  & 0.0144(02) & 0.1958(13) \\
		 \hline
		 J500 & 0.08157(17)  & 0.08157(17)  & 0.02983(10) & 13.964(31) & 0.0941(06) & 0.1966(14) \\
		 J501 & 0.06590(23)  & 0.08796(24)  & 0.02855(15) & 13.984(49) & 0.0673(06) & 0.1952(16) \\
		\hline
	\end{tabular} 
	\caption{Pseudoscalar masses in lattice units, including finite-size corrections. Estimates of the gluonic observable $t_0/a^2$ and the two dimensionless variables $\phi_2$ and $\phi_4$ used in the extrapolation to the physical point. \label{tab:ps}
	}
\end{table}

%% file: tables/tab_aLD.tex
\begin{table}[!p]
\centering
\vskip 0.1in
\renewcommand{\arraystretch}{1.1}  
\begin{tabular}{
		l|
		@{\hskip 0.5em}l
		@{\hskip 0.5em}l
		@{\hskip 0.5em}|
		@{\hskip 0.5em}l
		@{\hskip 0.5em}l
		@{\hskip 0.5em}|
		@{\hskip 0.5em}S[table-format=1.10]
		@{\hskip 0.5em}S[table-format=1.10]
}%
\hline
        & \multicolumn{2}{c|@{\hskip 0.5em}}{$\aLDf{3}{3}$ - Set 2} 
        & \multicolumn{2}{c|@{\hskip 0.5em}}{$\textstyle\frac{1}{3}\aLDf{8}{8}$ - Set 2}
        & \multicolumn{2}{c}{${\textstyle \frac{4}{9}}\aLDf{\rm c}{\rm c}$ - Set 2} \\
\hline
id      &        \quad${\scriptstyle(LL)}$   &       \quad${\scriptstyle(LC)}$     &
                 \quad${\scriptstyle(LL)}$   &       \quad${\scriptstyle(LC)}$     &     
                 ${\scriptstyle(LL)}$   &       ${\scriptstyle(LC)}$    \\
\hline
A653 & 202.5(2.5)          & 206.2(2.6)           & 73.33(93)           & 74.56(94)            & 0.007403(46)    & 0.006947(33)     \\
A654 & 216.3(3.4)          & 219.6(3.4)           & 51.94(82)           & 53.23(82)            & 0.00836(12)     & 0.007856(96)     \\
\hline
H101 & 220.5(2.1)          & 222.7(2.1)           & 79.60(81)           & 80.31(80)            & 0.008232(99)    & 0.007294(81)     \\
H102 & 236.1(3.8)          & 238.2(3.8)           & 64.4(1.1)           & 65.0(1.0)            & 0.00881(14)     & 0.00781(11)      \\
H105 & 241.3(9.9)          & 243(10)              & 51.7(1.7)           & 52.3(1.6)            &{--}             &{--}              \\
N101 & 238.7(4.1)          & 241.2(3.8)           & 49.49(77)           & 50.45(73)            & 0.00973(10)     & 0.008624(85)     \\
C101 & 265.0(3.6)          & 266.9(4.0)           & 42.9(1.5)           & 44.1(1.5)            & 0.010180(96)    & 0.009021(77)     \\
C102 & 254.9(6.3)          & 253.6(7.3)           & 38.7(1.6)           & 39.5(1.6)            &{--}             &{--}              \\
D150 & 312.0(7.8)          & 313.4(7.8)           & 35.7(2.9)           & 36.7(3.0)            &{--}             &{--}              \\
\hline
B450 & 225.0(2.8)          & 226.8(3.0)           & 80.8(1.0)           & 81.4(1.1)            & 0.008530(77)    & 0.007462(62)     \\
S400 & 247.5(3.5)          & 248.7(3.5)           & 65.66(82)           & 66.19(81)            & 0.00896(13)     & 0.00794(11)      \\
N452 & 243.6(2.1)          & 242.8(2.1)           & 66.00(47)           & 66.32(46)            &{--}             &{--}              \\
N451 & 255.5(3.0)          & 256.5(3.0)           & 55.33(75)           & 55.85(73)            &{--}             &{--}              \\
D450 & 270.6(2.3)          & 271.6(2.3)           & 45.6(1.1)           & 46.7(1.1)            & 0.010836(95)    & 0.009566(75)     \\
D451 & 263.2(3.2)          & 264.4(3.1)           & 40.6(1.3)           & 41.2(1.2)            &{--}             &{--}              \\
D452 & 299.1(4.4)          & 299.8(4.4)           & 38.0(2.0)           & 39.1(2.0)            & 0.011301(94)    & 0.009963(74)     \\
\hline
H200 & 234.5(4.3)          & 235.0(4.3)           & 83.5(1.5)           & 83.7(1.5)            &{--}             &{--}              \\
N202 & 239.7(2.9)          & 240.7(2.9)           & 85.9(1.1)           & 86.2(1.1)            & 0.00885(13)     & 0.00786(11)      \\
N203 & 257.1(3.2)          & 257.8(3.2)           & 68.88(76)           & 69.12(76)            & 0.00964(12)     & 0.00855(10)      \\
N200 & 265.5(4.6)          & 265.9(4.6)           & 56.5(1.1)           & 57.1(1.1)            & 0.010523(95)    & 0.009364(80)     \\
D251 & 265.4(2.2)          & 265.8(2.2)           &\quad{--}             &\quad{--}                  &{--}             &{--}              \\
D200 & 289.2(3.0)          & 290.1(3.2)           & 46.1(1.6)           & 46.5(1.6)            & 0.011620(98)    & 0.010354(82)     \\
D201 & 281.8(4.5)          & 282.3(4.5)           & 42.7(2.0)           & 43.4(2.0)            &{--}             &{--}              \\
E250 & 341.9(3.2)          & 342.1(3.2)           & 37.9(2.2)           & 38.4(2.2)            & 0.012067(86)    & 0.010778(69)     \\
\hline
N300 & 232.6(3.3)          & 232.7(3.2)           & 83.0(1.2)           & 83.0(1.2)            & 0.00885(17)     & 0.00810(15)      \\
J307 & 252.9(3.9)          & 253.3(3.9)           & 90.7(1.5)           & 90.9(1.5)            &{--}             &{--}              \\
N302 & 248.2(4.6)          & 248.9(4.7)           & 67.2(1.2)           & 67.4(1.2)            & 0.01005(10)     & 0.009213(88)     \\
J306 & 258.4(5.1)          & 259.3(5.0)           &\quad{--}             &\quad{--}    &{--}             &{--}              \\
J303 & 274.4(4.5)          & 274.9(4.5)           & 53.8(1.2)           & 54.1(1.2)            & 0.01083(12)     & 0.01004(11)      \\
J304 & 262.2(4.7)          & 262.4(4.7)           & 45.6(1.4)           & 46.0(1.4)            &{--}             &{--}              \\
E300 & 310.7(5.0)          & 311.4(5.0)           & 45.8(2.2)           & 46.3(2.1)            & 0.012204(78)    & 0.011226(71)     \\
F300 & 350.4(9.7)          & 350.5(9.7)           &\quad{--}             &\quad{--} &{--}             &{--}              \\
\hline
J500 & 253.4(3.3)          & 254.6(4.0)           & 90.7(1.2)           & 91.1(1.4)            & 0.00827(25)     & 0.00780(23)      \\
J501 & 265.5(6.1)          & 263.3(6.8)           & 68.3(1.4)           & 68.2(1.5)            &{--}             &{--}              \\
\hline
\end{tabular} 
\caption{Values of the long-distance isovector, isoscalar and charm-connected contributions
	in units of $10^{-10}$, for the local-local (${\scriptstyle\rm LL}$) and for the local-conserved (${\scriptstyle\rm LC}$) discretizations of the correlation function, as described in the main text.  
	The finite-size correction to $(m_\pi L)^\mathrm{ref}$ has been applied to the isovector contribution. \label{tab:aLD}}
\end{table}

%% file: tables/tab_fv_ref.tex
\begin{table}[!ph]
	\centering
	\vskip 0.1in
	\renewcommand{\arraystretch}{1.1}  
	\begin{tabular}{
			c|
			S[table-format=-2.6]
			S[table-format=-2.6]
			S[table-format=-1.10]
			|S[table-format=-3.7]
			}
	\hline
	id &     {HP\&MLL}      &   {HP}       &    {Kaon}    &  total   \\
	\hline
	 A653 & -7.60(29)  & -7.23(39)  & {--}           & -7.60(29)  \\
	 A654 & \phantom{-}3.03(14)   & \phantom{-}2.66(15)   & \phantom{-}0.541(39)      & \phantom{-}3.57(14)   \\
	 \hline
	 H101 & -10.70(33) & -10.15(52) & {--}           & -10.70(33) \\
	 H102 & -4.27(25)  & -4.14(19)  & -1.186(90)     & -5.45(27)  \\
	 H105 & \phantom{-}4.07(27)   & \phantom{-}3.79(43)   & \phantom{-}0.322(24)      & \phantom{-}4.39(27)   \\
	 N101 & -6.59(20)  & -6.36(28)  & -0.336(22)     & -6.93(20)  \\
	 C101 & -2.53(13)  & -2.48(10)  & -0.0230(13)    & -2.56(13)  \\
	 C102 & -2.74(14)  & -2.66(12)  & -0.0163(20)    & -2.75(14)  \\
	 D150 & \phantom{-}11.33(98)  & \phantom{-}10.33(34)  & \phantom{-}0.000941(42)   & \phantom{-}11.33(98)  \\
	 \hline
	 B450 & -7.21(21)  & -6.89(35)  & {--}           & -7.21(21)  \\
	 S400 & -0.278(11) & -0.392(28) & -0.1142(94)    & -0.392(14) \\
	 N452 & -7.73(16)  & -7.50(31)  & -1.91(14)      & -9.64(24)  \\
	 N451 & -5.54(12)  & -5.39(19)  & -0.363(24)     & -5.90(13)  \\
	 D450 & -6.50(16)  & -6.29(18)  & -0.0357(19)    & -6.54(16)  \\
	 D451 & -6.45(14)  & -6.22(18)  & -0.0218(12)    & -6.47(14)  \\
	 D452 & \phantom{-}6.85(38)   & \phantom{-}6.36(19)   & \phantom{-}0.00296(14)    & \phantom{-}6.85(38)   \\
	 \hline
	 H200 & -0.793(34) & -0.974(57) & {--}           & -0.793(34) \\
	 N202 & -11.08(27) & -10.62(48) & {--}           & -11.08(27) \\
	 N203 & -5.64(14)  & -5.49(24)  & -1.39(10)      & -7.04(17)  \\
	 N200 & -1.057(34) & -1.082(59) & -0.0842(57)    & -1.141(34) \\
	 D251 & -7.07(11)  & -6.87(23)  & -0.378(24)     & -7.45(12)  \\
	 D200 & \phantom{-}1.365(62)  & \phantom{-}1.267(59)  & \phantom{-}0.00585(31)    & \phantom{-}1.371(62)  \\
	 D201 & \phantom{-}1.329(55)  & \phantom{-}1.228(42)  & \phantom{-}0.00405(21)    & \phantom{-}1.333(55)  \\
	 E250 & \phantom{-}3.351(99)  & \phantom{-}3.164(58)  & \phantom{-}0.0000909(33)  & \phantom{-}3.351(99)  \\
	 \hline
	 N300 & -5.95(19)  & -5.71(26)  & {--}           & -5.95(19)  \\
	 J307 & -11.95(50) & -11.45(79) & {--}           & -11.95(50) \\
	 N302 & \phantom{-}0.857(43)  & \phantom{-}0.711(36)  & \phantom{-}0.227(18)      & \phantom{-}1.084(46)  \\
	 J306 & -5.81(23)  & -5.65(35)  & -1.340(96)     & -7.15(25)  \\
	 J303 & \phantom{-}1.317(39)  & \phantom{-}1.216(45)  & \phantom{-}0.0498(31)     & \phantom{-}1.367(40)  \\
	 J304 & \phantom{-}0.794(23)  & \phantom{-}0.709(27)  & \phantom{-}0.01557(99)    & \phantom{-}0.810(23)  \\
	 E300 & \phantom{-}0.715(13)  & \phantom{-}0.651(15)  & \phantom{-}0.000621(29)   & \phantom{-}0.716(13)  \\
	 F300 & -0.491(18) & -0.480(15) & -0.0000146(20) & -0.491(18) \\
	 \hline
	 J500 & -7.65(32)  & -7.27(32)  & {--}           & -7.65(32)  \\
	 J501 & \phantom{-}0.572(22)  & \phantom{-}0.455(31)  & \phantom{-}0.1218(82)     & \phantom{-}0.694(24)  \\
	\hline
	\end{tabular}
	\caption{%
		Overview of finite-volume corrections to $(m_\pi L)^\mathrm{ref}$ using
		$f_\pi$ to set the scale.
		The column denoted by HP\&MLL gives the correction using the Hansen-Patella
		formalism for time separations smaller than $t^\ast$ and using the MLL beyond that point.
		The column denoted by HP uses only the Hansen-Patella formalism.
		The column ``Kaon'' gives the correction from the Kaon, which is included in the
		pion correction on SU(3) symmetric ensembles. 
		The total is computed by the sum of the columns ``HP\&MLL'' and ``Kaon''
		and enters the numbers for $\aLDf{3}{3}$ in \cref{tab:aLD}.
		All uncertainties are statistical. \label{tab:fv_ref}
	}%
\end{table}

%% file: amuld.bbl
\providecommand{\href}[2]{#2}\begingroup\raggedright\begin{thebibliography}{100}

\bibitem{Aoyama:2020ynm}
T.~Aoyama et~al., \emph{{The anomalous magnetic moment of the muon in the
  Standard Model}},
  \href{https://doi.org/10.1016/j.physrep.2020.07.006}{\emph{Phys. Rept.}
  {\bfseries 887} (2020) 1} [\href{https://arxiv.org/abs/2006.04822}{{\ttfamily
  2006.04822}}].

\bibitem{Davier:2017zfy}
M.~Davier, A.~Hoecker, B.~Malaescu and Z.~Zhang, \emph{{Reevaluation of the
  hadronic vacuum polarisation contributions to the Standard Model predictions
  of the muon $g-2$ and ${\alpha (m_Z^2)}$ using newest hadronic cross-section
  data}}, \href{https://doi.org/10.1140/epjc/s10052-017-5161-6}{\emph{Eur.
  Phys. J. C} {\bfseries 77} (2017) 827}
  [\href{https://arxiv.org/abs/1706.09436}{{\ttfamily 1706.09436}}].

\bibitem{Keshavarzi:2018mgv}
A.~Keshavarzi, D.~Nomura and T.~Teubner, \emph{{The muon $g-2$ and
  $\alpha(M_Z^2)$: a new data-based analysis}},
  \href{https://doi.org/10.1103/PhysRevD.97.114025}{\emph{Phys. Rev. D}
  {\bfseries 97} (2018) 114025}
  [\href{https://arxiv.org/abs/1802.02995}{{\ttfamily 1802.02995}}].

\bibitem{Colangelo:2018mtw}
G.~Colangelo, M.~Hoferichter and P.~Stoffer, \emph{{Two-pion contribution to
  hadronic vacuum polarization}},
  \href{https://doi.org/10.1007/JHEP02(2019)006}{\emph{JHEP} {\bfseries 02}
  (2019) 006} [\href{https://arxiv.org/abs/1810.00007}{{\ttfamily
  1810.00007}}].

\bibitem{Hoferichter:2019mqg}
M.~Hoferichter, B.-L. Hoid and B.~Kubis, \emph{{Three-pion contribution to
  hadronic vacuum polarization}},
  \href{https://doi.org/10.1007/JHEP08(2019)137}{\emph{JHEP} {\bfseries 08}
  (2019) 137} [\href{https://arxiv.org/abs/1907.01556}{{\ttfamily
  1907.01556}}].

\bibitem{Davier:2019can}
M.~Davier, A.~Hoecker, B.~Malaescu and Z.~Zhang, \emph{{A new evaluation of the
  hadronic vacuum polarisation contributions to the muon anomalous magnetic
  moment and to $\alpha(m_Z^2)$}},
  \href{https://doi.org/10.1140/epjc/s10052-020-7792-2}{\emph{Eur. Phys. J. C}
  {\bfseries 80} (2020) 241}
  [\href{https://arxiv.org/abs/1908.00921}{{\ttfamily 1908.00921}}].

\bibitem{Keshavarzi:2019abf}
A.~Keshavarzi, D.~Nomura and T.~Teubner, \emph{{$g-2$ of charged leptons,
  $\alpha (M^2_Z)$, and the hyperfine splitting of muonium}},
  \href{https://doi.org/10.1103/PhysRevD.101.014029}{\emph{Phys. Rev. D}
  {\bfseries 101} (2020) 014029}
  [\href{https://arxiv.org/abs/1911.00367}{{\ttfamily 1911.00367}}].

\bibitem{Wittig:2023pcl}
H.~Wittig, \emph{{Progress on $(g-2)_\mu$ from Lattice QCD}},  in \emph{{57th
  Rencontres de Moriond on Electroweak Interactions and Unified Theories}}, 6,
  2023, \href{https://arxiv.org/abs/2306.04165}{{\ttfamily 2306.04165}}.

\bibitem{Kuberski:2023qgx}
S.~Kuberski, \emph{{Muon $g-2$: Lattice calculations of the hadronic vacuum
  polarization}}, \href{https://doi.org/10.22323/1.453.0125}{\emph{PoS}
  {\bfseries LATTICE2023} (2024) 125}
  [\href{https://arxiv.org/abs/2312.13753}{{\ttfamily 2312.13753}}].

\bibitem{CMD-3:2023alj}
{\scshape CMD-3} collaboration, F.~V. Ignatov et~al., \emph{{Measurement of the
  $e^+e^-\to{\pi}^+{\pi}^-$ cross section from threshold to 1.2~GeV with the
  CMD-3 detector}},
  \href{https://doi.org/10.1103/PhysRevD.109.112002}{\emph{Phys. Rev. D}
  {\bfseries 109} (2024) 112002}
  [\href{https://arxiv.org/abs/2302.08834}{{\ttfamily 2302.08834}}].

\bibitem{CMD-3:2023rfe}
{\scshape CMD-3} collaboration, F.~V. Ignatov et~al., \emph{{Measurement of the
  Pion Form Factor with CMD-3 Detector and its Implication to the Hadronic
  Contribution to Muon $(g-2)$}},
  \href{https://doi.org/10.1103/PhysRevLett.132.231903}{\emph{Phys. Rev. Lett.}
  {\bfseries 132} (2024) 231903}
  [\href{https://arxiv.org/abs/2309.12910}{{\ttfamily 2309.12910}}].

\bibitem{Muong-2:2021ojo}
{\scshape Muon $g-2$} collaboration, B.~Abi et~al., \emph{{Measurement of the
  Positive Muon Anomalous Magnetic Moment to 0.46 ppm}},
  \href{https://doi.org/10.1103/PhysRevLett.126.141801}{\emph{Phys. Rev. Lett.}
  {\bfseries 126} (2021) 141801}
  [\href{https://arxiv.org/abs/2104.03281}{{\ttfamily 2104.03281}}].

\bibitem{Muong-2:2023cdq}
{\scshape Muon $g-2$} collaboration, D.~P. Aguillard et~al., \emph{{Measurement
  of the Positive Muon Anomalous Magnetic Moment to 0.20~ppm}},
  \href{https://doi.org/10.1103/PhysRevLett.131.161802}{\emph{Phys. Rev. Lett.}
  {\bfseries 131} (2023) 161802}
  [\href{https://arxiv.org/abs/2308.06230}{{\ttfamily 2308.06230}}].

\bibitem{Borsanyi:2020mff}
S.~Borsanyi et~al., \emph{{Leading hadronic contribution to the muon magnetic
  moment from lattice QCD}},
  \href{https://doi.org/10.1038/s41586-021-03418-1}{\emph{Nature} {\bfseries
  593} (2021) 51} [\href{https://arxiv.org/abs/2002.12347}{{\ttfamily
  2002.12347}}].

\bibitem{Lehner:2020crt}
C.~Lehner and A.~S. Meyer, \emph{{Consistency of hadronic vacuum polarization
  between lattice QCD and the R-ratio}},
  \href{https://doi.org/10.1103/PhysRevD.101.074515}{\emph{Phys. Rev. D}
  {\bfseries 101} (2020) 074515}
  [\href{https://arxiv.org/abs/2003.04177}{{\ttfamily 2003.04177}}].

\bibitem{Wang:2022lkq}
{\scshape $\chi$QCD} collaboration, G.~Wang, T.~Draper, K.-F. Liu and Y.-B.
  Yang, \emph{{Muon g-2 with overlap valence fermions}},
  \href{https://doi.org/10.1103/PhysRevD.107.034513}{\emph{Phys. Rev. D}
  {\bfseries 107} (2023) 034513}
  [\href{https://arxiv.org/abs/2204.01280}{{\ttfamily 2204.01280}}].

\bibitem{Aubin:2022hgm}
C.~Aubin, T.~Blum, M.~Golterman and S.~Peris, \emph{{Muon anomalous magnetic
  moment with staggered fermions: Is the lattice spacing small enough?}},
  \href{https://doi.org/10.1103/PhysRevD.106.054503}{\emph{Phys. Rev. D}
  {\bfseries 106} (2022) 054503}
  [\href{https://arxiv.org/abs/2204.12256}{{\ttfamily 2204.12256}}].

\bibitem{Ce:2022kxy}
M.~C\`e, A.~G\'erardin, G.~von Hippel, R.~J. Hudspith, S.~Kuberski, H.~B. Meyer
  et~al., \emph{{Window observable for the hadronic vacuum polarization
  contribution to the muon g-2 from lattice QCD}},
  \href{https://doi.org/10.1103/PhysRevD.106.114502}{\emph{Phys. Rev. D}
  {\bfseries 106} (2022) 114502}
  [\href{https://arxiv.org/abs/2206.06582}{{\ttfamily 2206.06582}}].

\bibitem{ExtendedTwistedMass:2022jpw}
{\scshape Extended Twisted Mass} collaboration, C.~Alexandrou et~al.,
  \emph{{Lattice calculation of the short and intermediate time-distance
  hadronic vacuum polarization contributions to the muon magnetic moment using
  twisted-mass fermions}},
  \href{https://doi.org/10.1103/PhysRevD.107.074506}{\emph{Phys. Rev. D}
  {\bfseries 107} (2023) 074506}
  [\href{https://arxiv.org/abs/2206.15084}{{\ttfamily 2206.15084}}].

\bibitem{FermilabLatticeHPQCD:2023jof}
{\scshape Fermilab Lattice, HPQCD, MILC} collaboration, A.~Bazavov et~al.,
  \emph{{Light-quark connected intermediate-window contributions to the muon
  g-2 hadronic vacuum polarization from lattice QCD}},
  \href{https://doi.org/10.1103/PhysRevD.107.114514}{\emph{Phys. Rev. D}
  {\bfseries 107} (2023) 114514}
  [\href{https://arxiv.org/abs/2301.08274}{{\ttfamily 2301.08274}}].

\bibitem{RBC:2023pvn}
{\scshape RBC, UKQCD} collaboration, T.~Blum et~al., \emph{{Update of Euclidean
  windows of the hadronic vacuum polarization}},
  \href{https://doi.org/10.1103/PhysRevD.108.054507}{\emph{Phys. Rev. D}
  {\bfseries 108} (2023) 054507}
  [\href{https://arxiv.org/abs/2301.08696}{{\ttfamily 2301.08696}}].

\bibitem{Kuberski:2024bcj}
S.~Kuberski, M.~C\`e, G.~von Hippel, H.~B. Meyer, K.~Ottnad, A.~Risch et~al.,
  \emph{{Hadronic vacuum polarization in the muon g \ensuremath{-} 2: the
  short-distance contribution from lattice QCD}},
  \href{https://doi.org/10.1007/JHEP03(2024)172}{\emph{JHEP} {\bfseries 03}
  (2024) 172} [\href{https://arxiv.org/abs/2401.11895}{{\ttfamily
  2401.11895}}].

\bibitem{Boito:2022dry}
D.~Boito, M.~Golterman, K.~Maltman and S.~Peris, \emph{{Data-based
  determination of the isospin-limit light-quark-connected contribution to the
  anomalous magnetic moment of the muon}},
  \href{https://doi.org/10.1103/PhysRevD.107.074001}{\emph{Phys. Rev. D}
  {\bfseries 107} (2023) 074001}
  [\href{https://arxiv.org/abs/2211.11055}{{\ttfamily 2211.11055}}].

\bibitem{Benton:2024kwp}
G.~Benton, D.~Boito, M.~Golterman, A.~Keshavarzi, K.~Maltman and S.~Peris,
  \emph{{Data-driven results for light-quark connected and
  strange-plus-disconnected hadronic g\ensuremath{-}2 short- and long-distance
  windows}}, \href{https://doi.org/10.1103/PhysRevD.111.034018}{\emph{Phys.
  Rev. D} {\bfseries 111} (2025) 034018}
  [\href{https://arxiv.org/abs/2411.06637}{{\ttfamily 2411.06637}}].

\bibitem{Boccaletti:2024guq}
A.~Boccaletti et~al., \emph{{High precision calculation of the hadronic vacuum
  polarisation contribution to the muon anomaly}},
  \href{https://arxiv.org/abs/2407.10913}{{\ttfamily 2407.10913}}.

\bibitem{Bernecker:2011gh}
D.~Bernecker and H.~B. Meyer, \emph{{Vector Correlators in Lattice QCD: Methods
  and applications}},
  \href{https://doi.org/10.1140/epja/i2011-11148-6}{\emph{Eur. Phys. J. A}
  {\bfseries 47} (2011) 148} [\href{https://arxiv.org/abs/1107.4388}{{\ttfamily
  1107.4388}}].

\bibitem{Blum:2018mom}
{\scshape RBC, UKQCD} collaboration, T.~Blum, P.~A. Boyle, V.~G{\"u}lpers,
  T.~Izubuchi, L.~Jin, C.~Jung et~al., \emph{{Calculation of the hadronic
  vacuum polarization contribution to the muon anomalous magnetic moment}},
  \href{https://doi.org/10.1103/PhysRevLett.121.022003}{\emph{Phys. Rev. Lett.}
  {\bfseries 121} (2018) 022003}
  [\href{https://arxiv.org/abs/1801.07224}{{\ttfamily 1801.07224}}].

\bibitem{Bruno:2014jqa}
M.~Bruno et~al., \emph{{Simulation of QCD with N$\_{f} =$ 2 $+$ 1 flavors of
  non-perturbatively improved Wilson fermions}},
  \href{https://doi.org/10.1007/JHEP02(2015)043}{\emph{JHEP} {\bfseries 02}
  (2015) 043} [\href{https://arxiv.org/abs/1411.3982}{{\ttfamily 1411.3982}}].

\bibitem{Bali:2016umi}
G.~S. Bali, E.~E. Scholz, J.~Simeth and W.~Söldner, \emph{{Lattice simulations
  with $N_f=2+1$ improved Wilson fermions at a fixed strange quark mass}},
  \href{https://doi.org/10.1103/PhysRevD.94.074501}{\emph{Phys. Rev. D}
  {\bfseries 94} (2016) 074501}
  [\href{https://arxiv.org/abs/1606.09039}{{\ttfamily 1606.09039}}].

\bibitem{Bulava:2013cta}
J.~Bulava and S.~Schaefer, \emph{{Improvement of $N_f=3$ lattice QCD with
  Wilson fermions and tree-level improved gauge action}},
  \href{https://doi.org/10.1016/j.nuclphysb.2013.05.019}{\emph{Nucl. Phys. B}
  {\bfseries 874} (2013) 188}
  [\href{https://arxiv.org/abs/1304.7093}{{\ttfamily 1304.7093}}].

\bibitem{Clark:2006fx}
M.~Clark and A.~Kennedy, \emph{{Accelerating dynamical fermion computations
  using the rational hybrid Monte Carlo (RHMC) algorithm with multiple
  pseudofermion fields}},
  \href{https://doi.org/10.1103/PhysRevLett.98.051601}{\emph{Phys. Rev. Lett.}
  {\bfseries 98} (2007) 051601}
  [\href{https://arxiv.org/abs/hep-lat/0608015}{{\ttfamily hep-lat/0608015}}].

\bibitem{Luscher:2012av}
M.~L{\"u}scher and S.~Schaefer, \emph{{Lattice QCD with open boundary
  conditions and twisted-mass reweighting}},
  \href{https://doi.org/10.1016/j.cpc.2012.10.003}{\emph{Comput. Phys. Commun.}
  {\bfseries 184} (2013) 519}
  [\href{https://arxiv.org/abs/1206.2809}{{\ttfamily 1206.2809}}].

\bibitem{Mohler:2020txx}
D.~Mohler and S.~Schaefer, \emph{{Remarks on strange-quark simulations with
  Wilson fermions}},
  \href{https://doi.org/10.1103/PhysRevD.102.074506}{\emph{Phys. Rev. D}
  {\bfseries 102} (2020) 074506}
  [\href{https://arxiv.org/abs/2003.13359}{{\ttfamily 2003.13359}}].

\bibitem{Kuberski:2023zky}
S.~Kuberski, \emph{{Low-mode deflation for twisted-mass and RHMC reweighting in
  lattice QCD}}, \href{https://doi.org/10.1016/j.cpc.2024.109173}{\emph{Comput.
  Phys. Commun.} {\bfseries 300} (2024) 109173}
  [\href{https://arxiv.org/abs/2306.02385}{{\ttfamily 2306.02385}}].

\bibitem{Gerardin:2019rua}
A.~G\'erardin, M.~C\`e, G.~von Hippel, B.~H\"orz, H.~B. Meyer, D.~Mohler
  et~al., \emph{{The leading hadronic contribution to $(g-2)_\mu$ from lattice
  QCD with $N_{\rm f}=2+1$ flavours of O($a$) improved Wilson quarks}},
  \href{https://doi.org/10.1103/PhysRevD.100.014510}{\emph{Phys. Rev. D}
  {\bfseries 100} (2019) 014510}
  [\href{https://arxiv.org/abs/1904.03120}{{\ttfamily 1904.03120}}].

\bibitem{Ce:2022eix}
M.~C\`e, A.~G\'erardin, G.~von Hippel, H.~B. Meyer, K.~Miura, K.~Ottnad et~al.,
  \emph{{The hadronic running of the electromagnetic coupling and the
  electroweak mixing angle from lattice QCD}},
  \href{https://doi.org/10.1007/JHEP08(2022)220}{\emph{JHEP} {\bfseries 08}
  (2022) 220} [\href{https://arxiv.org/abs/2203.08676}{{\ttfamily
  2203.08676}}].

\bibitem{Strassberger:2021tsu}
B.~Strassberger et~al., \emph{{Scale setting for CLS 2+1 simulations}},
  \href{https://doi.org/10.22323/1.396.0135}{\emph{PoS} {\bfseries LATTICE2021}
  (2022) 135} [\href{https://arxiv.org/abs/2112.06696}{{\ttfamily
  2112.06696}}].

\bibitem{RQCD:2022xux}
{\scshape RQCD} collaboration, G.~S. Bali, S.~Collins, P.~Georg, D.~Jenkins,
  P.~Korcyl, A.~Sch\"afer et~al., \emph{{Scale setting and the light baryon
  spectrum in N$_{f}$ = 2 + 1 QCD with Wilson fermions}},
  \href{https://doi.org/10.1007/JHEP05(2023)035}{\emph{JHEP} {\bfseries 05}
  (2023) 035} [\href{https://arxiv.org/abs/2211.03744}{{\ttfamily
  2211.03744}}].

\bibitem{Gerardin:2018kpy}
A.~G{\'e}rardin, T.~Harris and H.~B. Meyer, \emph{{Nonperturbative
  renormalization and $O(a)$-improvement of the nonsinglet vector current with
  $N_f=2+1$ Wilson fermions and tree-level Symanzik improved gauge action}},
  \href{https://doi.org/10.1103/PhysRevD.99.014519}{\emph{Phys. Rev. D}
  {\bfseries 99} (2019) 014519}
  [\href{https://arxiv.org/abs/1811.08209}{{\ttfamily 1811.08209}}].

\bibitem{Heitger:2020zaq}
{\scshape ALPHA} collaboration, J.~Heitger and F.~Joswig, \emph{{The
  renormalised $\mathrm{O}(a)$ improved vector current in three-flavour lattice
  QCD with Wilson quarks}},
  \href{https://doi.org/10.1140/epjc/s10052-021-09037-4}{\emph{Eur. Phys. J. C}
  {\bfseries 81} (2021) 254}
  [\href{https://arxiv.org/abs/2010.09539}{{\ttfamily 2010.09539}}].

\bibitem{Bali:2023sdi}
{\scshape RQCD} collaboration, G.~S. Bali, S.~Collins, S.~Heybrock,
  M.~L\"offler, R.~R\"odl, W.~S\"oldner et~al., \emph{{Octet baryon isovector
  charges from Nf=2+1 lattice QCD}},
  \href{https://doi.org/10.1103/PhysRevD.108.034512}{\emph{Phys. Rev. D}
  {\bfseries 108} (2023) 034512}
  [\href{https://arxiv.org/abs/2305.04717}{{\ttfamily 2305.04717}}].

\bibitem{Harris:20XX}
T.~Harris and H.~B. Meyer, to be published.

\bibitem{Fritzsch:2018zym}
P.~Fritzsch, \emph{{Mass-improvement of the vector current in three-flavor
  QCD}}, \href{https://doi.org/10.1007/JHEP06(2018)015}{\emph{JHEP} {\bfseries
  06} (2018) 015} [\href{https://arxiv.org/abs/1805.07401}{{\ttfamily
  1805.07401}}].

\bibitem{Giusti:2004yp}
L.~Giusti, P.~Hernandez, M.~Laine, P.~Weisz and H.~Wittig, \emph{{Low-energy
  couplings of QCD from current correlators near the chiral limit}},
  \href{https://doi.org/10.1088/1126-6708/2004/04/013}{\emph{JHEP} {\bfseries
  04} (2004) 013} [\href{https://arxiv.org/abs/hep-lat/0402002}{{\ttfamily
  hep-lat/0402002}}].

\bibitem{DeGrand:2004qw}
T.~A. DeGrand and S.~Schaefer, \emph{{Improving meson two point functions in
  lattice QCD}}, \href{https://doi.org/10.1016/j.cpc.2004.02.006}{\emph{Comput.
  Phys. Commun.} {\bfseries 159} (2004) 185}
  [\href{https://arxiv.org/abs/hep-lat/0401011}{{\ttfamily hep-lat/0401011}}].

\bibitem{Jansen:2008wv}
{\scshape ETM} collaboration, K.~Jansen, C.~Michael and C.~Urbach, \emph{{The
  eta-prime meson from lattice QCD}},
  \href{https://doi.org/10.1140/epjc/s10052-008-0764-6}{\emph{Eur. Phys. J. C}
  {\bfseries 58} (2008) 261} [\href{https://arxiv.org/abs/0804.3871}{{\ttfamily
  0804.3871}}].

\bibitem{Giusti:2019kff}
L.~Giusti, T.~Harris, A.~Nada and S.~Schaefer, \emph{{Frequency-splitting
  estimators of single-propagator traces}},
  \href{https://doi.org/10.1140/epjc/s10052-019-7049-0}{\emph{Eur. Phys. J. C}
  {\bfseries 79} (2019) 586}
  [\href{https://arxiv.org/abs/1903.10447}{{\ttfamily 1903.10447}}].

\bibitem{Stathopoulos:2013aci}
A.~Stathopoulos, J.~Laeuchli and K.~Orginos, \emph{{Hierarchical probing for
  estimating the trace of the matrix inverse on toroidal lattices}},
  \href{https://arxiv.org/abs/1302.4018}{{\ttfamily 1302.4018}}.

\bibitem{Gulpers:2013uca}
V.~G{\"u}lpers, G.~von Hippel and H.~Wittig, \emph{{Scalar pion form factor in
  two-flavor lattice QCD}},
  \href{https://doi.org/10.1103/PhysRevD.89.094503}{\emph{Phys. Rev. D}
  {\bfseries 89} (2014) 094503}
  [\href{https://arxiv.org/abs/1309.2104}{{\ttfamily 1309.2104}}].

\bibitem{LehnerBounding2016}
C.~Lehner, \emph{The hadronic vacuum polarization contribution to the muon
  anomalous magnetic moment},  in \emph{{RBRC Workshop on Lattice Gauge
  Theories}}, 2016,
  \href{https://indico.bnl.gov/event/1628/contributions/2819/}{https://indico.bnl.gov/event/1628/contributions/2819/}.

\bibitem{Borsanyi:2016lpl}
S.~Borsanyi, Z.~Fodor, T.~Kawanai, S.~Krieg, L.~Lellouch, R.~Malak et~al.,
  \emph{{Slope and curvature of the hadronic vacuum polarization at vanishing
  virtuality from lattice QCD}},
  \href{https://doi.org/10.1103/PhysRevD.96.074507}{\emph{Phys. Rev. D}
  {\bfseries 96} (2017) 074507}
  [\href{https://arxiv.org/abs/1612.02364}{{\ttfamily 1612.02364}}].

\bibitem{DellaMorte:2017dyu}
M.~Della~Morte, A.~Francis, V.~G{\"u}lpers, G.~Herdo{\'\i}za, G.~von Hippel,
  H.~Horch et~al., \emph{{The hadronic vacuum polarization contribution to the
  muon $g-2$ from lattice QCD}},
  \href{https://doi.org/10.1007/JHEP10(2017)020}{\emph{JHEP} {\bfseries 10}
  (2017) 020} [\href{https://arxiv.org/abs/1705.01775}{{\ttfamily
  1705.01775}}].

\bibitem{Andersen:2018mau}
C.~Andersen, J.~Bulava, B.~H\"orz and C.~Morningstar, \emph{{The $I=1$
  pion-pion scattering amplitude and timelike pion form factor from $N_{\rm f}
  = 2+1$ lattice QCD}},
  \href{https://doi.org/10.1016/j.nuclphysb.2018.12.018}{\emph{Nucl. Phys. B}
  {\bfseries 939} (2019) 145}
  [\href{https://arxiv.org/abs/1808.05007}{{\ttfamily 1808.05007}}].

\bibitem{Paul:2021pjz}
S.~Paul, A.~D. Hanlon, B.~H\"orz, D.~Mohler, C.~Morningstar and H.~Wittig,
  \emph{{I=1 $\pi$-$\pi$ scattering at the physical point}},
  \href{https://doi.org/10.22323/1.396.0551}{\emph{PoS} {\bfseries LATTICE2021}
  (2022) 551} [\href{https://arxiv.org/abs/2112.07385}{{\ttfamily
  2112.07385}}].

\bibitem{Paul:2023ksa}
S.~Paul, A.~D. Hanlon, B.~H\"orz, D.~Mohler, C.~Morningstar and H.~Wittig,
  \emph{{The long-distance behaviour of the vector-vector correlator from
  $\pi\pi$ scattering}}, \href{https://doi.org/10.22323/1.430.0073}{\emph{PoS}
  {\bfseries LATTICE2022} (2023) 073}.

\bibitem{Colangelo:2021moe}
G.~Colangelo, M.~Hoferichter, B.~Kubis, M.~Niehus and J.~R. de~Elvira,
  \emph{{Chiral extrapolation of hadronic vacuum polarization}},
  \href{https://doi.org/10.1016/j.physletb.2021.136852}{\emph{Phys. Lett. B}
  {\bfseries 825} (2022) 136852}
  [\href{https://arxiv.org/abs/2110.05493}{{\ttfamily 2110.05493}}].

\bibitem{Husung:2019ytz}
N.~Husung, P.~Marquard and R.~Sommer, \emph{{Asymptotic behavior of cutoff
  effects in Yang\textendash{}Mills theory and in Wilson\textquoteright{}s
  lattice QCD}},
  \href{https://doi.org/10.1140/epjc/s10052-020-7685-4}{\emph{Eur. Phys. J. C}
  {\bfseries 80} (2020) 200}
  [\href{https://arxiv.org/abs/1912.08498}{{\ttfamily 1912.08498}}].

\bibitem{Husung:2021mfl}
N.~Husung, P.~Marquard and R.~Sommer, \emph{{The asymptotic approach to the
  continuum of lattice QCD spectral observables}},
  \href{https://doi.org/10.1016/j.physletb.2022.137069}{\emph{Phys. Lett. B}
  {\bfseries 829} (2022) 137069}
  [\href{https://arxiv.org/abs/2111.02347}{{\ttfamily 2111.02347}}].

\bibitem{Husung:2024cgc}
N.~Husung, \emph{{Lattice artifacts of local fermion bilinears up to
  $\mathrm{O}(a^2)$}},  \href{https://arxiv.org/abs/2409.00776}{{\ttfamily
  2409.00776}}.

\bibitem{Colangelo:2005gd}
G.~Colangelo, S.~D{\"u}rr and C.~Haefeli, \emph{{Finite volume effects for
  meson masses and decay constants}},
  \href{https://doi.org/10.1016/j.nuclphysb.2005.05.015}{\emph{Nucl. Phys. B}
  {\bfseries 721} (2005) 136}
  [\href{https://arxiv.org/abs/hep-lat/0503014}{{\ttfamily hep-lat/0503014}}].

\bibitem{DallaBrida:2018tpn}
M.~Dalla~Brida, T.~Korzec, S.~Sint and P.~Vilaseca, \emph{{High precision
  renormalization of the flavour non-singlet Noether currents in lattice QCD
  with Wilson quarks}},
  \href{https://doi.org/10.1140/epjc/s10052-018-6514-5}{\emph{Eur. Phys. J. C}
  {\bfseries 79} (2019) 23} [\href{https://arxiv.org/abs/1808.09236}{{\ttfamily
  1808.09236}}].

\bibitem{Bali:2021qem}
{\scshape RQCD} collaboration, G.~S. Bali, V.~Braun, S.~Collins, A.~Sch\"afer
  and J.~Simeth, \emph{{Masses and decay constants of the \ensuremath{\eta} and
  \ensuremath{\eta}' mesons from lattice QCD}},
  \href{https://doi.org/10.1007/JHEP08(2021)137}{\emph{JHEP} {\bfseries 08}
  (2021) 137} [\href{https://arxiv.org/abs/2106.05398}{{\ttfamily
  2106.05398}}].

\bibitem{Gasser:1984gg}
J.~Gasser and H.~Leutwyler, \emph{{Chiral Perturbation Theory: Expansions in
  the Mass of the Strange Quark}},
  \href{https://doi.org/10.1016/0550-3213(85)90492-4}{\emph{Nucl. Phys. B}
  {\bfseries 250} (1985) 465}.

\bibitem{Baikov:2016tgj}
{Baikov, P. A. and Chetyrkin, K. G. and K\"uhn, J. H.}, \emph{{Five-Loop
  Running of the QCD coupling constant}},
  \href{https://doi.org/10.1103/PhysRevLett.118.082002}{\emph{Phys. Rev. Lett.}
  {\bfseries 118} (2017) 082002}
  [\href{https://arxiv.org/abs/1606.08659}{{\ttfamily 1606.08659}}].

\bibitem{Bruno:2017gxd}
{\scshape ALPHA} collaboration, M.~Bruno, M.~Dalla~Brida, P.~Fritzsch,
  T.~Korzec, A.~Ramos, S.~Schaefer et~al., \emph{{QCD Coupling from a
  Nonperturbative Determination of the Three-Flavor $\Lambda$ Parameter}},
  \href{https://doi.org/10.1103/PhysRevLett.119.102001}{\emph{Phys. Rev. Lett.}
  {\bfseries 119} (2017) 102001}
  [\href{https://arxiv.org/abs/1706.03821}{{\ttfamily 1706.03821}}].

\bibitem{Jay:2020jkz}
W.~I. Jay and E.~T. Neil, \emph{{Bayesian model averaging for analysis of
  lattice field theory results}},
  \href{https://doi.org/10.1103/PhysRevD.103.114502}{\emph{Phys. Rev. D}
  {\bfseries 103} (2021) 114502}
  [\href{https://arxiv.org/abs/2008.01069}{{\ttfamily 2008.01069}}].

\bibitem{Akaike1998}
H.~Akaike, \emph{Information theory and an extension of the maximum likelihood
  principle},  in \emph{Springer Series in Statistics}, p.~199, Springer New
  York, (1998), \href{https://doi.org/10.1007/978-1-4612-1694-0\_15}{{\ttfamily
  10.1007/978-1-4612-1694-0\_15}}.

\bibitem{Wolff:2003sm}
{\scshape ALPHA} collaboration, U.~Wolff, \emph{{Monte Carlo errors with less
  errors}}, \href{https://doi.org/10.1016/S0010-4655(03)00467-3}{\emph{Comput.
  Phys. Commun.} {\bfseries 156} (2004) 143}
  [\href{https://arxiv.org/abs/hep-lat/0306017}{{\ttfamily hep-lat/0306017}}].

\bibitem{Ramos:2018vgu}
A.~Ramos, \emph{{Automatic differentiation for error analysis of Monte Carlo
  data}}, \href{https://doi.org/10.1016/j.cpc.2018.12.020}{\emph{Comput. Phys.
  Commun.} {\bfseries 238} (2019) 19}
  [\href{https://arxiv.org/abs/1809.01289}{{\ttfamily 1809.01289}}].

\bibitem{Joswig:2022qfe}
F.~Joswig, S.~Kuberski, J.~T. Kuhlmann and J.~Neuendorf, \emph{{pyerrors: A
  python framework for error analysis of Monte Carlo data}},
  \href{https://doi.org/10.1016/j.cpc.2023.108750}{\emph{Comput. Phys. Commun.}
  {\bfseries 288} (2023) 108750}
  [\href{https://arxiv.org/abs/2209.14371}{{\ttfamily 2209.14371}}].

\bibitem{Hansen:2019rbh}
M.~T. Hansen and A.~Patella, \emph{{Finite-volume effects in
  $(g-2)^{\text{HVP,LO}}_\mu$}},
  \href{https://doi.org/10.1103/PhysRevLett.123.172001}{\emph{Phys. Rev. Lett.}
  {\bfseries 123} (2019) 172001}
  [\href{https://arxiv.org/abs/1904.10010}{{\ttfamily 1904.10010}}].

\bibitem{Hansen:2020whp}
M.~T. Hansen and A.~Patella, \emph{{Finite-volume and thermal effects in the
  leading-HVP contribution to muonic ($g-2$)}},
  \href{https://doi.org/10.1007/JHEP10(2020)029}{\emph{JHEP} {\bfseries 10}
  (2020) 029} [\href{https://arxiv.org/abs/2004.03935}{{\ttfamily
  2004.03935}}].

\bibitem{Meyer:2011um}
H.~B. Meyer, \emph{{Lattice QCD and the Timelike Pion Form Factor}},
  \href{https://doi.org/10.1103/PhysRevLett.107.072002}{\emph{Phys. Rev. Lett.}
  {\bfseries 107} (2011) 072002}
  [\href{https://arxiv.org/abs/1105.1892}{{\ttfamily 1105.1892}}].

\bibitem{DellaMorte:2010aq}
M.~Della~Morte and A.~J{\"u}ttner, \emph{{Quark disconnected diagrams in chiral
  perturbation theory}},
  \href{https://doi.org/10.1007/JHEP11(2010)154}{\emph{JHEP} {\bfseries 1011}
  (2010) 154} [\href{https://arxiv.org/abs/1009.3783}{{\ttfamily 1009.3783}}].

\bibitem{Francis:2013fzp}
A.~Francis, B.~J{\"a}ger, H.~B. Meyer and H.~Wittig, \emph{{A new
  representation of the Adler function for lattice QCD}},
  \href{https://doi.org/10.1103/PhysRevD.88.054502}{\emph{Phys. Rev. D}
  {\bfseries 88} (2013) 054502}
  [\href{https://arxiv.org/abs/1306.2532}{{\ttfamily 1306.2532}}].

\bibitem{pion_ff_to_be_published}
D.~Djukanovic, A.~D. Hanlon, B.~H{\"o}rz, S.~Kuberski, H.~B. Meyer, N.~Miller
  et~al., \emph{{The pion form factor on an $N_f = 2 + 1$ O($a$)-improved
  Wilson fermions physical point ensemble}},  in preparation.

\bibitem{ParticleDataGroup:2024cfk}
{\scshape Particle Data Group} collaboration, S.~Navas et~al., \emph{{Review of
  particle physics}},
  \href{https://doi.org/10.1103/PhysRevD.110.030001}{\emph{Phys. Rev. D}
  {\bfseries 110} (2024) 030001}.

\bibitem{Aubin:2019usy}
C.~Aubin, T.~Blum, C.~Tu, M.~Golterman, C.~Jung and S.~Peris, \emph{{Light
  quark vacuum polarization at the physical point and contribution to the muon
  $g-2$}}, \href{https://doi.org/10.1103/PhysRevD.101.014503}{\emph{Phys. Rev.
  D} {\bfseries 101} (2020) 014503}
  [\href{https://arxiv.org/abs/1905.09307}{{\ttfamily 1905.09307}}].

\bibitem{SND:2023gan}
{\scshape SND} collaboration, M.~N. Achasov et~al., \emph{{Study of the process
  $e^+e^-\to{\omega}{\pi}^0\to{\pi}^+{\pi}^-{\pi}^0{\pi}^0$ in the energy range
  1.05{}\textendash{}2.00 GeV with SND}},
  \href{https://doi.org/10.1103/PhysRevD.108.092012}{\emph{Phys. Rev. D}
  {\bfseries 108} (2023) 092012}
  [\href{https://arxiv.org/abs/2309.00280}{{\ttfamily 2309.00280}}].

\bibitem{Wess:1971yu}
J.~Wess and B.~Zumino, \emph{{Consequences of anomalous Ward identities}},
  \href{https://doi.org/10.1016/0370-2693(71)90582-X}{\emph{Phys. Lett. B}
  {\bfseries 37} (1971) 95}.

\bibitem{Witten:1983tw}
E.~Witten, \emph{{Global Aspects of Current Algebra}},
  \href{https://doi.org/10.1016/0550-3213(83)90063-9}{\emph{Nucl. Phys. B}
  {\bfseries 223} (1983) 422}.

\bibitem{Yan:2024gwp}
H.~Yan, M.~Mai, M.~Garofalo, U.-G. Mei\ss{}ner, C.~Liu, L.~Liu et~al.,
  \emph{{\ensuremath{\omega} Meson from Lattice QCD}},
  \href{https://doi.org/10.1103/PhysRevLett.133.211906}{\emph{Phys. Rev. Lett.}
  {\bfseries 133} (2024) 211906}
  [\href{https://arxiv.org/abs/2407.16659}{{\ttfamily 2407.16659}}].

\bibitem{Bruno:2019nzm}
M.~Bruno, T.~Izubuchi, C.~Lehner and A.~S. Meyer, \emph{{Exclusive Channel
  Study of the Muon HVP}},
  \href{https://doi.org/10.22323/1.363.0239}{\emph{PoS} {\bfseries LATTICE2019}
  (2019) 239} [\href{https://arxiv.org/abs/1910.11745}{{\ttfamily
  1910.11745}}].

\bibitem{MILC:2010hzw}
{\scshape MILC} collaboration, A.~Bazavov et~al., \emph{{Results for light
  pseudoscalar mesons}}, \href{https://doi.org/10.22323/1.105.0074}{\emph{PoS}
  {\bfseries LATTICE2010} (2010) 074}
  [\href{https://arxiv.org/abs/1012.0868}{{\ttfamily 1012.0868}}].

\bibitem{Dowdall:2013rya}
R.~J. Dowdall, C.~T.~H. Davies, G.~P. Lepage and C.~McNeile, \emph{{$V_{us}$
  from $\pi$ and $K$ decay constants in full lattice QCD with physical $u$,
  $d$, $s$ and $c$ quarks}},
  \href{https://doi.org/10.1103/PhysRevD.88.074504}{\emph{Phys. Rev. D}
  {\bfseries 88} (2013) 074504}
  [\href{https://arxiv.org/abs/1303.1670}{{\ttfamily 1303.1670}}].

\bibitem{FlavourLatticeAveragingGroupFLAG:2021npn}
{\scshape Flavour Lattice Averaging Group (FLAG)} collaboration, Y.~Aoki
  et~al., \emph{{FLAG Review 2021}},
  \href{https://doi.org/10.1140/epjc/s10052-022-10536-1}{\emph{Eur. Phys. J. C}
  {\bfseries 82} (2022) 869}
  [\href{https://arxiv.org/abs/2111.09849}{{\ttfamily 2111.09849}}].

\bibitem{Bruno:2014ufa}
{\scshape ALPHA} collaboration, M.~Bruno, J.~Finkenrath, F.~Knechtli, B.~Leder
  and R.~Sommer, \emph{{Effects of Heavy Sea Quarks at Low Energies}},
  \href{https://doi.org/10.1103/PhysRevLett.114.102001}{\emph{Phys. Rev. Lett.}
  {\bfseries 114} (2015) 102001}
  [\href{https://arxiv.org/abs/1410.8374}{{\ttfamily 1410.8374}}].

\bibitem{Knechtli:2017xgy}
{\scshape ALPHA} collaboration, F.~Knechtli, T.~Korzec, B.~Leder and G.~Moir,
  \emph{{Power corrections from decoupling of the charm quark}},
  \href{https://doi.org/10.1016/j.physletb.2017.10.025}{\emph{Phys. Lett. B}
  {\bfseries 774} (2017) 649}
  [\href{https://arxiv.org/abs/1706.04982}{{\ttfamily 1706.04982}}].

\bibitem{Blum:2024drk}
{\scshape RBC, UKQCD} collaboration, T.~Blum et~al., \emph{{The long-distance
  window of the hadronic vacuum polarization for the muon $g-2$}},
  \href{https://arxiv.org/abs/2410.20590}{{\ttfamily 2410.20590}}.

\bibitem{Bazavov:2024eou}
{\scshape Fermilab Lattice, HPQCD, MILC} collaboration, A.~Bazavov et~al.,
  \emph{{Hadronic vacuum polarization for the muon $g-2$ from lattice QCD:
  Long-distance and full light-quark connected contribution}},
  \href{https://arxiv.org/abs/2412.18491}{{\ttfamily 2412.18491}}.

\bibitem{ExtendedTwistedMassCollaborationETMC:2024xdf}
{\scshape Extended Twisted Mass Collaboration (ETMC)} collaboration,
  C.~Alexandrou et~al., \emph{{Strange and charm quark contributions to the
  muon anomalous magnetic moment in lattice QCD with twisted-mass fermions}},
  \href{https://arxiv.org/abs/2411.08852}{{\ttfamily 2411.08852}}.

\bibitem{Hudspith:2024kzk}
R.~J. Hudspith, M.~F.~M. Lutz and D.~Mohler, \emph{{Precise Omega baryons from
  lattice QCD}},  \href{https://arxiv.org/abs/2404.02769}{{\ttfamily
  2404.02769}}.

\bibitem{Lingscheid2024}
S.~Lingscheid, ``{Nucleon mass and scale setting in lattice QCD using O($a$)
  improved $N_f = 2 + 1$ Wilson fermions}.'' Master Thesis, Johannes Gutenberg
  University Mainz, 2024.

\bibitem{MILC:2015tqx}
{\scshape MILC} collaboration, A.~Bazavov et~al., \emph{{Gradient flow and
  scale setting on MILC HISQ ensembles}},
  \href{https://doi.org/10.1103/PhysRevD.93.094510}{\emph{Phys. Rev. D}
  {\bfseries 93} (2016) 094510}
  [\href{https://arxiv.org/abs/1503.02769}{{\ttfamily 1503.02769}}].

\bibitem{ExtendedTwistedMass:2021qui}
{\scshape Extended Twisted Mass} collaboration, C.~Alexandrou et~al.,
  \emph{{Ratio of kaon and pion leptonic decay constants with $N_f=2+1+1$
  Wilson-clover twisted-mass fermions}},
  \href{https://doi.org/10.1103/PhysRevD.104.074520}{\emph{Phys. Rev. D}
  {\bfseries 104} (2021) 074520}
  [\href{https://arxiv.org/abs/2104.06747}{{\ttfamily 2104.06747}}].

\bibitem{Shintani:2019wai}
{\scshape PACS} collaboration, E.~Shintani and Y.~Kuramashi, \emph{{Hadronic
  vacuum polarization contribution to the muon $g-2$ with 2+1 flavor lattice
  QCD on a larger than (10 fm)$^4$ lattice at the physical point}},
  \href{https://doi.org/10.1103/PhysRevD.100.034517}{\emph{Phys. Rev. D}
  {\bfseries 100} (2019) 034517}
  [\href{https://arxiv.org/abs/1902.00885}{{\ttfamily 1902.00885}}].

\bibitem{Davies:2019efs}
{\scshape Fermilab Lattice, HPQCD, MILC} collaboration, C.~T.~H. Davies et~al.,
  \emph{{Hadronic-vacuum-polarization contribution to the
  muon\textquoteright{}s anomalous magnetic moment from four-flavor lattice
  QCD}}, \href{https://doi.org/10.1103/PhysRevD.101.034512}{\emph{Phys. Rev. D}
  {\bfseries 101} (2020) 034512}
  [\href{https://arxiv.org/abs/1902.04223}{{\ttfamily 1902.04223}}].

\bibitem{Giusti:2019xct}
{\scshape European Twisted Mass} collaboration, D.~Giusti, V.~Lubicz,
  G.~Martinelli, F.~Sanfilippo and S.~Simula, \emph{{Electromagnetic and strong
  isospin-breaking corrections to the muon $g - 2$ from Lattice QCD+QED}},
  \href{https://doi.org/10.1103/PhysRevD.99.114502}{\emph{Phys. Rev. D}
  {\bfseries 99} (2019) 114502}
  [\href{https://arxiv.org/abs/1901.10462}{{\ttfamily 1901.10462}}].

\bibitem{Hayakawa:2008an}
M.~Hayakawa and S.~Uno, \emph{{QED in finite volume and finite size scaling
  effect on electromagnetic properties of hadrons}},
  \href{https://doi.org/10.1143/PTP.120.413}{\emph{Prog. Theor. Phys.}
  {\bfseries 120} (2008) 413}
  [\href{https://arxiv.org/abs/0804.2044}{{\ttfamily 0804.2044}}].

\bibitem{deDivitiis:2013xla}
{\scshape RM123} collaboration, G.~M. de~Divitiis, R.~Frezzotti, V.~Lubicz,
  G.~Martinelli, R.~Petronzio, G.~C. Rossi et~al., \emph{{Leading isospin
  breaking effects on the lattice}},
  \href{https://doi.org/10.1103/PhysRevD.87.114505}{\emph{Phys. Rev. D}
  {\bfseries 87} (2013) 114505}
  [\href{https://arxiv.org/abs/1303.4896}{{\ttfamily 1303.4896}}].

\bibitem{Risch:2021hty}
A.~Risch and H.~Wittig, \emph{{Leading isospin breaking effects in the HVP
  contribution to $a_{\mu}$ and to the running of $\alpha$}},
  \href{https://doi.org/10.22323/1.396.0106}{\emph{PoS} {\bfseries LATTICE2021}
  (2022) 106} [\href{https://arxiv.org/abs/2112.00878}{{\ttfamily
  2112.00878}}].

\bibitem{Risch:2021nfs}
A.~Risch, \emph{{Isospin breaking effects in hadronic matrix elements on the
  lattice}}, Ph.D. thesis, Mainz U., 2021.
\newblock \href{https://doi.org/10.25358/openscience-6314}{{\ttfamily
  10.25358/openscience-6314}}.

\bibitem{Chao:2023lxw}
E.-H. Chao, H.~B. Meyer and J.~Parrino, \emph{{Coordinate-space calculation of
  QED corrections to the hadronic vacuum polarization contribution to
  $(g-2)_\mu$}}, \href{https://doi.org/10.22323/1.453.0256}{\emph{PoS}
  {\bfseries LATTICE2023} (2024) 256}
  [\href{https://arxiv.org/abs/2310.20556}{{\ttfamily 2310.20556}}].

\bibitem{Parrino:2025afq}
J.~Parrino, V.~Biloshytskyi, E.-H. Chao, H.~B. Meyer and V.~Pascalutsa,
  \emph{{Computing the UV-finite electromagnetic corrections to the hadronic
  vacuum polarization in the muon $(g-2)$ from lattice QCD}},
  \href{https://arxiv.org/abs/2501.03192}{{\ttfamily 2501.03192}}.

\bibitem{Biloshytskyi:2022ets}
V.~Biloshytskyi, E.-H. Chao, A.~G\'erardin, J.~R. Green, F.~Hagelstein, H.~B.
  Meyer et~al., \emph{{Forward light-by-light scattering and electromagnetic
  correction to hadronic vacuum polarization}},
  \href{https://doi.org/10.1007/JHEP03(2023)194}{\emph{JHEP} {\bfseries 03}
  (2023) 194} [\href{https://arxiv.org/abs/2209.02149}{{\ttfamily
  2209.02149}}].

\bibitem{Chao:2021tvp}
E.-H. Chao, R.~J. Hudspith, A.~G\'erardin, J.~R. Green, H.~B. Meyer and
  K.~Ottnad, \emph{{Hadronic light-by-light contribution to $(g-2)_\mu $ from
  lattice QCD: a complete calculation}},
  \href{https://doi.org/10.1140/epjc/s10052-021-09455-4}{\emph{Eur. Phys. J. C}
  {\bfseries 81} (2021) 651}
  [\href{https://arxiv.org/abs/2104.02632}{{\ttfamily 2104.02632}}].

\bibitem{Hoferichter:2023sli}
M.~Hoferichter, G.~Colangelo, B.-L. Hoid, B.~Kubis, J.~R. de~Elvira, D.~Schuh
  et~al., \emph{{Phenomenological Estimate of Isospin Breaking in Hadronic
  Vacuum Polarization}},
  \href{https://doi.org/10.1103/PhysRevLett.131.161905}{\emph{Phys. Rev. Lett.}
  {\bfseries 131} (2023) 161905}
  [\href{https://arxiv.org/abs/2307.02532}{{\ttfamily 2307.02532}}].

\bibitem{Bijnens:2016hgx}
J.~Bijnens and J.~Relefors, \emph{{Pion light-by-light contributions to the
  muon $g-2$}}, \href{https://doi.org/10.1007/JHEP09(2016)113}{\emph{JHEP}
  {\bfseries 09} (2016) 113}
  [\href{https://arxiv.org/abs/1608.01454}{{\ttfamily 1608.01454}}].

\bibitem{ParrinoLAT24}
V.~Biloshytskyi, E.-H. Chao, D.~Erb, F.~Hagelstein, H.~B. Meyer, J.~Parrino
  et~al., ``{UV-finite QED correction to the hadronic vacuum polarization
  contribution to $(g-2)_\mu$}.''
  \url{https://conference.ippp.dur.ac.uk/event/1265/contributions/7439/attachments/5627/7348/lattice2024_parrino.pdf},
  2024.

\bibitem{Bennett:2006fi}
{\scshape Muon $g-2$} collaboration, G.~W. Bennett et~al., \emph{{Final Report
  of the Muon E821 Anomalous Magnetic Moment Measurement at BNL}},
  \href{https://doi.org/10.1103/PhysRevD.73.072003}{\emph{Phys. Rev. D}
  {\bfseries 73} (2006) 072003}
  [\href{https://arxiv.org/abs/hep-ex/0602035}{{\ttfamily hep-ex/0602035}}].

\bibitem{Segner:2023igh}
A.~M. Segner, A.~Risch and H.~Wittig, \emph{{Precision Determination of Baryon
  Masses including Isospin-breaking}},
  \href{https://doi.org/10.22323/1.453.0044}{\emph{PoS} {\bfseries LATTICE2023}
  (2023) 044} [\href{https://arxiv.org/abs/2312.09065}{{\ttfamily
  2312.09065}}].

\bibitem{harris2020array}
C.~R. Harris, K.~J. Millman, S.~J. van~der Walt, R.~Gommers, P.~Virtanen,
  D.~Cournapeau et~al., \emph{Array programming with {NumPy}},
  \href{https://doi.org/10.1038/s41586-020-2649-2}{\emph{Nature} {\bfseries
  585} (2020) 357}.

\bibitem{maclaurin2015autograd}
D.~Maclaurin, D.~Duvenaud and R.~P. Adams, \emph{Autograd: Effortless gradients
  in numpy},  in \emph{ICML 2015 AutoML Workshop}, vol.~238, p.~5, 2015.

\bibitem{Hunter:2007}
J.~D. Hunter, \emph{Matplotlib: A 2d graphics environment},
  \href{https://doi.org/10.1109/MCSE.2007.55}{\emph{Computing in Science \&
  Engineering} {\bfseries 9} (2007) 90}.

\bibitem{gnuplot}
T.~Williams, C.~Kelley and {many others}, ``Gnuplot 4.6: an interactive
  plotting program.'' \url{http://gnuplot.sourceforge.net/}, April, 2013.

\bibitem{Tanabashi:2018oca}
{\scshape Particle Data Group} collaboration, M.~Tanabashi et~al.,
  \emph{{Review of Particle Physics}},
  \href{https://doi.org/10.1103/PhysRevD.98.030001}{\emph{Phys. Rev. D}
  {\bfseries 98} (2018) 030001}.

\bibitem{Bazavov:2017lyh}
A.~Bazavov et~al., \emph{{$B$- and $D$-meson leptonic decay constants from
  four-flavor lattice QCD}},
  \href{https://doi.org/10.1103/PhysRevD.98.074512}{\emph{Phys. Rev. D}
  {\bfseries 98} (2018) 074512}
  [\href{https://arxiv.org/abs/1712.09262}{{\ttfamily 1712.09262}}].

\bibitem{Miller:2020xhy}
N.~Miller et~al., \emph{{$F_K / F_\pi$ from M\"obius Domain-Wall fermions
  solved on gradient-flowed HISQ ensembles}},
  \href{https://doi.org/10.1103/PhysRevD.102.034507}{\emph{Phys. Rev. D}
  {\bfseries 102} (2020) 034507}
  [\href{https://arxiv.org/abs/2005.04795}{{\ttfamily 2005.04795}}].

\bibitem{Borsanyi:2012zs}
S.~Borsanyi, S.~D{\"u}rr, Z.~Fodor, C.~Hoelbling, S.~D. Katz et~al.,
  \emph{{High-precision scale setting in lattice QCD}},
  \href{https://doi.org/10.1007/JHEP09(2012)010}{\emph{JHEP} {\bfseries 1209}
  (2012) 010} [\href{https://arxiv.org/abs/1203.4469}{{\ttfamily 1203.4469}}].

\bibitem{Blum:2002ii}
T.~Blum, \emph{{Lattice calculation of the lowest order hadronic contribution
  to the muon anomalous magnetic moment}},
  \href{https://doi.org/10.1103/PhysRevLett.91.052001}{\emph{Phys. Rev. Lett.}
  {\bfseries 91} (2003) 052001}
  [\href{https://arxiv.org/abs/hep-lat/0212018}{{\ttfamily hep-lat/0212018}}].

\bibitem{Blossier:2010vz}
{\scshape Alpha} collaboration, B.~Blossier et~al., \emph{{HQET at order $1/m$:
  II. Spectroscopy in the quenched approximation}},
  \href{https://doi.org/10.1007/JHEP05(2010)074}{\emph{JHEP} {\bfseries 05}
  (2010) 074} [\href{https://arxiv.org/abs/1004.2661}{{\ttfamily 1004.2661}}].

\bibitem{DeGrand:1988vx}
T.~A. DeGrand, \emph{{A Conditioning Technique for Matrix Inversion for Wilson
  Fermions}}, \href{https://doi.org/10.1016/0010-4655(88)90180-4}{\emph{Comput.
  Phys. Commun.} {\bfseries 52} (1988) 161}.

\bibitem{Giusti:2008vb}
L.~Giusti and M.~L{\"u}scher, \emph{{Chiral symmetry breaking and the
  Banks--Casher relation in lattice QCD with Wilson quarks}},
  \href{https://doi.org/10.1088/1126-6708/2009/03/013}{\emph{JHEP} {\bfseries
  03} (2009) 013} [\href{https://arxiv.org/abs/0812.3638}{{\ttfamily
  0812.3638}}].

\bibitem{slepc-toms}
V.~Hernandez, J.~E. Roman and V.~Vidal, \emph{{SLEPc}: A scalable and flexible
  toolkit for the solution of eigenvalue problems},
  \href{https://doi.org/10.1145/1089014.1089019}{\emph{{ACM} Trans. Math.
  Software} {\bfseries 31} (2005) 351}.

\bibitem{slepc-manual}
J.~E. Roman, C.~Campos, E.~Romero and A.~Tomas, \emph{{SLEPc} users manual},
  Tech. Rep. DSIC-II/24/02 - Revision 3.14, D. Sistemes Inform\`atics i
  Computaci\'o, Universitat Polit\`ecnica de Val\`encia, 2020.

\bibitem{petsc-user-ref}
S.~Balay, S.~Abhyankar, M.~F. Adams, J.~Brown, P.~Brune, K.~Buschelman et~al.,
  \emph{{PETS}c users manual},  Tech. Rep. ANL-95/11 - Revision 3.11, Argonne
  National Laboratory, 2019.
\newblock \href{https://doi.org/10.2172/2205494}{{\ttfamily 10.2172/2205494}}.

\bibitem{petsc-efficient}
S.~Balay, W.~D. Gropp, L.~C. McInnes and B.~F. Smith, \emph{Efficient
  management of parallelism in object oriented numerical software libraries},
  in \emph{Modern Software Tools in Scientific Computing}, E.~Arge, A.~M.
  Bruaset and H.~P. Langtangen, eds., pp.~163--202, Birkh{\"{a}}user Press,
  1997, \href{https://doi.org/10.1007/978-1-4612-1986-6\_8}{{\ttfamily
  10.1007/978-1-4612-1986-6\_8}}.

\bibitem{Luscher:2007se}
M.~L{\"u}scher, \emph{{Local coherence and deflation of the low quark modes in
  lattice QCD}},
  \href{https://doi.org/10.1088/1126-6708/2007/07/081}{\emph{JHEP} {\bfseries
  07} (2007) 081} [\href{https://arxiv.org/abs/0706.2298}{{\ttfamily
  0706.2298}}].

\bibitem{openQCD}
M.~Lüscher and S.~Schaefer, ``\texttt{openQCD}.''
  \url{http://luscher.web.cern.ch/luscher/openQCD/}.

\bibitem{ETM:2008zte}
{\scshape ETM} collaboration, P.~Boucaud et~al., \emph{{Dynamical Twisted Mass
  Fermions with Light Quarks: Simulation and Analysis Details}},
  \href{https://doi.org/10.1016/j.cpc.2008.06.013}{\emph{Comput. Phys. Commun.}
  {\bfseries 179} (2008) 695}
  [\href{https://arxiv.org/abs/0803.0224}{{\ttfamily 0803.0224}}].

\bibitem{Bali:2009hu}
G.~S. Bali, S.~Collins and A.~Sch{\"a}fer, \emph{{Effective noise reduction
  techniques for disconnected loops in Lattice QCD}},
  \href{https://doi.org/10.1016/j.cpc.2010.05.008}{\emph{Comput. Phys. Commun.}
  {\bfseries 181} (2010) 1570}
  [\href{https://arxiv.org/abs/0910.3970}{{\ttfamily 0910.3970}}].

\bibitem{Blum:2012uh}
T.~Blum, T.~Izubuchi and E.~Shintani, \emph{{New class of variance-reduction
  techniques using lattice symmetries}},
  \href{https://doi.org/10.1103/PhysRevD.88.094503}{\emph{Phys. Rev. D}
  {\bfseries 88} (2013) 094503}
  [\href{https://arxiv.org/abs/1208.4349}{{\ttfamily 1208.4349}}].

\bibitem{Michael:1982gb}
C.~Michael and I.~Teasdale, \emph{{Extracting Glueball Masses From Lattice
  {QCD}}}, \href{https://doi.org/10.1016/0550-3213(83)90674-0}{\emph{Nucl.
  Phys. B} {\bfseries 215} (1983) 433}.

\bibitem{Luscher:1990ck}
M.~L{\"u}scher and U.~Wolff, \emph{{How to Calculate the Elastic Scattering
  Matrix in Two-dimensional Quantum Field Theories by Numerical Simulation}},
  \href{https://doi.org/10.1016/0550-3213(90)90540-T}{\emph{Nucl. Phys. B}
  {\bfseries 339} (1990) 222}.

\bibitem{Blossier:2009kd}
{\scshape ALPHA} collaboration, B.~Blossier, M.~Della~Morte, G.~von Hippel,
  T.~Mendes and R.~Sommer, \emph{{On the generalized eigenvalue method for
  energies and matrix elements in lattice field theory}},
  \href{https://doi.org/10.1088/1126-6708/2009/04/094}{\emph{JHEP} {\bfseries
  04} (2009) 094} [\href{https://arxiv.org/abs/0902.1265}{{\ttfamily
  0902.1265}}].

\bibitem{Blossier:2010mk}
{\scshape ALPHA} collaboration, B.~Blossier, M.~Della~Morte, N.~Garron, G.~von
  Hippel, T.~Mendes, H.~Simma et~al., \emph{{HQET at order 1/m: III. Decay
  constants in the quenched approximation}},
  \href{https://doi.org/10.1007/JHEP12(2010)039}{\emph{JHEP} {\bfseries 12}
  (2010) 039} [\href{https://arxiv.org/abs/1006.5816}{{\ttfamily 1006.5816}}].

\bibitem{Neil:2022joj}
E.~T. Neil and J.~W. Sitison, \emph{{Improved information criteria for Bayesian
  model averaging in lattice field theory}},
  \href{https://doi.org/10.1103/PhysRevD.109.014510}{\emph{Phys. Rev. D}
  {\bfseries 109} (2024) 014510}
  [\href{https://arxiv.org/abs/2208.14983}{{\ttfamily 2208.14983}}].

\bibitem{Erben:2019nmx}
F.~Erben, J.~R. Green, D.~Mohler and H.~Wittig, \emph{{Rho Resonance, Timelike
  Pion Form Factor, and Implications for Lattice Studies of the Hadronic Vacuum
  Polarisation}},
  \href{https://doi.org/10.1103/PhysRevD.101.054504}{\emph{Phys. Rev. D}
  {\bfseries 101} (2020) 054504}
  [\href{https://arxiv.org/abs/1910.01083}{{\ttfamily 1910.01083}}].

\end{thebibliography}\endgroup
